\documentclass[twocolumn,twocolappendix]{aastex631}

\usepackage{amsmath}
\shorttitle{SMA chemical survey of massive star-forming regions}
\shortauthors{Law et al.}
\graphicspath{{./}{figures/}}

\begin{document}

\title{A Wideband Chemical Survey of Massive Star-forming Regions at Subarcsecond Resolution with the Submillimeter Array}

\correspondingauthor{Charles J.\ Law}
\email{cjl8rd@virginia.edu}

\author[0000-0003-1413-1776]{Charles J.\ Law}
\altaffiliation{NASA Hubble Fellowship Program Sagan Fellow}
\affiliation{Department of Astronomy, University of Virginia, Charlottesville, VA 22904, USA}


\author[0000-0003-2384-6589]{Qizhou Zhang}
\affiliation{Center for Astrophysics $|$ Harvard \& Smithsonian, 60 Garden St., Cambridge, MA 02138, USA}

\author[0009-0005-9407-9278]{Arielle C. Frommer}
\affiliation{Center for Astrophysics $|$ Harvard \& Smithsonian, 60 Garden St., Cambridge, MA 02138, USA}

\author[0000-0001-8798-1347]{Karin I. \"Oberg}
\affiliation{Center for Astrophysics $|$ Harvard \& Smithsonian, 60 Garden St., Cambridge, MA 02138, USA}

\author[0000-0003-1480-4643]{Roberto Galv\'an-Madrid}
\affiliation{Instituto de Radioastronomía y Astrofísica, Universidad Nacional Autónoma de México, Morelia, Michoacán 58089, México}

\author{Eric Keto}
\affiliation{Center for Astrophysics $|$ Harvard \& Smithsonian, 60 Garden St., Cambridge, MA 02138, USA}

\author[0000-0003-2300-2626]{Hauyu Baobab Liu}
\affiliation{Department of Physics, National Sun Yat-Sen University, No. 70, Lien-Hai Road, Kaohsiung City 80424, Taiwan, R.O.C.}
\affiliation{Center of Astronomy and Gravitation, National Taiwan Normal University, Taipei 116, Taiwan}

\author[0000-0002-3412-4306]{Paul T. P. Ho}
\affiliation{Institute of Astronomy and Astrophysics, Academia Sinica, 11F of Astronomy--Mathematics Building, AS/NTU No. 1, Sec. 4, Roosevelt Rd., Taipei 10617, Taiwan, R.O.C.} 

\author[0000-0001-8446-3026]{Andr\'es F. Izquierdo}
\altaffiliation{NASA Hubble Fellowship Program Sagan Fellow}
\affiliation{European Southern Observatory, Karl-Schwarzschild-Straße 2, 85748 Garching bei München, Germany}
\affiliation{Leiden Observatory, Leiden University, 2300 RA Leiden, The Netherlands}
\affiliation{Department of Astronomy, University of Florida, Gainesville, FL 32611, USA}

\author[0000-0003-2076-8001]{L. Ilsedore Cleeves}
\affiliation{Department of Astronomy, University of Virginia, Charlottesville, VA 22904, USA}



\begin{abstract}
Massive star-forming regions exhibit a rich chemistry with complex gas distributions, especially on small scales. While surveys have yielded constraints on typical gas conditions, they often have coarse spatial resolution and limited bandwidths. Thus, to establish an interpretative framework for these efforts, detailed observations that simultaneously provide high sensitivity, spatial resolution, and large bandwidths for a subset of diverse sources are needed. Here, we present wideband~(${\approx}$32~GHz)~Submillimeter Array observations of four high-mass star-forming regions (G28.20-0.05,~G20.08-0.14~N,~G35.58-0.03,~W33~Main) at subarcsecond resolution, where we detect and spatially-resolve 100s of lines from over 60 molecules, including many complex organic molecules (COMs). The chemical richness of our sample is consistent with an evolutionary sequence from the line-rich hot cores and HC~\ion{H}{2}~regions of G28.20-0.05 and G20.08-0.14~N to the more chemically-modest UC~\ion{H}{2}~regions in G35.58-0.03, followed by the molecule-poor H~II~region W33~Main. We detect lines across a range of excitation conditions (E$_{\rm{u}}{\approx}$20~to~${\gtrsim}$800~K) and from numerous isotopologues, which enables robust estimates of gas properties. We derive nearly constant COM column density ratios that agree with literature values in other low- and high-mass protostellar cores, supporting the idea that COM abundances are set during the pre-stellar phase. In all regions, we identify spatial offsets among different molecular families, due to a combination of source physical structure and chemistry. In particular, we find potential evidence of carbon grain sublimation in G28.20-0.05 and identify an elemental oxygen gradient and rich sulfur-chemistry in G35.58-0.03. Overall, these results demonstrate that the SMA’s wide bandwidth is a powerful tool to untangle the complex molecular gas structures associated with massive star formation. 
\end{abstract}


\section{Introduction} \label{sec:intro}

The formation of massive stars (${>}$8~$M_{\odot}$) remains poorly understood due, in large part, to a variety of observational challenges, including their relative rarity, obscured nature, high gas optical depths, fast evolutionary timescales, and large distances (several to tens of kpcs) compared to their low-mass counterparts \citep[e.g.,][]{Beuther07, Tan14, Motte18}. Classically, a massive young stellar object (MYSO) is formed via gravitational collapse, which is then followed by the appearance of a compact, dense, and hot molecular core (HMC) \citep[e.g.,][]{Kurtz00}. These cores then continue to evolve into hypercompact (HC) \ion{H}{2} regions with electron densities of ${\sim}10^6$~cm$^{-3}$ and sizes of ${<}$0.05 pc \citep[e.g.,][]{Hoare07} and then into ultracompact (UC) \ion{H}{2} regions with electron densities of ${\sim}10^4$~cm$^{-3}$ and sizes of $\sim$0.1 pc as the UV radiation from the massive central star breaks out into the nearby environment. Ultimately, continued expansion will result in a fully-fledged \ion{H}{2} region, before the surrounding material is destroyed or dispersed by powerful stellar winds \citep[e.g.,][]{Kurtz00, Churchwell02}. However, in reality, the separation of these stages is not distinct and instead, several of these phases often exist at the same time within one region as multiple massive young stars are being formed in close proximity \citep[e.g.,][]{Motte22}. 

The emergence of young \ion{H}{2} regions is marked by bright infrared emission \citep[e.g.,][]{Csengeri17}, a variety of molecular masers \citep[e.g.,][]{Genzel77, Menten91, Argon00}, and broad (FWHM~$\gtrsim$~40~km~s$^{-1}$) radio recombination lines (RRLs) from ionized gas \citep[e.g.,][]{Sewilo04, Keto08, Sewilo11}. Hot cores trace individual massive protostars or low-order multiples and are identified via emission from complex organic molecules (COMs), which are either direct products of sublimated ice grain mantles or formed via gas-phase reactions \citep[e.g.,][]{Herbst09}. Given the multi-component and hierarchical nature of most massive star formation, warm (few 100~K) molecular gas and hot (${\sim}10^4$~K) ionized gas co-exist in relatively small regions. This often results in dynamical interactions between the two, e.g., in the form of smaller-scale outflows in ionized gas driving larger-scale molecular outflows. On the smallest scales, observational evidence now exists that massive protostars host Keplerian accretion disks in either molecular \citep[e.g.,][]{Ilee18, Sanna19, Lu22} or ionized gas \citep{Maud18, JimSerra20, GalvanMadrid23} depending on the mass of the central object.

The chemical complexity associated with massive star-forming regions thus provides an opportunity to directly trace the underlying physical gas conditions. However, this, in turn, requires: high angular resolution to localize distinct but contemporaneous evolutionary phases within a single star-forming region; fine velocity resolution to separate multiple gas kinematic components; and sufficiently large bandwidth to cover multiple molecular tracers and transitions across a wide range of excitation conditions to constrain gas physical conditions. Due to high gas densities, many commonly-observed dense gas tracers are optically thick, and it is thus also vital to simultaneously observe multiple isotopologues.

Only with such observations is it possible to confront a variety of outstanding questions, namely, what is the origin and frequency of spatial offsets between molecules of different chemical families \citep[e.g.,][]{Blake87, Wyrowski99, Allen17, Tercero18, Busch22, Mininni23}? What types of molecules serve as the best tracers for gas physical conditions, e.g., UV fields, temperature, density, the presence of shocks \citep[e.g.,][]{vanGelder22, Brouillet22}? How are hot cores related to the HC and UC \ion{H}{2} region evolutionary phases and how does the presence of nearby \ion{H}{2} regions influence hot core chemistry, e.g., via external heating \citep[e.g.,][]{Blake96, Zapata11, Wilkins22}? Which types of COM chemistry are driven by ice sublimation versus gas-phase formation \citep[e.g.,][]{Garrod08, Codella17, vant_Hoff_carbon_grains}; and how should recent observations of constant COM abundance ratios across a wide range of sources from pre-stellar cores to high-mass protostars be interpreted, e.g., is pre-stellar ice chemistry largely dominant in COM production \citep[][]{Scibelli21,Nazari22_NCOMS, Chen23}?

With these questions in mind, we present initial results from a wideband chemical survey of four massive star-forming regions at subarcsecond resolution with the Submillimeter Array (SMA). Overall, we detect and spatially-resolve 100s of lines from over 60 molecules, including multiple COMs and isotopologues. In Section \ref{sec:sources}, we briefly describe the properties of the four star-forming regions that comprise our survey and summarize the SMA observations in Section \ref{sec:observations_overview}. In Section \ref{sec:results}, we present our results, including line detection statistics, maps of molecular gas morphologies, and derived gas properties. In Section \ref{sec:discussion}, we discuss the origins of the observed emission, including spatial variations of gas conditions, COM abundance ratios, and how our sources compare to other low- and high-mass protostars. We summarize our conclusions in Section \ref{sec:conlcusions}.

\section{Source Description}
\label{sec:sources}

Our survey targeted four massive star-forming regions within ${\approx}$12~kpc with significant reservoirs of molecular gas and bolometric luminosities of a few-to-several times 10$^{5}$~L$_{\rm{\odot}}$. Our sources also span a range of evolutionary stages from young hot cores and HC~\ion{H}{2} regions to more evolved \ion{H}{2} regions. Table \ref{tab:source_prop} provides a summary of source properties. Below, we briefly describe each of the sources:

\textit{G28.20-0.05} (hereafter `G28'): G28 is an HC~\ion{H}{2} at a distance of 5.7~kpc \citep{Fish03, Law_Chi_22} surrounded by an infalling molecular gas torus with a central ionized core and an expanding shell of molecular gas driven by a wide-angle outflow \citep{Sollins05_G28, Klaassen09, Klaassen11}. Recent high-resolution ALMA observations show that G28 takes the form of a ring of ionized gas with a radius of ${\approx}$2000~au, while H30$\alpha$ kinematics indicate the presence of a rotating, ionized disk wind launched by the central massive protostar \citep{Law_Chi_22}. This ionized disk wind is driving a larger-scale molecular outflow observed in SiO emission along the NE-SW direction out to a distance of $\sim$0.2~pc \citep{Gorai23}. G28 appears to be forming in an isolated environment with no additional protostellar cores identified in its vicinity (${\sim}$0.1 to 0.4~pc) \citep{Law_Chi_22}. G28 has also been detected in several RRLs with broad line widths \citep{Sewilo04, Keto08, Sewilo08} and in multiple molecular masers \citep{Menten91, Han98, Argon00}, indicating its relative youth. A variety of hot-core-tracing molecules were observed toward G28 in early SMA observations \citep{Qin08}, and the presence of complex molecular gas distributions, including emission from COMs, was confirmed in subsequent ALMA observations \citep{Gorai23}. In particular, \citet{Gorai23} identified at least two distinct hot molecular cores (HMC1, HMC2) with high gas densities (n$_{\rm{H}} \sim 10^9$~cm$^{-3}$) and inferred excitation temperatures of ${\approx}$270-300~K. HMC1 is located along the millimeter continuum ring and is thought to be closest in projection to the protostar, while HMC2, located on the other side of the ring, instead traces an externally-heated concentration of molecular gas \citep{Gorai23}.

\textit{G20.08-0.14 N} (hereafter `G20'): G20 is a proto-OB-cluster at a distance of 12.3~kpc \citep{Fish03, Anderson09} comprising one UC~\ion{H}{2} and two HC~\ion{H}{2} regions first identified in cm observations \citep{Wood89}. Molecular masers \citep{Ho83_OH_maser, Hofner96, Walsh98} and broad-line RRLs \citep{Garay85, Sewilo04} have been detected toward G20. These UC~\ion{H}{2} and HC~\ion{H}{2} regions are surrounded by dense molecular gas \citep{Plume92, GalvanMadrid09} and are embedded within a rotating and contracting molecular cloud \citep[e.g.,][]{Klaassen07}. A hierarchical accretion flow that extends from parsec-scales down onto the individual UC~\ion{H}{2} and HC~\ion{H}{2} regions has also been identified \citep{GalvanMadrid09}. Emission from several hot core molecules is present in the brightest of these \ion{H}{2} regions \citep[region `A' in][]{GalvanMadrid09} and likely traces photoevaporation at the inner edge of a rotationally-flattened molecular accretion flow \citep{GalvanMadrid09, Xu13_G20}. There is also evidence of a collimated SiO outflow along the north-east to south-west direction that is centered at this bright \ion{H}{2} region \citep{Xu13_G20}. No molecular line emission was detected around the other two \ion{H}{2} regions \citep[regions `B' and `C' in][]{GalvanMadrid09}.

\textit{G35.58-0.03} (hereafter `G35'): G35 is a high-mass star-forming region at a distance of 10.2~kpc \citep{Fish03, Watson03} that consists of two closely separated (${\sim}$1$^{\prime \prime}$) UC~\ion{H}{2} regions, the western G35.578-0.030 and eastern G35.578-0.031, identified in cm observations \citep{Kurtz94}. For simplicity, we subsequently refer to these UC~\ion{H}{2} regions as `G35 West' and `G35 East', respectively. Water and OH masers have been detected in G35 \citep{Forster89, dBuizer05, Fish05} along with emisison from broad-line RRLs \citep{Zhang14}. There is a thick dust and infalling gas envelope surrounding the central UC~\ion{H}{2} regions in G35 \citep{Kurtz99, Zhang14}. Previous H30$\alpha$ observations also indicate the presence of an ionized outflow, which, in turn, is driving a collimated, bipolar molecular outflow along the east-west direction. The mass-loss rate associated with this outflow is nearly an order of magnitude less than that of the infall rate, which implies that the mass of the central proto-O-star is still increasing \citep{Zhang14}. Previous SMA observations, which did not resolve the two central UC~\ion{H}{2} regions, detected several hot-core-tracing molecules in G35 \citep{Zhang14}.

\textit{W33 Main} (hereafter `W33'; also G012.81-0.19): W33 Main is one of three molecular clumps in the larger W33 complex \citep[e.g.,][]{Schuller09, Immer14, Lin16}, which was first discovered as a thermal radio source \citep{Westerhout58} and is located at a distance of 2.4~kpc in the Scutum spiral arm of the Galaxy \citep{Immer13}. Water, methanol, and ammonia masers have been detected in W33~Main \citep[e.g.,][]{Genzel77, Menten86, Haschick90,Immer13, Tursun22} along with bright infrared emission \citep[e.g.,][]{Dyck77, Molinari10}, indicating the presence of a cluster of proto-OB stars. Several star clusters are located near W33~Main \citep{Morales13} and it is the only source in the W33 complex to show strong radio continuum \citep{Haschick83, Hoare12, Khan22}, which indicates that it is an evolved \ion{H}{2} region. Sub-millimeter observations revealed a compact continuum source \citep{Contreras13}, which was later resolved into a U-shaped feature \citep[`W33 Main-Central' in][]{Immer14} along with several additional nearby cores \citep{Immer14, Armante24}. Emission from SiO and H30$\alpha$ is only detected toward W33~Main, but none is observed from complex molecules, unlike other nearby regions within the W33 region \citep{Immer14}. This relative lack of emission from hot-core-tracing molecules is likely due to their destruction close to the central heating source(s) in W33~Main \citep{Immer14}.

\begin{deluxetable*}{cccccccc}
\tablecaption{Source Properties\label{tab:source_prop}}
\tablewidth{0pt}
\tablehead{
\colhead{Source} & \colhead{dist.}  & \colhead{D$_{\rm{GC}}$} & \colhead{v$_{\rm{sys}}$} & \colhead{L$_{\rm{bol}}$} & \colhead{M$_{*}$} & \colhead{M$_{\rm{env,gas}}$} & \colhead{Region}  \\ 
\colhead{} & \colhead{(kpc)}  & \colhead{(kpc)} & \colhead{(km~s$^{-1}$)} & \colhead{(L$_{\odot}$)} & \colhead{(M$_{\odot}$)} & \colhead{(M$_{\odot}$)} & \colhead{Type}
}
\startdata
G28.20-0.05 & 5.7$^{+0.5}_{-0.8}$~$^{[1]}$ & 4.6~$^{[1]}$ & 95.4~$^{[2]}$ & 1.4$\times10^5$~$^{[3]}$ & 40~$^{[3]}$ & 25~$^{[4]}$ & HC~\ion{H}{2}$+$HMC~$^{[4,5]}$ \\
G20.08-0.14 N & 12.3$^{+0.7}_{-1.2}$~$^{[1]}$ & 5.3~$^{[1]}$ & 42.0~$^{[6]}$ & 6.6~$\times10^5$~$^{[6,7]}$ & 34~$^{[6]}$ & 35--95~$^{[6]}$ & UC$+$HC~\ion{H}{2}~$^{[7]}$  \\
G35.58-0.03 & 10.2$^{+0.6}_{-0.6}$~$^{[8]}$ & 6.0~$^{[9]}$ & 52.5~$^{[8]}$ & 2.9--8.4$\times10^5$~$^{[10]}$ & 24--64~$^{[10]}$ & 277--441~$^{[10]}$ & UC~\ion{H}{2}~$^{[9]}$ \\
W33 Main & 2.4$^{+0.17}_{-0.15}$~$^{[11]}$ & 5.4~$^{[11]}$ & 34.1~$^{[11]}$ & 4.5~$\times 10^5$~$^{[12]}$ & 189~$^{[13]}$ & 3965~$^{[12]}$ & \ion{H}{2}$^{[14]}$ \\
\enddata
\tablecomments{References are: 1.~\citet{Fish03}; 2.~\citet{Qin08}; 3.~\citet{Law_Chi_22}; 4.~\citet{Gorai23}; 5.~\citet{Walsh03}; 6.~\citet{GalvanMadrid09}; 7.~\citet{Wood89}; 8.~\citet{Watson03}; 9.~\citet{Kurtz94}; 10.~\citet{Liu_SOMA_19};  11.~\citet{Immer13}; 12.~\citet{Immer14}; 13.~\citet{Beilis22}; 14.~\citet{Haschick83}. \\ \\
Distances were derived from \ion{H}{1} and H$_2$CO absorption studies in all sources except W33, which instead was computed via trigonometric parallaxes of water masers. Systemic velocities were measured using observed molecular lines. For all regions, bolometric luminosities were inferred from SED modeling, as were stellar masses with the exception of W33, where M$_*$ was derived via detailed 3D hydrodynamic and line profile simulations of [\ion{Ne}{2}] emission line data. Envelope masses were estimated from the dust emission in each region by assuming a gas-to-dust ratio of 100, with the exception of G35, where the reported M$_{\rm{env,gas}}$ values correspond to the range predicted by the five best-fit radiative transfer SED models.}
\end{deluxetable*}

\section{Observations}
\label{sec:observations_overview}

We observed each of the four massive star-forming regions with the SMA\footnote{The Submillimeter Array is a joint project between the Smithsonian Astrophysical Observatory and the Academia Sinica Institute of Astronomy and Astrophysics and is funded by the Smithsonian Institution and the Academia Sinica.} \citep{Ho04_SMA} using both the compact (COM) and very extended (VEX) array configurations between 12 March 2019 and 27 August 2019. The COM observations had typical baselines of ${\approx}$70~m, while the VEX configuration had baselines of ${\approx}$500~m. For those sources in the COM configuration observed on the same night, observing time was split equally between them. Table \ref{tab:calb_details} shows a summary of the observations, including observing dates, SMA configuration, number of available antennas, and PWV.

All observations used both the 230~GHz and 345~GHz receivers with the SWARM correlator, which at the time, provided a total frequency coverage of 32~GHz. The two sidebands of the lower frequency receiver covered 208.948–216.910~GHz and 224.944–232.907~GHz, while the higher frequency receiver was tuned to cover 332.908–340.870~GHz and 348.906–356.868~GHz. The native spectral resolution of the observations was 140~kHz (${\approx}$0.12-0.20~km~s$^{-1}$), but to reduce data volume, we binned by a factor of four for initial calibration and reduction, followed by a further smoothing to a uniform velocity resolution of 1.12~km~s$^{-1}$ during imaging. 

We calibrated the SMA data using the MIR software package\footnote{\url{https://lweb.cfa.harvard.edu/~cqi/mircook.html}}. Passband calibrations were performed using observations of the bright quasars 3C279 or 3C84. Depending on the source and observation, gain calibration was achieved using one or more of the following quasars 1743-038, 1751+096, and nrao530, while flux calibrators included Callisto, Ganymede, Neptune, and Titan. Table \ref{tab:calb_details} shows a detailed listing of all calibrators. 

The calibrated visibilities were then exported into CASA \texttt{v6.3} \citep{McMullin_etal_2007,CASATeam20} for continuum subtraction and imaging. For all sources, line-free channels were manually identified and we subtracted the continuum using the \texttt{uvcontsub} task with a first-order polynomial. We then used the \texttt{tclean} task with a Briggs weighting of \texttt{robust}=0.5 to generate images. Each 8~GHz spectral window was imaged individually. Images were generated with channel spacings of 1.12~km~s$^{-1}$ to maximize SNR but ensure line profiles were sufficiently well-resolved. All images were made using the ‘multi-scale` deconvolver with pixel scales of [0,8,24,40,100,300] and were CLEANed down to a 3$\sigma$ level, where $\sigma$ was measured over five line-free channels of the dirty image. We also made 230~GHz and 345~GHz continuum images with the full bandwidth of the observations after flagging channels containing line emission using the same CLEANing parameters. Table \ref{tab:image_info} lists image cube properties for each source per sideband, including the synthesized beam size and measured RMS noise.

\begin{figure*}
\centering
\includegraphics[width=\linewidth]{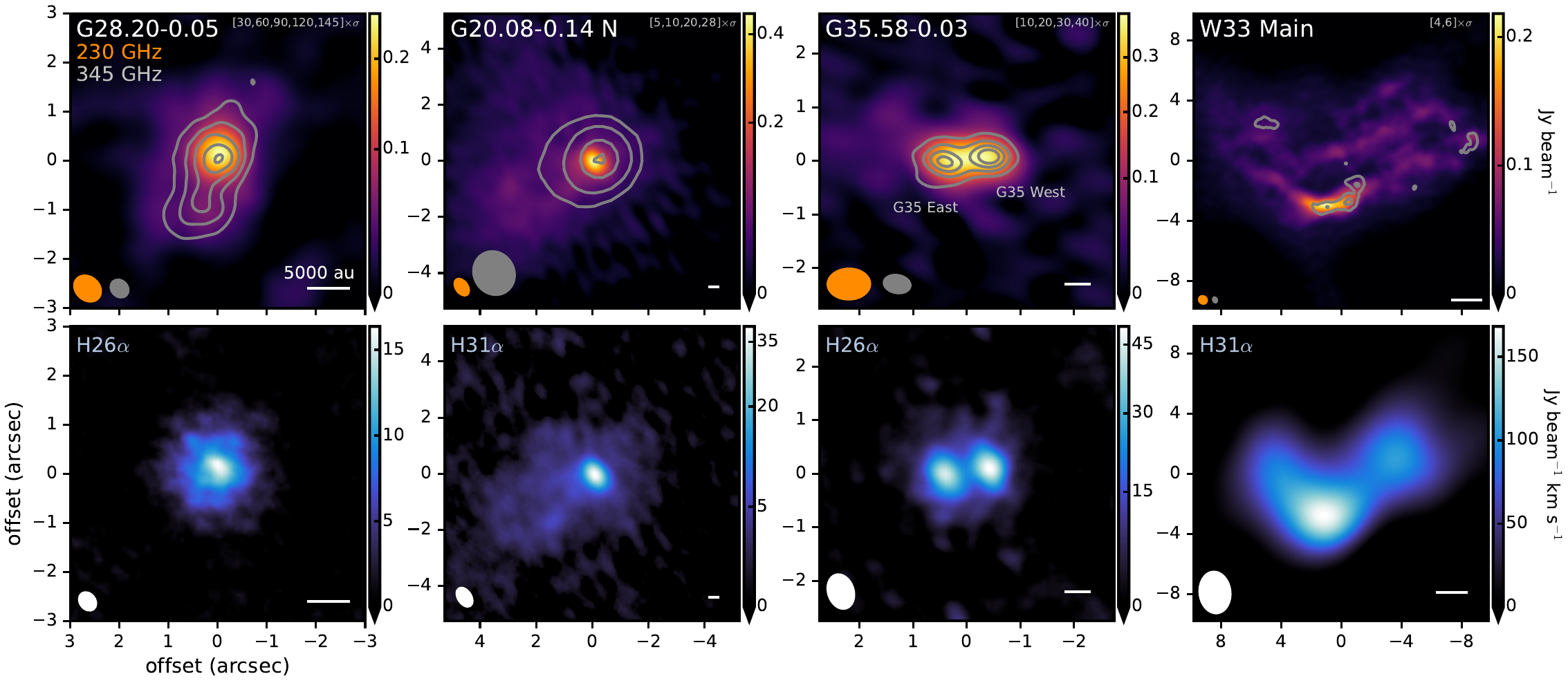}
\caption{Gallery of continuum images \textit{(top row)} and RRL integrated-intensity maps \textit{(bottom row)} for the SMA sample. Colorscale shows the 230~GHz continuum and gray contours show the 345~GHz continuum. The synthesized beam and a scale bar indicating 5000~au is shown in the lower left and right corner, respectively, of each panel.}
\label{fig:continuum_gallery}
\end{figure*}

Due to the range of on-source observing times, number of available antennas, weather conditions, and observed frequencies, image cubes exhibited a range of properties. Synthesized beam sizes range from 0\farcs5 to over 2$^{\prime \prime}$, which corresponds to physical scales of ${\approx}$2000-3000~au in G28, ${\approx}$5000-6000~au in W33, and ${\approx}$10,000-20,000~au in G20 and G35. The line RMS noise values are between 23~mJy~beam$^{-1}$ and 455~mJy~beam$^{-1}$. In particular, the high frequency image cubes of G20 have considerably coarser beam sizes than those of the lower frequencies, since several antennas associated with the 345~GHz receiver were flagged during calibration due to unstable phases during the VEX observations. W33 suffers from spatial filtering even in the combined COM+VEX data due its close distance and the extended nature of its emission (${\gtrsim}$10$^{\prime \prime}$) \citep[e.g.,][]{Immer14}. It is also the southernmost source in our sample, which resulted in an elongated beam shape. While we initially explored a range of \texttt{robust} values and \textit{uv}-tapers in the CLEANing of W33 image cubes, we did not find any significant differences in either the line detection statistics or emission morphology. Thus, for consistency, we decided to adopt the same imaging parameters as for the other sources and instead smoothed all line image cubes of W33 using \texttt{imsmooth} with a circular beam of 2$^{\prime \prime}$ to boost low SNR line detections. The continuum images and continuum-subtracted line emission image cubes as well as the calibrated measurement sets for all sources are publicly available on Zenodo doi: 10.5281/zenodo.13342640.

\begin{deluxetable*}{cccccccccc}
\tablecaption{Observational Details\label{tab:calb_details}}
\tablewidth{0pt}
\tablehead{
\colhead{Source} & \colhead{R.A.} & \colhead{Dec.} & \colhead{UT Date} & \colhead{Num.} & \colhead{SMA} & \colhead{$\tau_{\rm{225GHz}}$} & \multicolumn3c{Calibrators} \\ \cline{8-10}
\colhead{} & \colhead{(J2000)} & \colhead{(J2000)} & \colhead{} & \colhead{Ant.\tablenotemark{a}} & \colhead{config.} &\colhead{(mm)} & \colhead{Flux} & \colhead{Passband} & \colhead{Gain}
}
\startdata
G28.20-0.05  & 18:42:58.10 & $-$04:13:57.87 & 2019 Apr 25 & 6 & COM & 0.07 & Callisto & 3C279 & 1743-038,1751+096  \\
                         &                     &                      & 2019 Apr 26	& 7 & COM & 0.06 & Callisto & 3C279 & 1743-038,1751+096  \\
                         &                    &                        & 2019 Mar 12 & 8 & VEX & 0.06 & Ganymede & 3C279 & 1743-038,1751+096 \\
                          &                    &                       & 2019 Jun 15 & 8 & VEX & 0.10 & Callisto & 3C279 & 1743-038,1751+096 \\
G20.08-0.14 N &  18:28:10.40 & $-$11:28:49.00 & 2019 Apr 19 & 7 & COM & 0.04 & Callisto & 3C279 & 1743-038  \\
                           &                       &                           & 2019 Jun 14 & 8\tablenotemark{b} & VEX & 0.07 & Titan & 3C279 & 1743-038,nrao530 \\
G35.58-0.03 & 18:56:22.53 & $+$02:20:27.00 & 2019 Apr 25 & 6 & COM & 0.07 & Callisto & 3C279 & 1743-038,1751+096 \\
             &                        &                       & 2019 Aug 27 & 8  & COM & 0.09 & Neptune & 3C84 & 1743-038,1751+096\\ 
             &                       &                   & 2019 Mar 18 & 8 & VEX & 0.06 & Callisto & 3C279 & 1743-038,1751+096\\             
W33 Main &  18:14:13.67 & $-$17:55:40.00 & 2019 Apr 19 & 7 & COM & 0.04 & Callisto & 3C279 & 1743-038 \\
              &                      &                     & 2019 Mar 15 & 8 & VEX & 0.04 & Callisto & 3C279 & 1743-038,nrao530 \\
\enddata
\tablenotetext{a}{Number of antennas remaining after flagging.}
\tablenotetext{b}{An additional three antennas were flagged due to unstable phases in the 345~GHz receiver. This resulted in correspondingly degraded angular resolution in the final images (see Table \ref{tab:image_info}).}
\end{deluxetable*}

\begin{deluxetable*}{lccccc}[ht!]
\tablecaption{Image Cube Properties\label{tab:image_info}}
\tablehead{
\colhead{Frequency Range} & \colhead{Beam} & \colhead{Proj. Linear Res.}  & \colhead{$\delta v$} & \colhead{RMS\tablenotemark{a}}  \vspace{-0.15cm}\\
\colhead{(GHz)} & \colhead{ ($^{\prime \prime}$~$\times$~$^{\prime \prime}$, $\deg$)} & \colhead{(au~$\times$~au) }& \colhead{(km~s$^{-1}$)} & \colhead{(mJy~beam$^{-1}$)} }
\startdata
\textbf{G28.20-0.05:} \\
Cont., 230~GHz & 0.61~$\times$~0.49, 48.6 & 3500~$\times$~2800 & \ldots & 1.0 \\
Cont., 345~GHz & 0.41~$\times$~0.34, 44.6 & 2300~$\times$~1900 & \ldots & 1.3 \\
208.948--216.910 & 0.64~$\times$~0.51, 43.1 & 3600~$\times$~2900 & 1.12 & 23.2 & \\
224.944--232.907 & 0.62~$\times$~0.49, 43.6 & 3500~$\times$~2800 & 1.12 & 23.2 & \\
332.908--340.870 & 0.42~$\times$~0.36, 45.1 & 2400~$\times$~2100 & 1.12 & 62.5 & \\
348.906--356.868 & 0.42~$\times$~0.33, 36.8 & 2400~$\times$~1900 & 1.12 & 90.0 &\\ \\
\textbf{G20.08-0.14 N:}\\
Cont., 230~GHz & 0.69~$\times$~0.45, 35.0 & 8500~$\times$~5500 & \ldots & 10.5 \\
Cont., 345~GHz & 1.62~$\times$~1.46, 30.1 & 20000~$\times$~18000 & \ldots & 51.3 \\
208.948--216.910 & 0.74~$\times$~0.48, 33.3 & 9100~$\times$~5900 & 1.12 & 32.6 \\
224.944--232.907 & 0.71~$\times$~0.46, 33.6 & 8700~$\times$~5700 & 1.12 & 34.3 \\
332.908--340.870 & 1.72~$\times$~1.54, 22.2 & 21000~$\times$~19000 & 1.12 & 114.8 \\
348.906--356.868 & 1.67~$\times$~1.48, 31.1 & 21000~$\times$~18000 & 1.12 & 148.8 \\ \\
\textbf{G35.58-0.03:} \\
Cont., 230~GHz & 0.81~$\times$~0.59, $-$89.2 & 8300~$\times$~6000 & \ldots & 3.8 \\
Cont., 345~GHz & 0.52~$\times$~0.35, 79.9 & 5300~$\times$~3600 & \ldots & 7.0 \\
208.948--216.910 & 1.35~$\times$~1.27, $-$78.1 & 14000~$\times$~13000 & 1.12 & 32.4\\
224.944--232.907 & 1.25~$\times$~1.19, 13.6 & 13000~$\times$~12000 & 1.12 & 34.2 \\
332.908--340.870 & 0.77~$\times$~0.49, 11.89 & 7900~$\times$~5000 & 1.12 & 72.0 \\
348.906--356.868 & 0.65~$\times$~0.49, 20.3 & 6600~$\times$~5000 & 1.12 & 107.3 \\ \\
\textbf{W33 Main:} \\
Cont., 230~GHz & 0.60~$\times$~0.54, 48.9 & 1400~$\times$~1300 & \ldots & 7.0 \\
Cont., 345~GHz & 0.41~$\times$~0.30, 19.2 & 980~$\times$~720 & \ldots & 18.0 \\
208.948--216.910 & 2.82~$\times$~2.09, 6.7 & 6800~$\times$~5000 & 1.12 & 64.1 \\
224.944--232.907 & 2.69~$\times$~2.08, 6.8 & 6500~$\times$~5000 & 1.12 & 64.9\\
332.908--340.870 & 2.04~$\times$~2.03, 18.3 & 4900~$\times$~4900 & 1.12 & 373.2\\
348.906--356.868 & 2.04~$\times$~2.02, 18.6 & 4900~$\times$~4800 & 1.12 & 454.8 \\ 
\enddata
\tablenotetext{a}{The line RMS noise is calculated over five line-free channels of the dirty image, while the continuum RMS noise is computed in a circular region with a radius three times the beam major axis in an emission-free region of the dirty image.}
\tablecomments{All images were generated with a \texttt{robust} parameter of 0.5.}
\end{deluxetable*}

\section{Results} \label{sec:results}

\subsection{Continuum and RRL Morphology} \label{sec:emission_morphology}

All of our sources show strong emission from multiple RRLs, including H$\alpha$, H$\beta$, and H$\gamma$, while lines of H$\delta$ and He$\alpha$ were detected in at least two sources. These RRLs trace ionized gas and thus, the spatial distribution of RRL emission provides useful context for where the strongest radiation fields and hottest gas are located when later interpreting observations of the molecular gas.

Figure \ref{fig:continuum_gallery} shows continuum and H$\alpha$ integrated-intensity maps in our sample. We choose the particular RRL which has the highest angular resolution (H26$\alpha$ in G28 and G35; H31$\alpha$ in G20 and W33). In all cases, the continuum and RRLs have broadly similar emission morphologies, which is consistent with a significant fraction of the continuum emission originating from free-free emission. Given the powerful outflows present in these regions, a portion of this free-free emission may also originate from ionized jets \citep[e.g.,][]{Purser16}. Free-free contributions to the total 1.3~mm flux in our sample have been previously estimated to range from ${\sim}$25\% to 80\% depending on the particular source \citep[e.g.,][]{GalvanMadrid09, Zhang14, Immer14, Law_Chi_22,Armante24}. G35 is clearly resolved into two distinct UC~\ion{H}{2} regions in both RRL and continuum emission, while G28 only exhibits a central bright core at the location of its HC~\ion{H}{2} region. However, high-resolution ALMA observations show that this core in G28 is, in fact, an ionized ring that is not resolved with SMA \citep{Law_Chi_22}. Both G28 and G35 show larger scale diffuse continuum emission, likely due to contributions from thermal dust emission \citep{Hernandez14, Zhang14}. G20 has a central compact core with extended, diffuse continuum emission extending to the south-east. This core is tracing a known bright, molecule-rich HC~\ion{H}{2} region, while the diffuse emission traces a nearby, extended, but molecule-poor HC~\ion{H}{2} region \citep{GalvanMadrid09}. In contrast to G28 and G35, this diffuse emission component is also seen in H31$\alpha$ in G20, which suggests an extended reservoir of ionized gas, consistent with the presence of nearby \ion{H}{2} regions \citep{GalvanMadrid09}. W33 shows the most complex continuum structure of our sources with a bright U-shaped feature surrounded by diffuse, arc-like structures that are reminiscent of those seen in the massive star-forming region G10.6-0.4 \citep{Sollins05, Law_G10p6}, which have been interpreted as the ionized edges of clumps of molecular material. The RRL emission shows the same broad U-shaped-structure, but we cannot determine if similar arc-like structures are present in H31$\alpha$ on small scales due to the lower SNR and coarse beam sizes in the line image cubes of W33.

\subsection{Molecular Inventory} \label{sec:molecule_detections_HEADER}

\subsubsection{Summary of Molecule Detections} \label{sec:molecule_detections}

One of the primary goals of this work is to determine the chemical composition of each of our sources. To do so, we first aim to establish a comprehensive catalog of detected molecules. Here, we focus on those molecular transitions that are sufficiently bright to allow for an investigation of their spatial distribution and derivation of gas properties. Thus, we only consider transitions which have an absolute peak line intensity of ${\gtrsim}$3$\sigma$. We consider a molecule detected if it has one line with a velocity-integrated intensity exceeding 5$\sigma$ or at least two lines with total flux greater than 3$\sigma$. We excluded significantly blended lines during this process, which resulted in the removal of a few 10s of lines, or less than 5\% of total identified lines. In practice, due to our initial peak flux cut, the vast majority of molecules are securely detected via the presence of a single, bright (${>}$10$\sigma$) line (e.g., C$^{17}$O, $^{13}$CS, HCO$^+$) or numerous (10s of lines) lines at modest SNR (3-5$\sigma$). Molecule detections are compiled from spectra extracted within one beam at the continuum peaks of all sources, except for W33, where they represent the set of unique detections from spectra extracted at both the molecule-rich and RRL emission peak regions (see Appendix \ref{sec:appendix:w33}). Figures \ref{fig:SMA_spec_rx240} and \ref{fig:SMA_spec_rx345} show representative spectra from the 230~GHz and 345~GHz receivers, respectively, which illustrate the line-rich nature of our sample. Appendix \ref{sec:full_spectra} shows the full set of line identifications. Subsequently, we often classify molecule detections and gas properties according to molecular families, namely, those molecules containing oxygen (``O-bearing"), sulfur (``S-bearing"), or nitrogen (``N-bearing") atoms, as well as hydrocarbons, which consist of only hydrogen and carbon atoms.

\begin{figure*}[!p]
\centering
\includegraphics[width=0.975\linewidth]{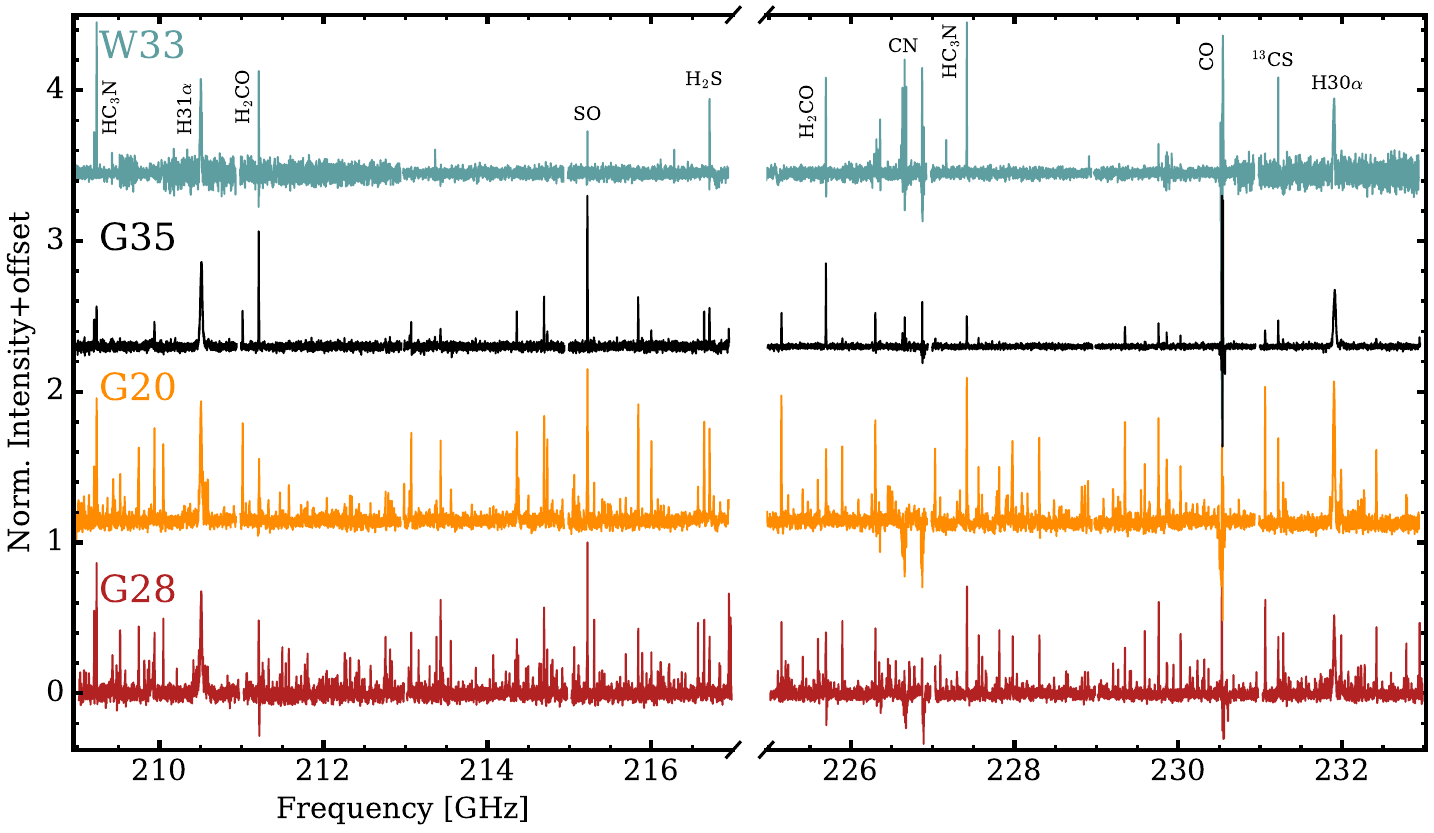}
\caption{SMA spectra from ${\approx}$209 to 233~GHz of our sample extracted within one beam at the continuum peak of each source. There are small gaps in spectral coverage at the edges of each sideband. Sources are ordered by evolutionary stage from top (most evolved) to bottom (youngest). A few selected lines are labeled.}
\label{fig:SMA_spec_rx240}
\end{figure*}

\begin{figure*}[!p]
\centering
\includegraphics[width=0.975\linewidth]{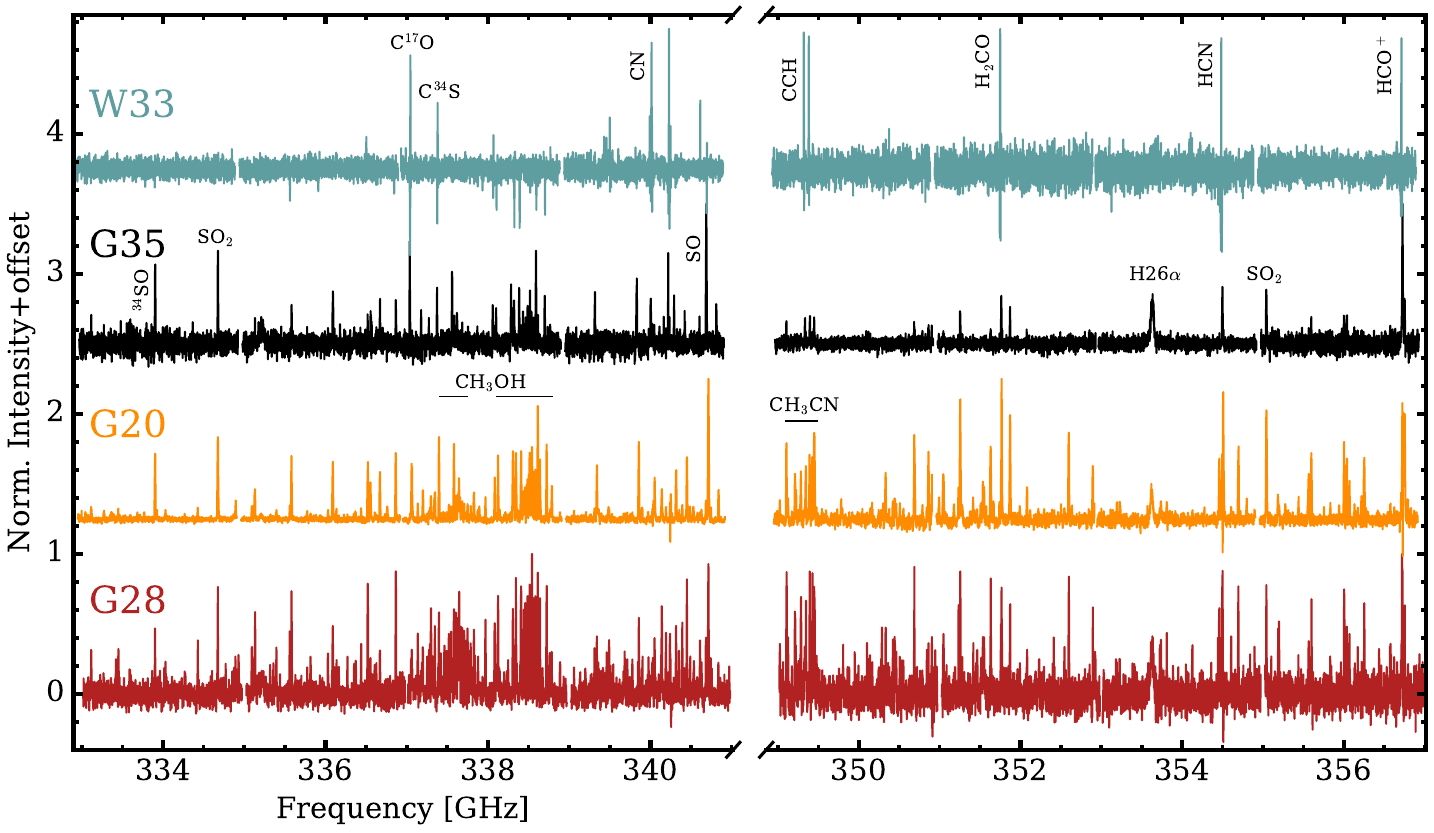}
\caption{SMA spectra from ${\approx}$333 to 357~GHz of our sample. Otherwise, as in Figure \ref{fig:SMA_spec_rx240}.}
\label{fig:SMA_spec_rx345}
\end{figure*}

\begin{figure*}[p!]
\centering
\includegraphics[width=0.8\linewidth]{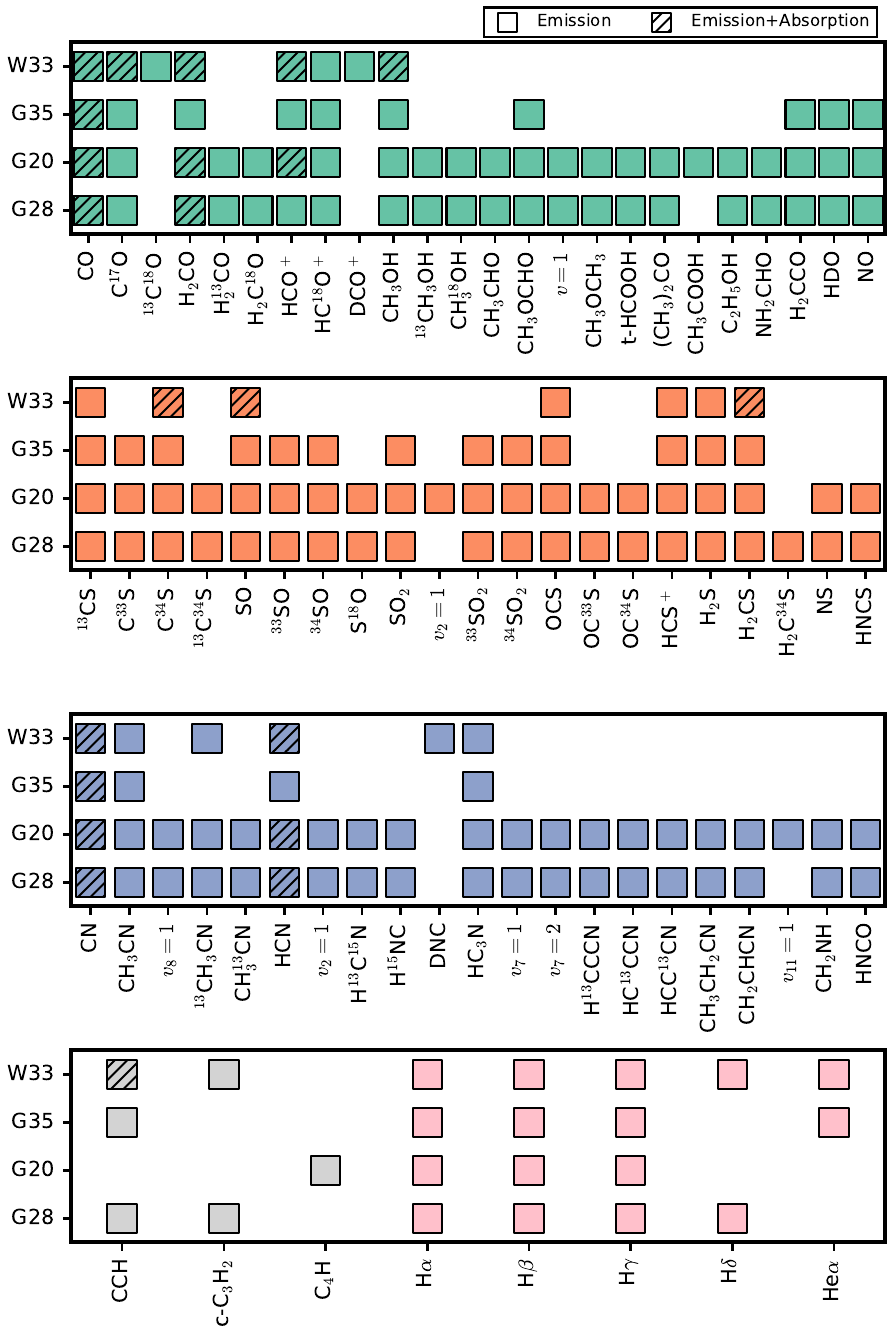}
\caption{Gallery of molecular and RRL detections in the SMA sample which indicate if the species is detected in emission (solid box) or in both emission and absorption (diagonal lines). No molecules were detected only in absorption. Molecules are divided and colored as O-bearing (green), S-bearing (orange), N-bearing (blue), and hydrocarbons (gray). The detected RRLs are colored in pink. All listed transitions are in the $v=0$ state, unless otherwise noted. Transitions with $v=1,2$, etc., are shown immediately to the right of the molecule name.}
\label{fig:molecule_detections}
\end{figure*}

Figure \ref{fig:molecule_detections} shows a summary of molecular detections per source. The majority of molecules are only detected in emission but a few (CO, H$_2$CO, CN, HCN, HCO$^+$) consistently show both emission and absorption components. The CO, and to a lesser extent, CN, line profiles are heavily influenced by the line-of-sight gas distribution as indicated by the presence of multiple distinct kinematic components, while H$_2$CO, HCO$^+$, and HCN show evidence of complex gas motions, i.e., inflows/outflows. W33 is the only source in which deuterated molecules (DCO$^+$, DNC) are detected and shows additional absorption signatures in several molecules (C$^{17}$O, CH$_3$OH, C$^{34}$S, SO, H$_2$CS, CCH) not seen in other sources (see Appendix \ref{sec:appendix:w33} for more details). The non-detection of certain molecules is, in some cases, likely due to strong line blending. For instance, while $^{13}$C$^{18}$O appears unique to W33, the single transition covered in our observations is blended with strong COM emission in G20 and G28, which is otherwise absent in W33 allowing us to only securely detect it here.

G28 and G20 have the most molecule detections (62 and 63, respectively), followed by the more modest G35 (28), and then the relatively molecule-poor W33 (23). G28 and G20 are chemically-rich in all types of molecules, including S-, O-, and N-bearing species. They also have numerous COM detections and appear similar in their overall chemical inventories. G35 has less than half of the total molecular detections of that of either G20 or G28 and does not show significant emission from large N- and O-bearing COMs, with H$_2$CCO and CH$_3$OCHO being the most complex molecules detected. Nonetheless, G35 does host a surprisingly rich sulfur chemistry that is nearly the same as G20 and G28 with the exception of a few rare isotopologues and hot gas tracers. W33 is the most molecule-poor source in our sample, with the simple COMs CH$_3$OH and CH$_3$CN being the most complex molecules detected. Emission from H$\alpha$, H$\beta$, and H$\gamma$ are detected in all sources, while H$\delta$ (G28, W33) and He$\alpha$ (G35, W33) are only detected in a subset of sources.

\subsubsection{Summary of Line Detections} \label{sec:line_detections}

In addition to establishing the molecular inventory of each source, our observations were designed to include multiple lines with a wide range of upper state energies (E$_{\rm{u}}$) to probe a variety of physical gas conditions. 

Figure \ref{fig:Eus} shows the distribution of detected lines as a function of E$_{\rm{u}}$. Overall, we detected ${\approx}$600, ${\approx}$400, ${\approx}$100, and ${\approx}$80 lines in G28, G20, G35, and W33, respectively. While G20 and G28 look similar in their overall line E$_{\rm{u}}$ distributions, there are a few notable differences. In G20, we often detect additional higher E$_{\rm{u}}$ lines relative to G28, particularly in the S-bearing species, while in G28, we consistently have more line detections of larger O-bearing COMs. For those species detected in all sources, we see the fewest number of lines on a per-molecule basis in W33, followed by G35, and those that are detected typically have lower E$_{\rm{u}}$ values. For instance, for CH$_3$OH, most detected lines in W33 and G35 are $E_{\rm{u}}<$ 300~K and $E_{\rm{u}}<$ 500~K, respectively, but in both G20 and G28, lines are consistently detected up to $E_{\rm{u}}\approx$800~K. A similar difference is seen in CH$_3$CN with only a few lower temperature transitions detected in G35 and W33, which taken together, indicates lower gas temperatures and/or efficient destruction (see Section \ref{sec:Trot_and_Ncol} for more details).

Figure \ref{fig:Eus} is not intended to provide an exhaustive line census, since there are numerous faint lines that do not meet our peak line intensity cutoff in G20 and G28 (see Appendix \ref{sec:full_spectra}), but rather is meant as a catalog of those lines that are minimally blended, sufficiently bright to map the emission morphology, and can be used to extract robust flux measurements for excitation analysis in the following Sections. While shown in Figure \ref{fig:Eus} for comparison, for this reason, we excluded W33 from all subsequent analysis due to its substantial spatial filtering (see Section \ref{sec:observations_overview}).

\begin{figure*}[p!]
\centering
\includegraphics[width=\linewidth]{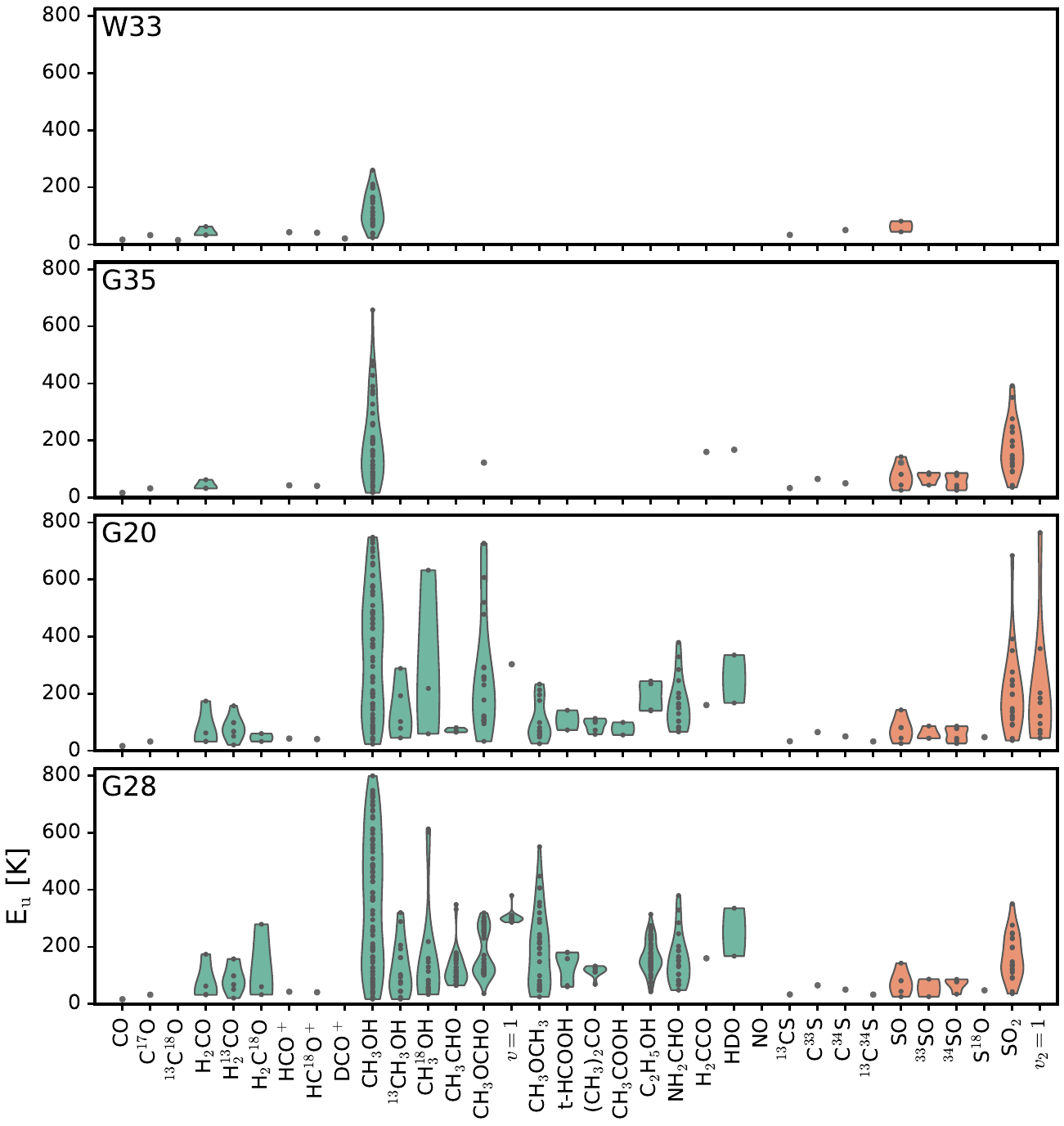}
\caption{Violin plot of detected transitions as a function of E$_{\rm{u}}$ for G28, G20, G35, and W33. The width of each violin shows a kernel density estimation of the underlying E$_{\rm{u}}$ distribution for that molecule. Line detections are compiled from beam-averaged spectra at the continuum peaks of all sources, except for W33, where they represent the set of unique detections at both the molecule-rich and RRL emission peak regions (see Appendix \ref{sec:full_spectra}). Molecules are colored as in Figure \ref{fig:molecule_detections}.}
\label{fig:Eus}
\end{figure*}

\setcounter{figure}{4}
\begin{figure*}[p!]
\centering
\includegraphics[width=\linewidth]{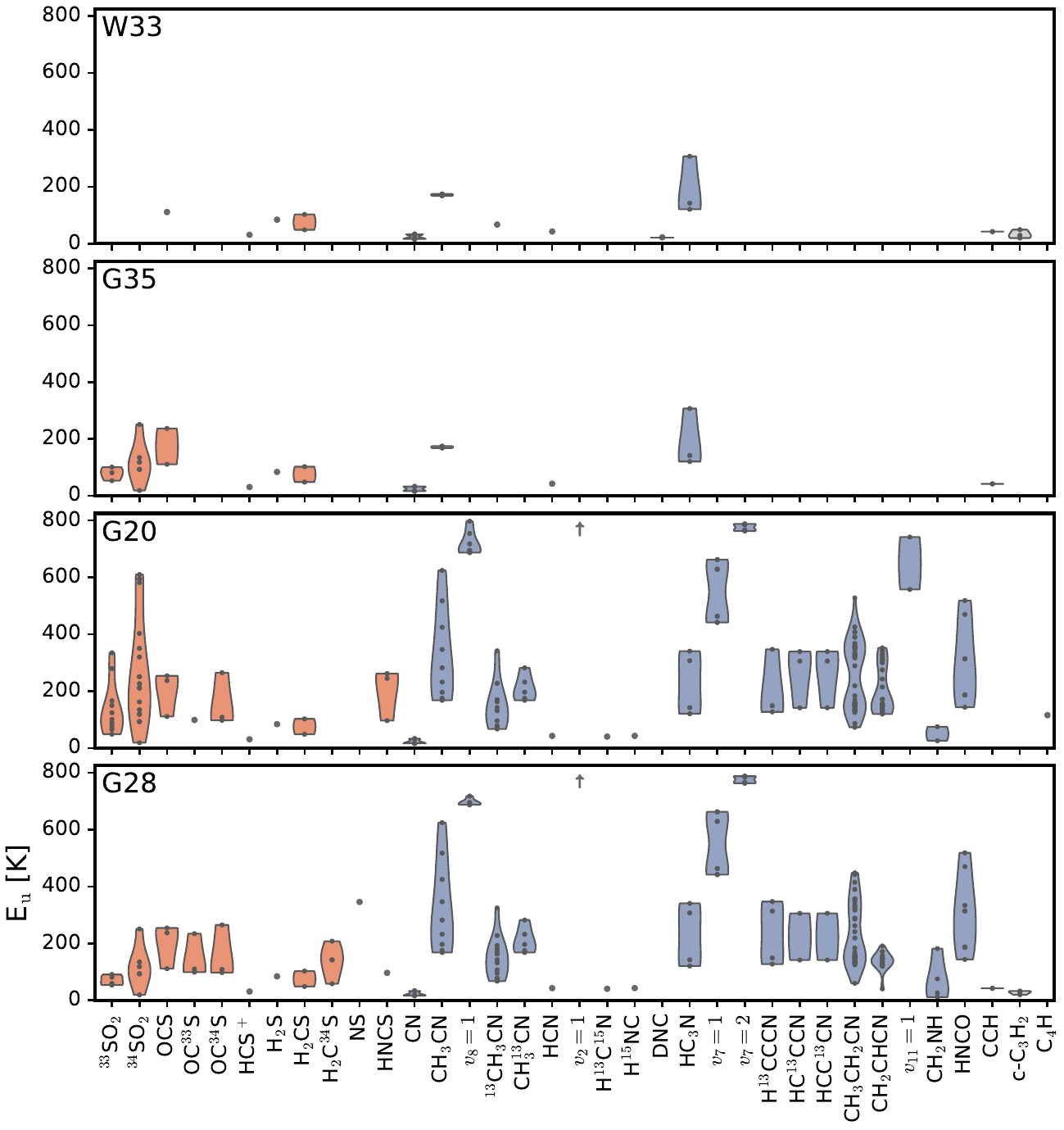}
\caption{\textit{(continued)} The detected HCN $v_2=1$ line, which has an E$_{\rm{u}}$=1067~K, is shown as a upward-facing arrow for visual clarity.}
\label{fig:Eus}
\end{figure*}

\subsection{Molecular Line Emission Morphology} \label{sec:molecule_morp}

\begin{figure*}[b]
\centering
\includegraphics[width=\linewidth]{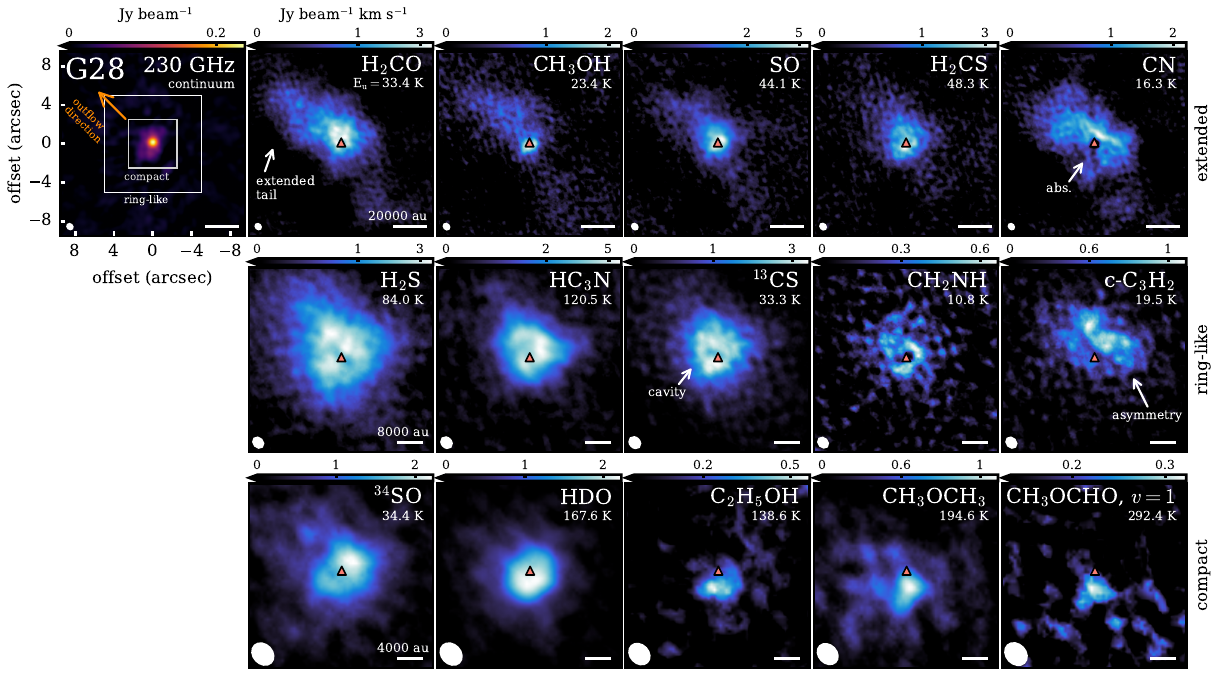}
\caption{Zeroth moment maps of representative lines demonstrating the multiple emission components that comprise G28. Salient emission features are labeled with arrows. The upper state energy of each transition is labeled in the upper right. The continuum peak is shown as a pink triangle and the synthesized beam is shown in the lower left of each panel. The boxed regions in the continuum image show zoom-ins with corresponding scale bars denoting 20000~au, 8000~au, or 4000~au depending on the spatial scale, respectively, in each panel. The outflow direction, as traced by SiO emission \citep{Gorai23}, is marked in the continuum image.}
\label{fig:g28_mom0_gallery_summary}
\end{figure*}

Nearly all of the detected molecular transitions are spatially resolved with the SMA. To visualize the line emission morphologies, we generated maps of velocity-integrated intensity, or ``zeroth moment maps," from the image cubes of each detected transition using the Python package \texttt{bettermoments} \citep{Teague18}. Figures \ref{fig:g28_mom0_gallery_summary}, \ref{fig:g35_mom0_gallery_summary}, and \ref{fig:g20_mom0_gallery_summary} show zeroth moment maps of representative molecular lines which are selected to highlight prominent morphological features of each source. These maps also demonstrate the complex, often molecule-specific, morphologies present across a wide range of scales from a few 1000~au up to ${\sim}$1~pc, depending on the source.

All sources show evidence of compact, hot-core-like chemistry occurring near to the central massive protostars in addition to a diffuse, extended emission component whose orientation matches that of known molecular outflows, identified in either SiO or high-velocity CO emission \citep{Xu13_G20, Zhang14, Gorai23}. Additional line emission substructures (e.g., rings, cavities, emission plateaus, asymmetries) are observed in all sources, highlighting the complex interplay between chemistry and physical gas conditions in these environments. Below, we provide a brief summary of each massive star-forming region, while Appendix \ref{sec:appendix_mom0_maps} shows zeroth moment map galleries of all detected molecules.

\textit{G28}: consists of several distinct emission components summarized in Figure \ref{fig:g28_mom0_gallery_summary}, which include: diffuse emission along the NE-SW direction extending out to ${\approx}$0.25~pc (e.g., H$_2$CO, CH$_3$OH, H$_2$CS, SO); a ring-like structure with a diameter of ${\approx}$8000~au with either a deep (e.g., CH$_2$NH) or shallow cavity (e.g., H$_2$S, HC$_3$N, $^{13}$CS) surrounded by a diffuse envelope out to ${\approx}$12000~au; and a compact (${\lesssim}$4000~au) emission component, which is either in the form of an unresolved core (e.g., C$_2$H$_5$OH, CH$_3$OCHO, $v=1$) or is spatially-resolved but with a mostly smooth emission profile (e.g., HDO, $^{34}$SO). In the latter case, there are, however, tentative brightness asymmetries, which suggest the presence of additional substructure on small scales. Several of the compact molecules also show a diffuse plateau of emission with a diameter of ${\lesssim}$6000~au (e.g., $^{34}$SO, HDO, CH$_3$OCH$_3$), which is directly interior to the molecular ring. Many molecules show the presence of several components simultaneously, e.g., SO, which has both extended, large-scale emission and a ring on intermediate scales. Large-scale asymmetries are also seen, such as in c-C$_3$H$_2$, which is spatially offset to the NW of the molecular ring and continuum peak. This ring is not azimuthally symmetric and shows several, distinct bright clumps of emission (e.g., CH$_3$OH, SO, HC$_3$N, $^{13}$CS). CN shows extended emission surrounding a deep but localized absorption region that is approximately coincident with the continuum peak and has a total size of ${\approx}$6000~au. This absorption manifests as an arc-like bowl in Figure \ref{fig:g28_mom0_gallery_summary}. Overall, these structures are consistent with those identified in a subset of lines in high-angular resolution ALMA observations \citep{Gorai23}, who attribute their origins to protostellar heating and outflow shocks as well as dense gas within a surrounding envelope.

\textit{G20}: exhibits two primary emission components, as shown in Figure \ref{fig:g20_mom0_gallery_summary}: extended emission (e.g., H$_2$CO, CH$_3$OH, SO) along the NE-SW direction with a total size of ${\approx}$1~pc and a compact core, which is spatially unresolved (${<}$8000~au) in most species (e.g., NH$_2$CHO, CH$_3$CH$_2$CN, SO$_2$, $v_2=1$) at the location of a known molecule-rich HC~\ion{H}{2} region. In the latter case, a few molecules (e.g., HC$_3$N, OC$^{34}$S) also show a diffuse plateau of emission on scales of ${\sim}$0.2~pc outside of this central core. The larger-scale, bipolar emission shows complex structure, i.e., in some molecules it is mostly symmetric along the NE-SW axis (e.g., SO, $^{13}$CS), while in others (e.g., H$_2$CO, CH$_3$OH), the emission is asymmetric with substantially brighter and more extended emission in the NE direction. Moreover, there is an asymmetry along the minor axis of this extended emission with diffuse emission located out to ${\sim}$0.3~pc toward the SE in the direction of another nearby \ion{H}{2} region. CN shows diffuse emission that extends beyond the large-scale emission of the other molecules and has an absorption feature co-located with the continuum. The stippling pattern present in Figure \ref{fig:g20_mom0_gallery_summary} is likely a signature of spatial filtering due to the large angular size of the CN emission.

\textit{G35}: comprises two major emission components, summarized in Figure \ref{fig:g35_mom0_gallery_summary}: diffuse, extended emission along the E-W direction (e.g., H$_2$CO, SO) with a total size of ${\approx}$0.7~pc and bright, compact (${\lesssim}$0.2~pc) emission, which traces the two central UC~\ion{H}{2} regions (e.g., HDO). This compact component is either centrally peaked (e.g., SO$_2$, NO), or is asymmetric and peaks on a single, but not always the same UC~\ion{H}{2} region (e.g., $^{13}$CS, CN, CH$_3$OH, CH$_3$OCHO). Both components are often present in the same molecule. CN shows extended emission, but along a different position angle than the extended emission of most other molecules, and an absorption region with a total size of ${\approx}$15000~au at the location of G35 East and West, which appears as a dark lane in Figure \ref{fig:g35_mom0_gallery_summary}.

\subsection{Line Peak Centroid Positions} \label{sec:peak_positions}

\begin{figure*}[h]
\centering
\includegraphics[width=\linewidth]{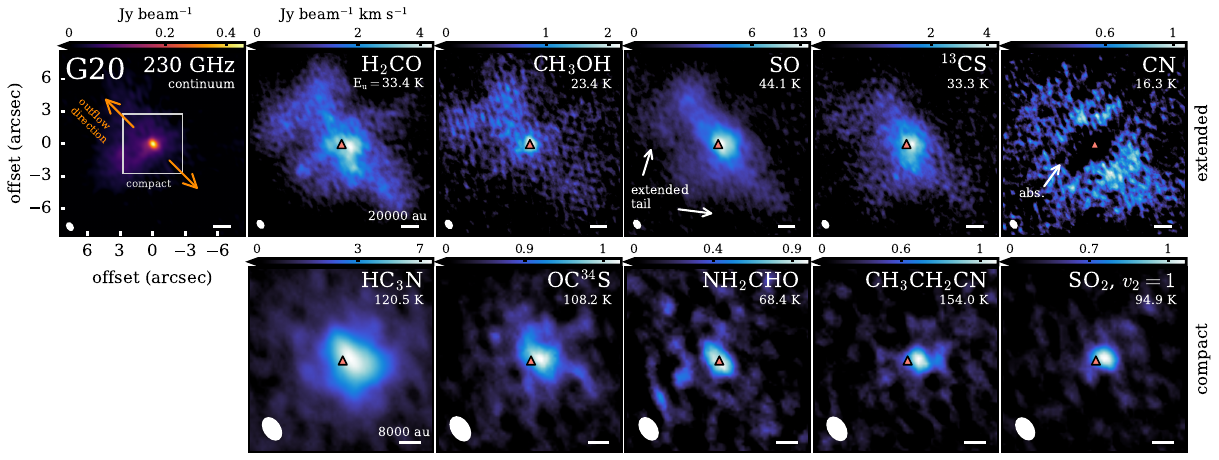}
\caption{Zeroth moment maps of representative lines in G20. The outflow direction, as traced by SiO emission \citep{Xu13_G20}, is marked in the continuum image. Otherwise, as in Figure \ref{fig:g28_mom0_gallery_summary}.}
\label{fig:g20_mom0_gallery_summary}
\end{figure*}

\begin{figure*}[t]
\centering
\includegraphics[width=\linewidth]{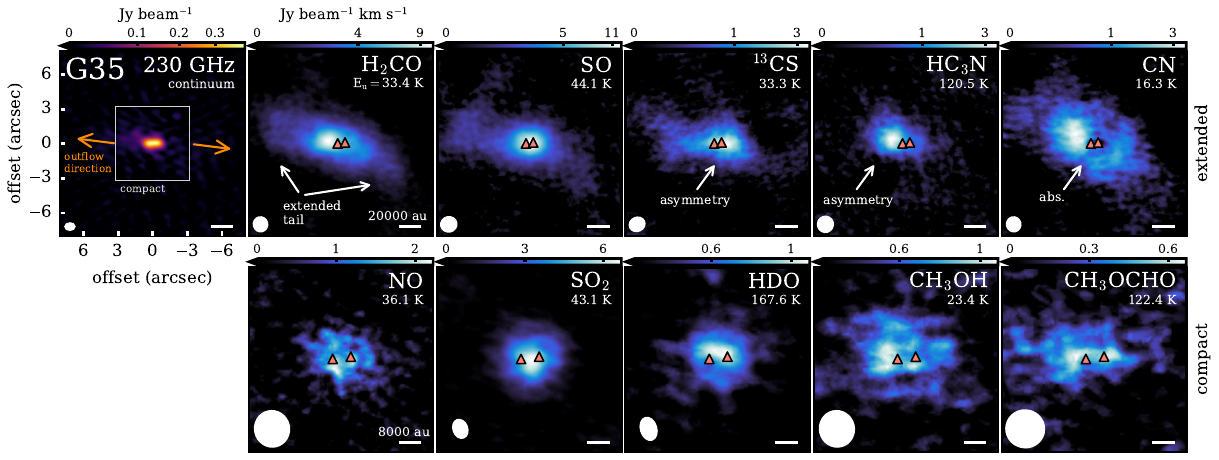}
\caption{Zeroth moment maps of representative lines in G35. The outflow direction, as traced by high-velocity $^{12}$CO emission \citep{Zhang14}, is marked in the continuum image. Otherwise, as in Figure \ref{fig:g28_mom0_gallery_summary}.}
\label{fig:g35_mom0_gallery_summary}
\end{figure*}

While Figures \ref{fig:g28_mom0_gallery_summary}-\ref{fig:g35_mom0_gallery_summary} provide a representative view of each source, we often have many 10s of detected transitions per molecule (see Section \ref{sec:line_detections}). Thus, we next aim to synthesize this combination of spatial and spectral information to most accurately trace the relative molecular distributions. To do so, we first computed the position of the peak intensity for every detected transition and then computed the median position for each molecule. We visually confirmed that these centroids reflected the overall distribution of line peak positions well. When computing the median position, we also experimented with more stringent selection cuts in line peak intensities (beyond that of the initial 3$\sigma$ criterion) but found that the centroid positions were not sensitive to the chosen thresholds. To better visualize the distributions, we also generated a kernel density estimation of the median position using the \texttt{gaussian\_kde} task in \texttt{scipy} \citep{Virtanen_etal_2020}.

\begin{figure*}[p!]
\centering
\includegraphics[width=\linewidth]{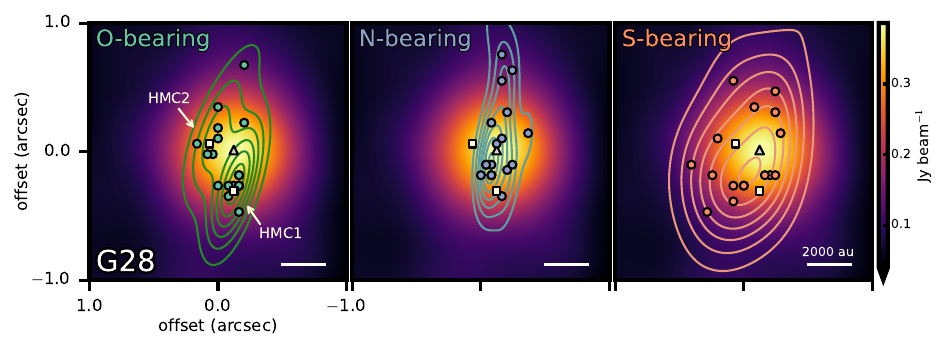}
\includegraphics[width=\linewidth]{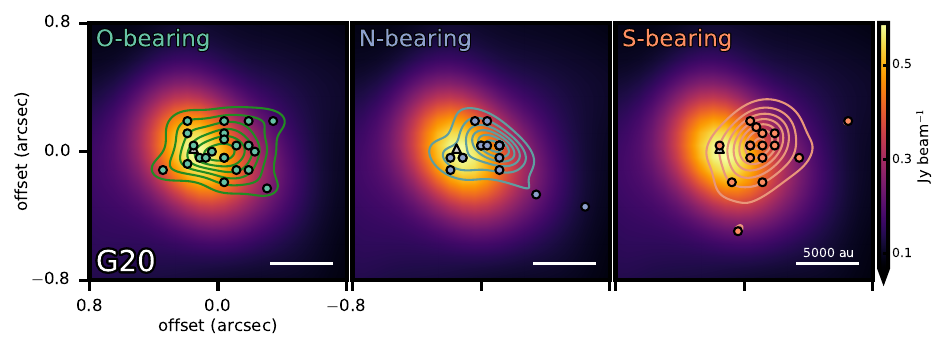}
\includegraphics[width=\linewidth]{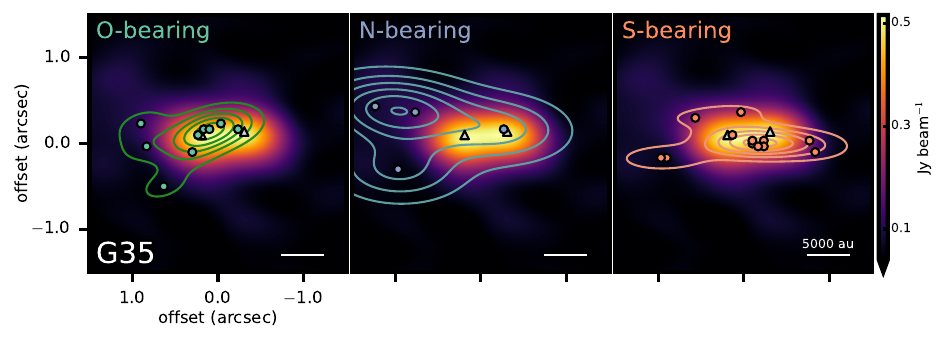}
\caption{Peak positions for different molecular families (\textit{columns}) are shown for each source (\textit{rows}). Individual points are the median line peak positions for all detected molecules and the contours shows the kernel density estimation. The 1.3~mm continuum is shown in colorscale. Continuum peaks are shown as gray triangles in all panels. In G28, the positions of HMC1 and HMC2 are denoted by white squares. In G35, the left-hand triangle corresponds to G35 East and the right-hand triangle marks G35 West. Only a few molecules peak far outside the continuum and are not shown here (but see Appendix \ref{sec:appendix_mom0_maps} for a full gallery of zeroth moment maps for each species).}
\label{fig:All_Median_Peaks}
\end{figure*}

Figure \ref{fig:All_Median_Peaks} shows a summary of the median positions of all detected molecules in our sample sub-dived into categories for O-, N-, and S-bearing species. All three sources show some degree of spatial segregation among different molecular families and with respect to the continuum peaks. These often take the form of relative spatial offsets where different types of molecules cluster at certain positions, while others exhibit more complex structures, such as ring- or shell-like distributions. Below, we summarize these results for each source, while we discuss the origins of the observed separations in Section \ref{sec:discussion_spatial_distr}.

\textit{G28}: The O-bearing molecules generally occupy two spatially distinct clusters in the SW and NE relative to the continuum peak. These two clusters correspond to the positions of the two hot molecular cores, HMC1 and HMC2, respectively, identified by \citet{Gorai23} and we subsequently adopt the same nomenclature. HMC1, in particular, shows emission from the majority of COMs and isotopologues as well as vibrationally-excited CH$_3$OCHO, which taken together, indicates it likely hosts the densest and hottest gas in G28. In contrast, the N-bearing species are confined to a narrow but elongated distribution along the N-S direction and between HMC1 and HMC2, i.e., near the continuum peak. The S-bearing molecules also show a mutually distinct distribution in the form of a ring around the continuum peak, encircling the majority of O- and N-bearing peak positions. Only a few molecules peak outside of the central core of continuum emission, including HCO$^+$, H$_2$S, HCS$^+$, CCH, and c-C$_3$H$_2$, which are not shown in Figure \ref{fig:All_Median_Peaks} (but see Appendix \ref{sec:appendix_mom0_maps} for zeroth moment maps of each of these molecules).

\textit{G20}: The O- and N-bearing molecules are largely co-spatial but are offset to the west of the continuum peak of the central bright HC~\ion{H}{2} region. The S-bearing molecules show a slightly further offset to the west than that of the O- or N-bearing molecules. Similar offsets were previously observed by \citet{GalvanMadrid09} in a few molecules (e.g., CH$_3$CN, SO$_2$, OCS). Beyond these broad offsets from the continuum emission, no distinct structures are apparent. However, G20 is the most distant source in our sample and hence, our observations probe relatively coarse physical scales (see Table \ref{tab:image_info}). Thus, the lack of apparent structure may simply reflect an observational limitation.

\textit{G35}: The O-bearing molecules generally trace the continuum emission, with peaks between or at one of the two central UC~\ion{H}{2} regions. A few O-bearing molecules, however, are offset from the continuum to the east. Besides CH$_3$CN, which peaks at G35 West, all other N-bearing molecules are asymmetric and peak outside of G35 East. S-bearing molecules are found in two distinct clusters, i.e., near the central UC~\ion{H}{2} regions or in a shell outside of both regions. Interestingly, only those S-bearing molecules containing oxygen are found close to either G35 West or East, while those that are carbon-bearing are located in this exterior shell (see Section \ref{sec:G35_elemental_gradient} for more details).

\subsection{Rotational Temperatures and Column Densities} \label{sec:Trot_and_Ncol}

\subsubsection{Fitting Process}

Next, we aim to infer the physical gas conditions on a per-molecule basis in each of our sources. The high number of lines detected across a wide range of excitation conditions enables a rotational diagram approach \citep[e.g.,][]{Goldsmith99}. We only considered those lines which show a well-defined profile, i.e., no absorption or multiple kinematic components, and excluded blended lines. These criteria resulted in the inability to fit lines from only a few molecules, i.e., CO, CN, HCO$^+$, HCN, H$_2$CO. We then fit all transitions with a single Gaussian profile and ensured fitting parameters were consistent for lines belonging to a single molecule, i.e., line FWHMs varied by no more than a factor of two and central velocities were within $\pm$2 velocity channels (${\approx}$5~km~s$^{-1}$) of the known systemic velocity. For each molecule, we further verified the initial line assignments (Section \ref{sec:line_detections}) by checking that the best-fit rotational diagram model did not predict lines at frequencies that were covered by our observations but where no emission is detected. We treated unresolved hyperfine line components from the same species (e.g., CCH, C$^{17}$O) as a single line by combining upper state level degeneracies and line intensities.

We used the Markov Chain Monte Carlo (MCMC) code \texttt{emcee} \citep{Foreman13} to generate posterior probability distributions for rotational temperature (T$_{\rm{rot}}$) and column density (N$_{\rm{col}}$). Following \citet{Law_G10p6}, we simultaneously fit and correct for line optical depths. We explored a relatively broad parameter space of 10$^{12}$~cm$^{-2} < $ N$_{\rm{col}} <$10$^{20}$cm$^{-2}$ and 50~K $<$ T$_{\rm{rot}} <$ 500~K to capture the diverse physical gas conditions expected in massive star-forming regions \citep[e.g.,][]{Moscadelli18, Gieser21, Motte22}. We ensured that the fitted T$_{\rm{rot}}$ and N$_{\rm{col}}$ values were not sharply peaked toward the edge of these priors. We adopted the 50th, 16th, and 84th percentiles from the marginalized posterior distributions as the best-fit values and uncertainties, respectively.

\begin{figure*}[p]
\centering
\includegraphics[width=\linewidth]{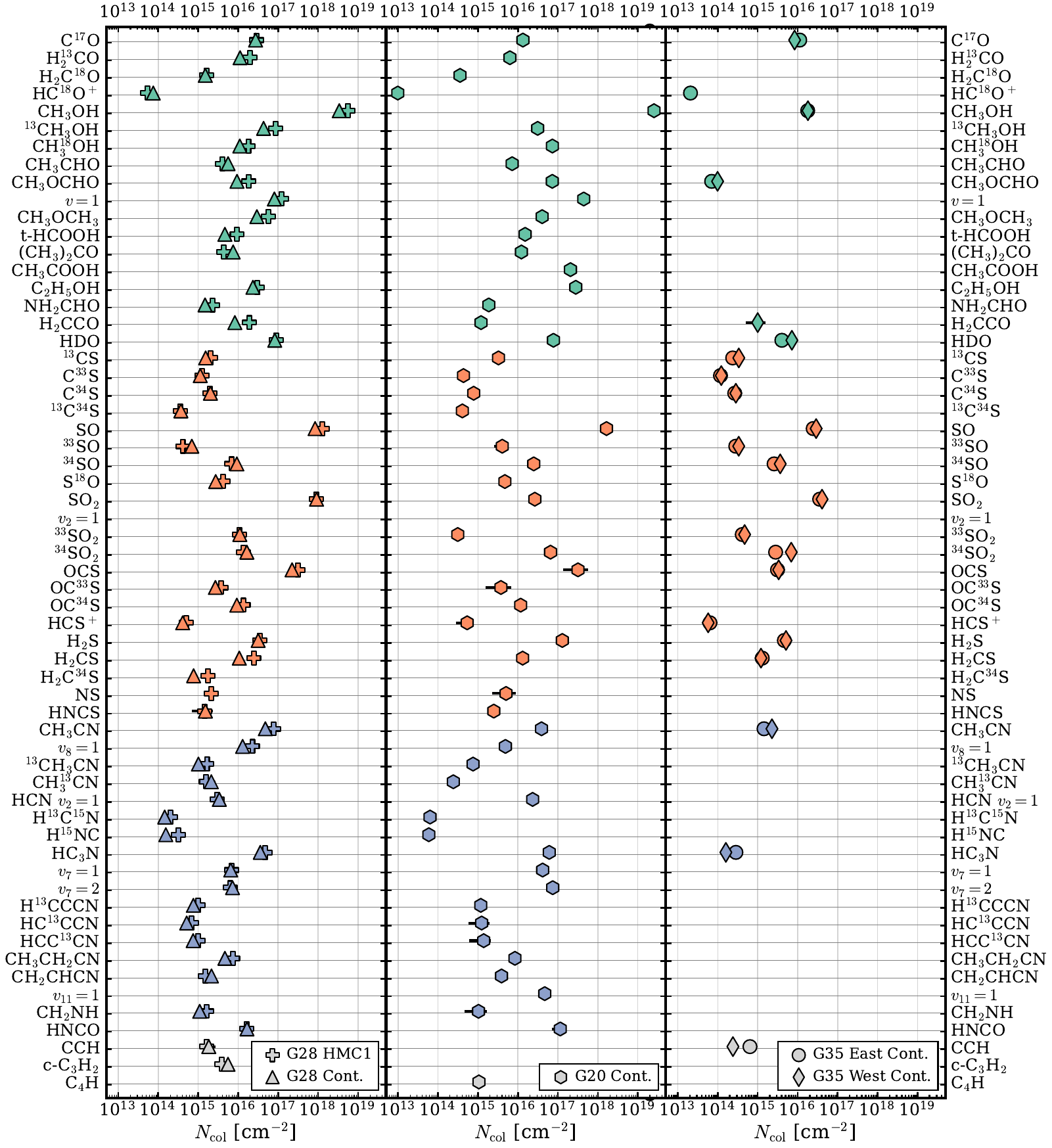}
\caption{Derived molecular column densities for selected positions in G28 (\textit{left}), G20 (\textit{middle}), and G35 (\textit{right}). Molecules are colored as in Figure \ref{fig:molecule_detections} and labels for the vibrational states correspond to the molecule listed immediately above. Due to the wide range in N$_{\rm{col}}$ shown, markers are larger than the uncertainties in most cases.}
\label{fig:Ncol_figure}
\end{figure*}

To ensure robust T$_{\rm{rot}}$ and N$_{\rm{col}}$ determinations, we only attempted to fit rotational diagrams for those species with at least two lines spanning a range of 30~K or more in E$_{\rm{u}}$. For those molecules that did not meet this criterion, or only had multiple low-SNR line detections resulting in poor T$_{\rm{rot}}$ constrains, we instead adopted a rotational temperature from a chemically-similar species, i.e., belonging to the same molecular family (see Section \ref{sec:molecule_detections}). Assumed T$_{\rm{rot}}$ values are marked by square brackets in Table \ref{tab:Ncol}. We excluded all highly optically-thick lines from the rotational diagram analysis. For molecules for which the main isotopologue lines are expected to be optically thick, we instead used minor isotopologues, when available, along with the appropriate isotopic elemental ratios to infer their column densities \citep[see Appendix \ref{sec:isotopic_ratio_used_appendix};][]{Wilson94, Milam05, Yu20, Yan23}. Given the high gas temperatures expected in our sources, we selected spectroscopic data for partition functions that included any non-negligible contributions from vibrational and torsional states. When such data was unavailable, we applied temperature-dependent corrections to the derived column densities. See Appendix \ref{sec:spec_data} for a detailed description of these corrections and a summary of the spectroscopic data used in this work. 

In G20, we only use those lines covered by the lower frequency receiver to derive rotational temperatures and column densities. This was done to avoid potentially significant beam dilution from the coarser angular resolution associated with data from the higher frequency receiver, as noted in Section \ref{sec:observations_overview}. However, a handful of molecules (e.g., C$^{17}$O, HC$^{18}$O$^+$, $^{33}$SO$_2$, H$_2$CCO, CH$_3^{13}$CN, C$^{33}$S, C$^{34}$S) only have transitions in the high frequency spectral windows. For these molecules, we also adopted the T$_{\rm{rot}}$ of a chemically-similar species to compute their column densities, but caution that these N$_{\rm{col}}$ values are likely underestimates. 

\tabletypesize{\scriptsize}
\setlength{\tabcolsep}{4pt}

\begin{deluxetable*}{lccccccccccccc}[p]
\tablecaption{Rotational Temperature and Column Density\label{tab:Ncol}}
\tablewidth{0pt}
\tablehead{
\colhead{} & \multicolumn2c{G28 HMC1} & \multicolumn2c{G28 Cont. Peak} &  \multicolumn2c{G20 Cont. Peak} &\multicolumn2c{G35 East Cont. Peak} & \multicolumn2c{G35 West Cont. Peak}    \\ \colhead{} & \colhead{T$_{\rm{rot}}$ (K)} & \colhead{N$_{\rm{col}}$ (cm$^{-2}$)} & \colhead{T$_{\rm{rot}}$ (K)} & \colhead{N$_{\rm{col}}$ (cm$^{-2}$)} & \colhead{T$_{\rm{rot}}$ (K)} & \colhead{N$_{\rm{col}}$ (cm$^{-2}$)} & \colhead{T$_{\rm{rot}}$ (K)} & \colhead{N$_{\rm{col}}$ (cm$^{-2}$)}  & \colhead{T$_{\rm{rot}}$ (K)} & \colhead{N$_{\rm{col}}$ (cm$^{-2}$)} }
\startdata
C$^{17}$O & [125] & ${\gtrsim}2.9\times$10$^{16}$ & [138] & ${\gtrsim}2.7\times$10$^{16}$ & [138] & ${\gtrsim}1.3\times$10$^{16}$ & [77] & ${\gtrsim}1.1\times$10$^{16}$ & [61] & ${\gtrsim}8.4\times$10$^{15}$\\
H$_2^{13}$CO & 181$^{+17}_{-14}$ & ${\gtrsim}2.0\times$10$^{16}$ & 165$^{+18}_{-14}$ & 1.1$^{+1.2}_{-1.0}\times$10$^{16}$ & 138$^{+45}_{-26}$ & 6.3$^{+8.7}_{-5.2}\times$10$^{15}$ & \ldots & \ldots & \ldots & \ldots\\
H$_2$C$^{18}$O & [181] & 1.6$^{+1.8}_{-1.4}\times$10$^{15}$ & [165] & 1.5$^{+1.7}_{-1.4}\times$10$^{15}$ & [138] & 3.6$^{+4.2}_{-3.0}\times$10$^{14}$ & \ldots & \ldots & \ldots & \ldots\\
HC$^{18}$O$^+$ & [125] & 5.3$^{+6.0}_{-4.7}\times$10$^{13}$ & [138] & 7.5$^{+8.6}_{-6.5}\times$10$^{13}$ & [138] & ${\gtrsim}9.9\times$10$^{12}$ & [77] & 2.1$^{+2.7}_{-1.4}\times$10$^{13}$ & \ldots & \ldots\\
CH$_3$OH\tablenotemark{a} & \ldots & 5.6$^{+6.4}_{-4.9}\times$10$^{18}$ & \ldots & 3.4$^{+3.6}_{-3.2}\times$10$^{18}$ & \ldots & 2.5$^{+2.8}_{-2.2}\times$10$^{19}$ & 77$^{+3}_{-3}$ & 1.8$^{+1.9}_{-1.7}\times$10$^{16}$ & 61$^{+4}_{-3}$ & ${\gtrsim}1.8\times$10$^{16}$\\
$^{13}$CH$_3$OH & 125$^{+3}_{-3}$ & 8.6$^{+8.8}_{-8.3}\times$10$^{16}$ & 138$^{+7}_{-6}$ & 4.3$^{+4.5}_{-4.2}\times$10$^{16}$ & [138] & 3.1$^{+3.6}_{-2.7}\times$10$^{16}$ & \ldots & \ldots & \ldots & \ldots\\
CH$_3^{18}$OH & 161$^{+17}_{-14}$ & 1.8$^{+2.1}_{-1.6}\times$10$^{16}$ & 129$^{+10}_{-8}$ & 1.1$^{+1.2}_{-1.0}\times$10$^{16}$ & [138] & 7.3$^{+8.1}_{-6.4}\times$10$^{16}$ & \ldots & \ldots & \ldots & \ldots\\
CH$_3$CHO & [103] & 4.1$^{+4.2}_{-3.9}\times$10$^{15}$ & [74] & 5.6$^{+5.7}_{-5.5}\times$10$^{15}$ & [228] & 7.2$^{+7.9}_{-6.5}\times$10$^{15}$ & \ldots & \ldots & \ldots & \ldots\\
CH$_3$OCHO & 103$^{+1}_{-1}$ & 1.9$^{+1.9}_{-1.8}\times$10$^{16}$ & 74$^{+2}_{-1}$ & 9.3$^{+9.8}_{-8.8}\times$10$^{15}$ & [228] & 7.2$^{+8.6}_{-5.7}\times$10$^{16}$ & [77] & 7.0$^{+7.6}_{-6.6}\times$10$^{13}$ & [61] & 9.9$^{+11.0}_{-8.7}\times$10$^{13}$\\
$v=1$ & [103] & 1.2$^{+1.2}_{-1.2}\times$10$^{17}$ & [74] & 8.1$^{+8.5}_{-7.7}\times$10$^{16}$ & [228] & 4.4$^{+4.6}_{-4.3}\times$10$^{17}$ & \ldots & \ldots & \ldots & \ldots\\
CH$_3$OCH$_3$ & 140$^{+5}_{-5}$ & 5.7$^{+5.9}_{-5.6}\times$10$^{16}$ & 149$^{+7}_{-7}$ & 2.9$^{+3.1}_{-2.9}\times$10$^{16}$ & 228$^{+43}_{-33}$ & 4.0$^{+5.5}_{-3.3}\times$10$^{16}$ & \ldots & \ldots & \ldots & \ldots\\
t-HCOOH & [125] & 9.4$^{+9.9}_{-8.9}\times$10$^{15}$ & [138] & 4.7$^{+4.9}_{-4.4}\times$10$^{15}$ & [138] & 1.5$^{+1.7}_{-1.3}\times$10$^{16}$ & \ldots & \ldots & \ldots & \ldots\\
C$_2$H$_5$OH & 59$^{+2}_{-2}$ & 3.0$^{+3.2}_{-2.8}\times$10$^{16}$ & 61$^{+4}_{-4}$ & 2.3$^{+2.6}_{-2.1}\times$10$^{16}$ & [228] & 2.8$^{+3.2}_{-2.4}\times$10$^{17}$ & \ldots & \ldots & \ldots & \ldots\\
(CH$_3$)$_2$CO & [140] & 4.4$^{+4.6}_{-4.1}\times$10$^{15}$ & [149] & 7.5$^{+8.0}_{-7.0}\times$10$^{15}$ & 51$^{+1}_{-1}$ & ${\gtrsim}1.2\times$10$^{16}$ & \ldots & \ldots & \ldots & \ldots\\
CH$_3$COOH & \ldots & \ldots & \ldots & \ldots & [228] & 2.1$^{+2.4}_{-1.7}\times$10$^{17}$ & \ldots & \ldots & \ldots & \ldots\\
NH$_2$CHO & 88$^{+6}_{-6}$ & ${\gtrsim}2.3\times$10$^{15}$ & 91$^{+10}_{-8}$ & 1.5$^{+1.6}_{-1.4}\times$10$^{15}$ & 55$^{+2}_{-2}$ & ${\gtrsim}1.9\times$10$^{15}$ & \ldots & \ldots & \ldots & \ldots\\
H$_2$CCO & [181] & ${\gtrsim}1.9\times$10$^{16}$ & [165] & ${\gtrsim}8.3\times$10$^{15}$ & [138] & ${\gtrsim}1.2\times$10$^{15}$ & \ldots & \ldots & [61] & 1.0$^{+1.6}_{-0.5}\times$10$^{15}$\\
HDO & 392$^{+75}_{-90}$ & 9.0$^{+10.0}_{-7.9}\times$10$^{16}$ & 381$^{+79}_{-90}$ & 8.2$^{+9.2}_{-7.3}\times$10$^{16}$ & [214] & ${\gtrsim}7.7\times$10$^{16}$ & [77] & 4.0$^{+4.3}_{-3.7}\times$10$^{15}$ & [61] & 7.2$^{+7.6}_{-6.7}\times$10$^{15}$\\
$^{13}$CS & [178] & ${\gtrsim}2.0\times$10$^{15}$ & [169] & 1.5$^{+1.6}_{-1.5}\times$10$^{15}$ & [132] & ${\gtrsim}3.3\times$10$^{15}$ & [122] & 2.4$^{+2.5}_{-2.2}\times$10$^{14}$ & [133] & 3.4$^{+3.6}_{-3.1}\times$10$^{14}$\\
C$^{33}$S & [178] & ${\gtrsim}1.2\times$10$^{15}$ & [169] & ${\gtrsim}1.1\times$10$^{15}$ & [132] & ${\gtrsim}4.3\times$10$^{14}$ & [122] & 1.2$^{+1.3}_{-1.0}\times$10$^{14}$ & [133] & 1.2$^{+1.4}_{-1.0}\times$10$^{14}$\\
C$^{34}$S & [178] & ${\gtrsim}2.0\times$10$^{15}$ & [169] & ${\gtrsim}2.0\times$10$^{15}$ & [132] & ${\gtrsim}7.8\times$10$^{14}$ & [122] & 2.7$^{+3.0}_{-2.4}\times$10$^{14}$ & [133] & 2.8$^{+3.0}_{-2.6}\times$10$^{14}$\\
$^{13}$C$^{34}$S & [178] & 3.6$^{+4.0}_{-3.2}\times$10$^{14}$ & [169] & 3.7$^{+4.0}_{-3.4}\times$10$^{14}$ & [132] & 4.1$^{+4.7}_{-3.4}\times$10$^{14}$ & \ldots & \ldots & \ldots & \ldots\\
SO\tablenotemark{a} & \ldots & 1.3$^{+1.6}_{-0.9}\times$10$^{18}$ & \ldots & 8.4$^{+10.0}_{-6.9}\times$10$^{17}$ & \ldots & 1.6$^{+2.0}_{-1.3}\times$10$^{18}$ & \ldots & 2.4$^{+2.8}_{-2.2}\times$10$^{16}$ & \ldots & 2.9$^{+3.3}_{-2.6}\times$10$^{16}$\\
$^{33}$SO & [178] & 4.1$^{+4.8}_{-3.4}\times$10$^{14}$ & [169] & 6.9$^{+8.1}_{-5.8}\times$10$^{14}$ & [132] & 4.1$^{+5.6}_{-2.6}\times$10$^{15}$ & [108] & 2.8$^{+3.2}_{-2.5}\times$10$^{14}$ & [169] & 3.3$^{+3.8}_{-3.0}\times$10$^{14}$\\
$^{34}$SO & [178] & ${\gtrsim}6.9\times$10$^{15}$ & [169] & ${\gtrsim}9.3\times$10$^{15}$ & [132] & ${\gtrsim}2.5\times$10$^{16}$ & [108] & 2.6$^{+2.6}_{-2.5}\times$10$^{15}$ & [169] & 3.7$^{+3.7}_{-3.6}\times$10$^{15}$\\
S$^{18}$O & [178] & 4.2$^{+5.2}_{-3.1}\times$10$^{15}$ & [169] & 2.7$^{+3.3}_{-2.3}\times$10$^{15}$ & [132] & 4.7$^{+5.7}_{-3.8}\times$10$^{15}$ & \ldots & \ldots & \ldots & \ldots\\
SO$_2$\tablenotemark{a} & \ldots & 9.0$^{+11.8}_{-6.4}\times$10$^{17}$ & \ldots & 9.1$^{+9.3}_{-9.0}\times$10$^{17}$ & \ldots & 2.6$^{+2.9}_{-2.4}\times$10$^{16}$ & \ldots & 3.5$^{+4.4}_{-2.6}\times$10$^{16}$ & \ldots & 4.1$^{+4.8}_{-3.3}\times$10$^{16}$\\
$^{33}$SO$_2$ & [178] & 1.1$^{+1.4}_{-0.8}\times$10$^{16}$ & [169] & 1.1$^{+1.1}_{-1.1}\times$10$^{16}$ & [132] & ${\gtrsim}3.1\times$10$^{14}$ & [108] & 4.1$^{+5.1}_{-3.0}\times$10$^{14}$ & [169] & 4.7$^{+5.5}_{-3.8}\times$10$^{14}$\\
$^{34}$SO$_2$ & [178] & 1.4$^{+1.4}_{-1.3}\times$10$^{16}$ & [169] & 1.7$^{+1.7}_{-1.6}\times$10$^{16}$ & 132$^{+12}_{-10}$ & 6.5$^{+6.9}_{-6.2}\times$10$^{16}$ & [108] & 2.8$^{+3.3}_{-2.4}\times$10$^{15}$ & [169] & 6.9$^{+8.5}_{-5.3}\times$10$^{15}$\\
OCS\tablenotemark{a} & \ldots & 3.2$^{+3.9}_{-2.3}\times$10$^{17}$ & \ldots & 2.2$^{+2.5}_{-2.0}\times$10$^{17}$ & \ldots & 3.2$^{+5.7}_{-1.3}\times$10$^{17}$ & 108$^{+40}_{-27}$ & 3.1$^{+3.7}_{-2.7}\times$10$^{15}$ & 169$^{+53}_{-34}$ & 3.3$^{+3.7}_{-3.1}\times$10$^{15}$\\
OC$^{33}$S & [178] & 3.8$^{+4.7}_{-2.8}\times$10$^{15}$ & [169] & 2.7$^{+3.0}_{-2.4}\times$10$^{15}$ & [132] & 3.8$^{+6.7}_{-1.6}\times$10$^{15}$ & \ldots & \ldots & \ldots & \ldots\\
OC$^{34}$S & 178$^{+22}_{-18}$ & 1.3$^{+1.4}_{-1.3}\times$10$^{16}$ & 169$^{+26}_{-20}$ & 9.3$^{+9.8}_{-8.7}\times$10$^{15}$ & [132] & 1.2$^{+1.5}_{-0.9}\times$10$^{16}$ & \ldots & \ldots & \ldots & \ldots\\
HCS$^+$ & [178] & 5.0$^{+5.7}_{-4.4}\times$10$^{14}$ & [169] & 4.1$^{+4.4}_{-3.7}\times$10$^{14}$ & [132] & 5.4$^{+7.7}_{-2.8}\times$10$^{14}$ & [122] & 6.4$^{+7.0}_{-5.7}\times$10$^{13}$ & [133] & 5.8$^{+7.0}_{-4.6}\times$10$^{13}$\\
H$_2$S & [178] & 3.5$^{+4.0}_{-3.0}\times$10$^{16}$ & [169] & 3.2$^{+3.6}_{-2.8}\times$10$^{16}$ & [214] & ${\gtrsim}1.3\times$10$^{17}$ & [122] & 4.6$^{+5.0}_{-4.3}\times$10$^{15}$ & [133] & 5.1$^{+5.5}_{-4.7}\times$10$^{15}$\\
H$_2$CS\tablenotemark{a} & \ldots & 2.4$^{+2.8}_{-2.2}\times$10$^{16}$ & \ldots & 1.1$^{+1.2}_{-0.9}\times$10$^{16}$ & [132] & ${\gtrsim}1.3\times$10$^{16}$ & 122$^{+54}_{-25}$ & 1.3$^{+1.7}_{-1.1}\times$10$^{15}$ & 133$^{+78}_{-30}$ & 1.2$^{+1.8}_{-1.0}\times$10$^{15}$\\
H$_2$C$^{34}$S & [178] & 1.8$^{+2.0}_{-1.6}\times$10$^{15}$ & [169] & 7.7$^{+8.6}_{-6.7}\times$10$^{14}$ & \ldots & \ldots & \ldots & \ldots & \ldots & \ldots\\
NS & [212] & 2.1$^{+2.4}_{-1.9}\times$10$^{15}$ & \ldots & \ldots & [155] & 5.0$^{+8.8}_{-2.3}\times$10$^{15}$ & \ldots & \ldots & \ldots & \ldots\\
HNCS & [178] & 1.5$^{+1.8}_{-1.1}\times$10$^{15}$ & [169] & 1.5$^{+2.3}_{-0.7}\times$10$^{15}$ & [132] & 2.5$^{+3.0}_{-2.0}\times$10$^{15}$ & \ldots & \ldots & \ldots & \ldots\\
CH$_3$CN\tablenotemark{a} & \ldots & 7.9$^{+8.5}_{-7.0}\times$10$^{16}$ & \ldots & 4.8$^{+5.2}_{-4.1}\times$10$^{16}$ & \ldots & 3.9$^{+4.8}_{-3.5}\times$10$^{16}$ & 71$^{+8}_{-5}$ & ${\gtrsim}1.4\times$10$^{15}$ & 62$^{+29}_{-1}$ & ${\gtrsim}2.3\times$10$^{15}$\\
$v_8=1$ & [212] & 2.3$^{+2.6}_{-2.0}\times$10$^{16}$ & [228] & 1.3$^{+1.5}_{-1.1}\times$10$^{16}$ & [155] & ${\gtrsim}4.9\times$10$^{15}$ & \ldots & \ldots & \ldots & \ldots\\
$^{13}$CH$_3$CN & 212$^{+29}_{-23}$ & 1.7$^{+1.8}_{-1.5}\times$10$^{15}$ & 228$^{+42}_{-28}$ & 1.0$^{+1.1}_{-0.9}\times$10$^{15}$ & 155$^{+32}_{-23}$ & 7.5$^{+9.3}_{-6.8}\times$10$^{14}$ & \ldots & \ldots & \ldots & \ldots\\
CH$_3^{13}$CN & [212] & 1.6$^{+1.8}_{-1.3}\times$10$^{15}$ & [228] & 2.1$^{+2.7}_{-1.6}\times$10$^{15}$ & [155] & ${\gtrsim}2.4\times$10$^{14}$ & \ldots & \ldots & \ldots & \ldots\\
HCN, $v_2=1$ & [304] & ${\gtrsim}3.0\times$10$^{15}$ & [299] & 3.4$^{+3.8}_{-3.0}\times$10$^{15}$ & [155] & 2.3$^{+2.7}_{-2.0}\times$10$^{16}$ & \ldots & \ldots & \ldots & \ldots\\
H$^{13}$C$^{15}$N & [233] & 2.0$^{+2.3}_{-1.7}\times$10$^{14}$ & [177] & 1.5$^{+1.5}_{-1.4}\times$10$^{14}$ & [155] & ${\gtrsim}6.3\times$10$^{13}$ & \ldots & \ldots & \ldots & \ldots\\
H$^{15}$NC & [233] & 3.2$^{+4.1}_{-2.3}\times$10$^{14}$ & [177] & 1.6$^{+1.7}_{-1.4}\times$10$^{14}$ & [155] & ${\gtrsim}5.9\times$10$^{13}$ & \ldots & \ldots & \ldots & \ldots\\
HC$_3$N\tablenotemark{a} & \ldots & 4.8$^{+6.0}_{-4.1}\times$10$^{16}$ & \ldots & 3.6$^{+3.8}_{-3.4}\times$10$^{16}$ & \ldots & 6.1$^{+6.4}_{-5.7}\times$10$^{16}$ & 146$^{+79}_{-43}$ & 2.8$^{+3.7}_{-2.4}\times$10$^{14}$ & [146] & 1.6$^{+1.7}_{-1.5}\times$10$^{14}$\\
$v_7=1$ & 304$^{+35}_{-30}$ & 6.8$^{+7.0}_{-6.5}\times$10$^{15}$ & 299$^{+33}_{-28}$ & 6.5$^{+6.8}_{-6.3}\times$10$^{15}$ & [155] & ${\gtrsim}4.2\times$10$^{16}$ & \ldots & \ldots & \ldots & \ldots\\
$v_7=2$ & [304] & 6.3$^{+6.6}_{-5.9}\times$10$^{15}$ & [299] & 7.2$^{+7.6}_{-6.8}\times$10$^{15}$ & [155] & 7.4$^{+7.7}_{-7.2}\times$10$^{16}$ & \ldots & \ldots & \ldots & \ldots\\
H$^{13}$CCCN & 233$^{+48}_{-37}$ & 1.0$^{+1.3}_{-0.9}\times$10$^{15}$ & 177$^{+18}_{-15}$ & 7.6$^{+8.1}_{-7.1}\times$10$^{14}$ & [155] & 1.2$^{+1.2}_{-1.1}\times$10$^{15}$ & \ldots & \ldots & \ldots & \ldots\\
HC$^{13}$CCN & [233] & 6.9$^{+7.1}_{-6.7}\times$10$^{14}$ & [177] & 5.1$^{+5.3}_{-5.0}\times$10$^{14}$ & [155] & 1.2$^{+1.9}_{-0.6}\times$10$^{15}$ & \ldots & \ldots & \ldots & \ldots\\
HCC$^{13}$CN & [233] & 9.8$^{+11.0}_{-8.4}\times$10$^{14}$ & [177] & 7.5$^{+8.3}_{-6.5}\times$10$^{14}$ & [155] & 1.4$^{+2.1}_{-0.6}\times$10$^{15}$ & \ldots & \ldots & \ldots & \ldots\\
CH$_3$CH$_2$CN & 52$^{+2}_{-1}$ & ${\gtrsim}7.3\times$10$^{15}$ & 51$^{+2}_{-1}$ & ${\gtrsim}4.7\times$10$^{15}$ & 98$^{+4}_{-4}$ & 8.4$^{+8.7}_{-8.2}\times$10$^{15}$ & \ldots & \ldots & \ldots & \ldots\\
CH$_2$CHCN & [52] & 1.5$^{+1.6}_{-1.4}\times$10$^{15}$ & [51] & 2.2$^{+2.2}_{-2.1}\times$10$^{15}$ & 129$^{+6}_{-6}$ & 3.9$^{+4.0}_{-3.8}\times$10$^{15}$ & \ldots & \ldots & \ldots & \ldots\\
$v_{11}=1$ & \ldots & \ldots & \ldots & \ldots & [129] & 4.7$^{+5.5}_{-3.8}\times$10$^{16}$ & \ldots & \ldots & \ldots & \ldots\\
CH$_2$NH & 139$^{+13}_{-11}$ & ${\gtrsim}1.6\times$10$^{15}$ & 75$^{+9}_{-6}$ & ${\gtrsim}1.1\times$10$^{15}$ & [138] & 1.0$^{+1.7}_{-0.5}\times$10$^{15}$ & \ldots & \ldots & \ldots & \ldots\\
HNCO & 179$^{+8}_{-6}$ & ${\gtrsim}1.6\times$10$^{16}$ & 167$^{+7}_{-7}$ & ${\gtrsim}1.7\times$10$^{16}$ & [138] & 1.1$^{+1.7}_{-0.7}\times$10$^{17}$ & \ldots & \ldots & \ldots & \ldots\\
CCH & [212] & ${\gtrsim}1.6\times$10$^{15}$ & [228] & 1.8$^{+2.7}_{-1.1}\times$10$^{15}$ & \ldots & \ldots & [71] & ${\gtrsim}6.4\times$10$^{14}$ & [62] & ${\gtrsim}2.4\times$10$^{14}$\\
c-C$_3$H$_2$ & [212] & 3.8$^{+4.2}_{-3.5}\times$10$^{15}$ & [228] & 5.6$^{+6.2}_{-5.0}\times$10$^{15}$ & \ldots & \ldots & \ldots & \ldots & \ldots & \ldots\\
C$_4$H & \ldots & \ldots & \ldots & \ldots & [155] & 1.1$^{+1.1}_{-1.0}\times$10$^{15}$ & \ldots & \ldots & \ldots & \ldots\\
\enddata
\vspace{-2mm}
\tablecomments{Column densities derived from species whose median line optical depth exceeds 0.5 are indicated by ${\gtrsim}$, as are those derived for molecules which only had transitions in the lower-resolution, high-frequency data of G20. These should collectively be treated as lower limits. Adopted rotational temperatures are indicated by brackets. Labels for the vibrational states correspond to the molecule listed immediately above.}
\vspace{-2.5mm}
\tablenotetext{a}{Column densities computed from minor isotopolouges when available assuming the appropriate isotopic elemental ratios (see Appendix \ref{sec:isotopic_ratio_used_appendix}).}\end{deluxetable*}

\tabletypesize{\normalsize}
\setlength{\tabcolsep}{6pt}

\subsubsection{Molecular Column Densities}

To probe the spatial variation expected across our sources, we chose a few points of interest for this analysis.  In G28, we picked HMC1 and the continuum peak, while in G35, we extracted values from both the G35 East and West UC~\ion{H}{2} regions. In G20, we chose the continuum peak of the central HC~\ion{H}{2} region. Table \ref{tab:Ncol} and Figure \ref{fig:Ncol_figure} provide a summary of the derived values. Table \ref{tab:Ncol} is also available in its entirety in machine-readable format.

We derived the highest number of rotational temperatures for G28, followed by G20, and then G35, where we only have a handful of T$_{\rm{rot}}$ constraints. We have a T$_{\rm{rot}}$ estimate from either CH$_3$CN or one its isotopologues, which provide a measure of physical gas temperatures, in each of our sources. G28 possesses the hottest (${\approx}$210~K) gas, followed by warm (${\approx}$160~K) gas in G20, and then relatively lukewarm (${\approx}$60-70~K) gas in G35. In terms of typical N$_{\rm{col}}$, both G20 and G28 have column densities that are several orders of magnitude greater than those seen in G35. Overall, G20 and G28 show comparable molecular column densities and relative N$_{\rm{col}}$ ratios among different species, which reflects their chemical similarity. 

We find modest evidence for spatial variation in the gas properties across our sources. We focus on G35 and G28, since we lack adequate physical resolution for a meaningfully analysis in G20 due to its far distance. While G35 East and West have consistent rotational temperatures, G35 West is generally enhanced in molecular column densities. The brightness asymmetries observed in the zeroth moment maps of a few molecules (Figure \ref{fig:g35_mom0_gallery_summary}) are reflected in N$_{\rm{col}}$ differences, such as CCH, HC$_3$N, which show higher N$_{\rm{col}}$ in G35 East. Nonetheless, these differences are typically no more than a factor of a few and both cores appear broadly chemically similar. In G28, HMC1 generally shows warmer rotational temperatures of ${\approx}$20-50~K and higher column densities by a factor of a few-to-several versus the continuum peak. These high temperatures and densities are consistent with its previous classification as a hot core \citep{Gorai23}. 

Each of our sources has previous estimates of rotational temperature and column density in at least a handful of the same molecules considered here (e.g., CH$_3$CN, CH$_3$OH, SO$_2$, H$_2$S). Our derived values are generally consistent with these prior inferences in G20 \citep{GalvanMadrid09, Xu13_G20} and G35 \citep{Zhang14}, which both used substantially lower resolution (a few arcseconds) SMA data, as well as those derived from high-resolution (${\approx}$0\farcs2) ALMA observations in G28 \citep{Gorai23}.

\section{Discussion} \label{sec:discussion}

\subsection{Chemical Richness and Evolutionary Stage} \label{sec:evol_stage}

In high-mass star-forming regions, the observed chemistry varies as a function of evolutionary stage \citep[e.g.,][]{Sanhueza12, Gerner14, Gerner15, Liu21_ATOMS}. Molecular complexity initially increases starting from the infrared dark cloud phase until the emergence of a hot core, where the molecular abundance, complexity, and detection rate tends to peak. Once more evolved \ion{H}{2} regions are formed, however, this detection rate then again decreases, as powerful stellar radiation destroys many molecular species.

Our sample follows these previously observed trends, which indicates that their relative chemical richness is consistent with an evolutionary sequence. G28 and G20 are the youngest regions with known HMCs and HC~\ion{H}{2} regions, followed by the UC~\ion{H}{2} regions in G35 and then the evolved \ion{H}{2} region W33, which shows the most advanced signs of star formation in our sample. G28 and G20 are the most chemically-rich with numerous COM detections and consistently show the highest column densities. In contrast, few COMs are detected in G35 with mean column density reductions of one-to-two orders of magnitude. W33 is notably molecule-poor and since it is the only source in our sample categorized as a full \ion{H}{2} region, this is likely attributable to much of its molecular material having already been largely destroyed or dispersed. This scenario is further supported by the detection of He$\alpha$ RRLs in only G35 and W33, which indicates that these sources have more massive zero-age-main-sequence stars driving their \ion{H}{2} regions due to the larger ionization potential of He compared to H.

Given that our survey sensitivity varies by no more than a factor of five among our sources (Table \ref{tab:image_info}), these discrepancies reflect true chemical and physical differences. Moreover, potential biases in the detection rates due to the different sources distances also cannot explain the observed trends. For instance, W33 is the closest source at 2.4~kpc but shows the fewest molecular detections, while G20, the furthest source at 12.3~kpc, shows a molecule-rich spectra.

\subsection{Spatial Distribution of Molecular Families} \label{sec:discussion_spatial_distr}

\subsubsection{Evidence for Carbon Grain Sublimation in G28}\label{sec:carbon_grain_sublimation_g28}

\begin{figure*}
\centering
\includegraphics[width=\linewidth]{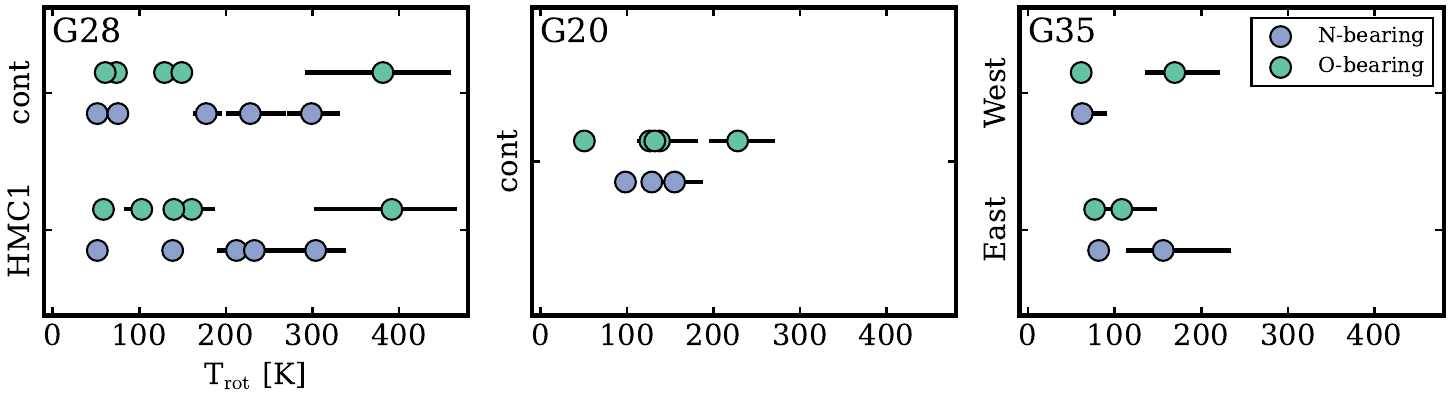}
\caption{Derived rotational temperatures of O-bearing (green) and N-bearing (blue) species in our sample. Molecules containing both N and O, i.e., HNCO and NH$_2$CHO, are excluded.}
\label{fig:Trot_figure}
\end{figure*}

Spatial offsets between N- and O-bearing COMs are commonly observed in massive star-forming regions. Orion~KL is the prototypical example, where the N-COMs peak at the hot core and the O-COMs are associated with the compact ridge \citep[e.g.,][]{Blake87, Friedel08}. Subsequent observations of larger samples show that N-COMs typically peak at the location of continuum maxima, while the O-COMs are offset from the location of the central protostar \citep[e.g.,][]{Jimenez_Serra12, Fayolle15, Allen17, Bogelund19, Csengeri19}. While the exact origins of these separations remains unclear, one explanation is the sublimation of refractory carbon grains within the so-called ``soot line" at ${\sim}$300~K near the central protostar \citep{vant_Hoff_carbon_grains}. This mechanism predicts that N-COMs will be located close to the warmest molecular gas, i.e., within the soot line, and thus have higher rotational temperatures than those of the O-COMs, which are desorbed uniformly within the water snowline (${\sim}$100~K). Here, we aim to see if such a scenario is operating in any of sources in our sample. 

G28 is the only source in our sample that shows significant spatial offsets between N- and O-bearing species. As shown in Figure \ref{fig:All_Median_Peaks}, N-bearing molecules are located at the continuum peak while the O-bearing molecules are offset at either the HMC1 or HMC2 positions. Moreover, the RRLs, which trace the hottest gas, also peak at the continuum position (see Figure \ref{fig:continuum_gallery}). We also compiled all derived rotational temperatures for both N- and O-bearing molecules in Figure \ref{fig:Trot_figure}. In G28, the N-bearing molecules potentially show warmer rotational temperatures than those of that are O-bearing, with median temperatures of ${\approx}$210~K and ${\approx}$140~K, respectively. This trend is primarily driven by the especially warm temperatures of H$^{13}$CCCN, HC$_3$N, v$_7=1$, and $^{13}$CH$_3$CN. Although we derive T$_{\rm{rot}}$ values for the N-bearing molecules that are less than 300~K, as predicted for grain sublimation, this is likely due to a beam dilution effect. At the resolution of our observations, most molecular emission is likely at least partially beam diluted and thus, it is the difference, rather the absolute values, of the rotational temperatures that points to potential carbon grain sublimation. Specifically, the hottest gas will be within the soot line, which is at least an order of magnitude more compact than the water snowline. Thus, we expect the N-COMs to be be more severely affected by beam dilution, and the fact that we still measure warmer T$_{\rm{rot}}$ values for a handful of N-bearing species implies a reservoir of even hotter gas. Further evidence of beam dilution is supported by the results of \citet{Gorai23}, who infer excitation temperatures\footnote{Due to the smaller bandwidth of the ALMA observations, \citet{Gorai23} are not able to derive rotational temperatures on a per-molecule basis. Instead, they adopt a single excitation temperature at each position that well-reproduces several high excitation (E$_{\rm{u}}\approx$350-500~K) transitions of HNCO, $^{13}$CH$_3$CN, and CH$_3$OCHO.} of ${\sim}$300~K and ${\sim}$250~K for HMC1 and HMC2, respectively, in ALMA observations of G28 in a 0\farcs2 beam. Taken together, carbon grain sublimation provides a plausible explanation for the observed spatial offsets and T$_{\rm{rot}}$ distributions of N- and O-bearing molecules in G28.

We cannot assess if a similar scenario is operating in the other sources in our sample. We do not detect a sufficient number of molecules in G35 or W33 to meaningfully compare relative spatial distributions. Due to the far distance of G20, we lack the effective spatial resolution to observe comparable separations (a few 1000~au) as were seen in G28. We also expect much more significant beam dilution in G20, making it difficult to properly interpret the T$_{\rm{rot}}$ distributions in Figure \ref{fig:Trot_figure} without additional high-angular resolution observations.

\subsubsection{An Elemental Oxygen Gradient in G35} \label{sec:G35_elemental_gradient}

\begin{figure}[h!]
\centering
\includegraphics[width=\linewidth]{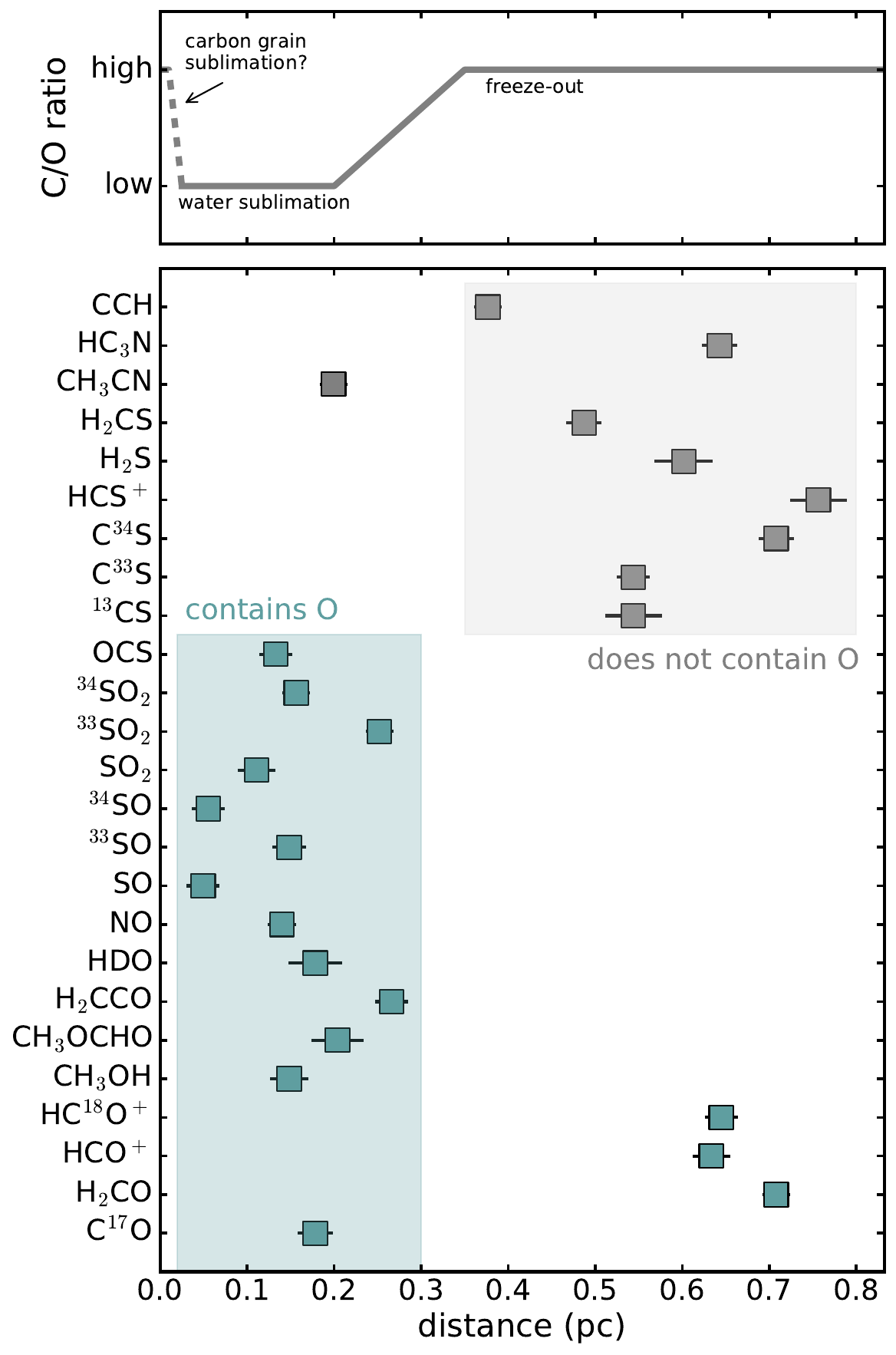}
\caption{Schematic of expected gas-phase C/O ratios (\textit{top}) and projected radial distance of molecular line peaks measured in G35 (\textit{bottom}). Molecules containing oxygen are shown in green and those without are colored in gray.}
\label{fig:G35_O_vs_C}
\end{figure}

While we do not observe any clear spatial separations among O-, N-, and S-bearing species in G35 in Figure \ref{fig:All_Median_Peaks}, we do identify a broad spatial difference between those molecules containing oxygen versus those lacking oxygen. We demonstrate this difference in Figure \ref{fig:G35_O_vs_C}, where we compute the radial distance of all detected molecules from a position between G35 East and West. The majority of molecules that contain oxygen are clustered at small radial distances (${<}$0.3~pc), while those that do not are located at larger distances (${\approx}$0.4-0.8~pc). These spatial offsets suggest the presence of a gas-phase elemental oxygen gradient across G35 where the regions close to G35 East and West are oxygen-rich and those at further distances are oxygen-poor. As noted in the previous Section, we are not able to probe the innermost regions (a few 1000~au) around the UC~\ion{H}{2} regions in G35, where we might expect carbon grain sublimation to perhaps boost the gas-phase C/O ratio on small scales, as illustrated in Figure \ref{fig:G35_O_vs_C}.

A possible explanation for the observed C/O gradient is vigorous thermal sublimation of O-rich ices, i.e., water ice, due to high gas temperatures near one or both of the central UC \ion{H}{2} regions, which, in turn, triggered the formation of abundant O-rich species in the gas-phase \citep[e.g.,][]{Bergin98, Fraser01, Rodgers03, Bjerkeli16, vanDishoeck21}. For instance, chemical modeling predicts that sudden ice evaporation induced by protostellar accretion bursts can form O-bearing organic molecules \citep{Taquet16}. This scenario is also consistent with the detection of gas-phase HDO in this inner 0.3~pc region (Figure \ref{fig:g35_mom0_gallery_summary}). Similar trends have been seen in protostellar envelopes \citep{vantHoff22}, disks around young eruptive stars \citep[][]{Tobin23, Lee24} and massive star-forming regions \citep[e.g.,][]{McGuire18, Csengeri19}, where gas-phase emission from HDO and O-rich COMs show similar spatial distributions. At larger distances, the temperatures are cooler and freeze-out timescales quicker, so the outermost regions would be the first regions where O-rich molecules have already begun freezing out. 

If we assume that the transition zone between O-rich and O-poor molecules corresponds to the water snowline at ${\approx}$0.3~pc, this would require a luminosity increase of ${\sim}$20-60$\times$ the current G35 luminosity, adopting typical dust temperatures and gas densities of massive star-forming regions \citep[e.g.,][]{Bisschop07, vantHoff22}. We can also approximately estimate the re-freeze-out timescale for water, which is on the order of a few thousands years \citep{Visser12}. This suggests a relatively recent event, such as an outburst, would be needed to drive this temporary luminosity increase. This scenario does not fully explain the relatively high detection rate of S-bearing molecules, relative to the O- or N-bearing molecules, in this source. This may be due to either an intrinsically rich sulfur chemistry in G35 or could point to an additional mechanism, e.g., shocks, outflows, outbursts, driving additional S-chemistry \citep[e.g.,][]{Pineau93, Gusdorf08, Liu12, Minh16, Towner24}. Although beyond the scope of this work, detailed chemical and physical modeling of G35 is required to verify the origins of these observed chemical trends. 

\begin{figure*}
\centering
\includegraphics[width=\linewidth]{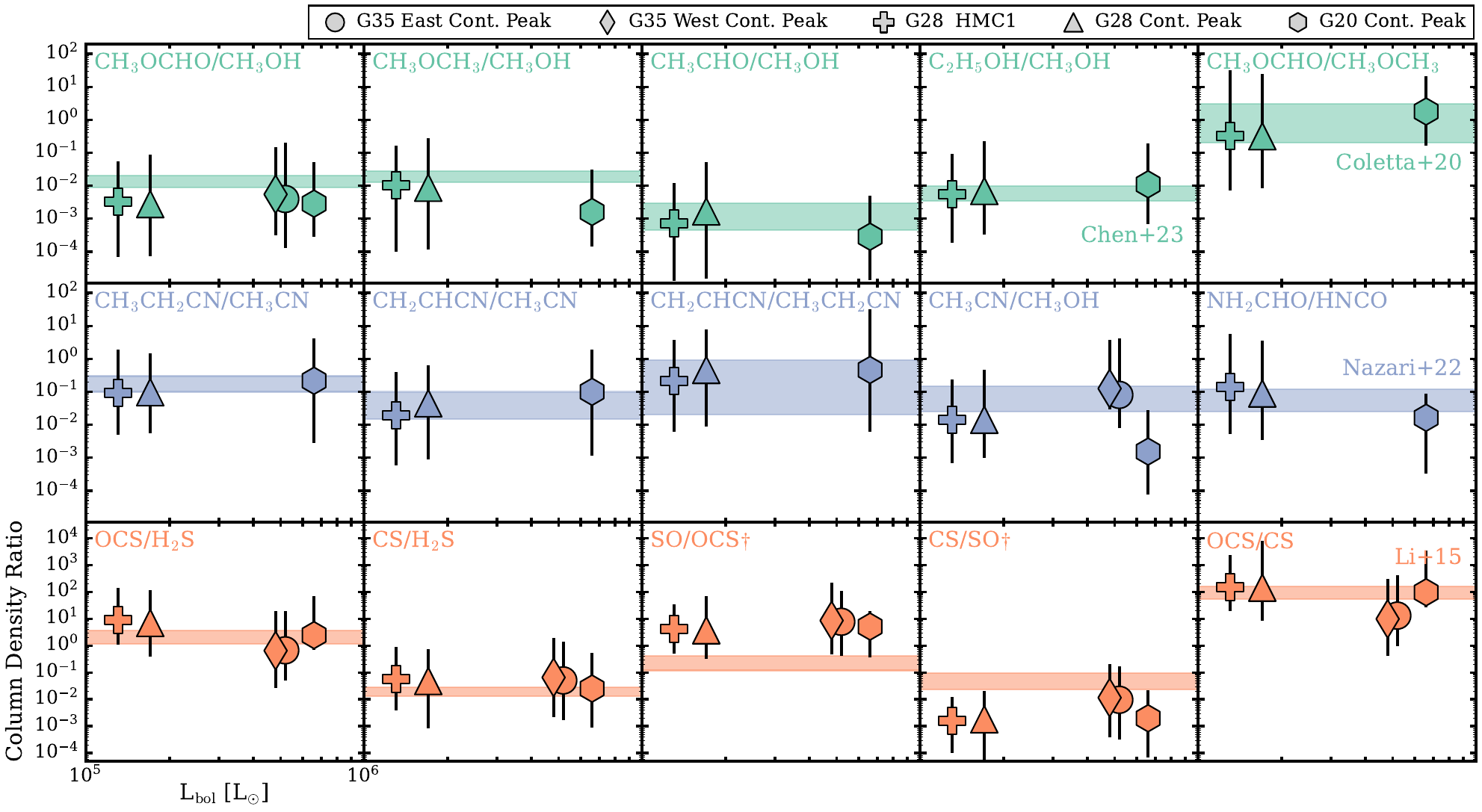}
\caption{Column density ratios of select molecules (\textit{columns}). Shaded bands correspond to the typical 1$\sigma$ range of abundance ratios observed in star-forming regions for O-, N-, and S-bearing species (\textit{rows}). Ratios from \citet{Nazari22_NCOMS, Chen23} include a sample of both high- and low-mass protostars, while \citet{Li15_S, Coletta20} were derived from only high-mass sources. The CS column densities are computed from $^{13}$CS, assuming the relevant isotopic elemental ratios \citep{Yan23}. The ratios marked by a $\dagger$ symbol indicate those derived using likely optically-thick SO. Bolometric luminosity uncertainties are on the order of ${\approx}$30-50\% \citep{Wood89, Liu19, Law_Chi_22}. Molecules are colored as in Figure \ref{fig:molecule_detections}.}
\label{fig:Ncol_ratio_figure}
\end{figure*}

\subsection{Chemical Origins of COMs}

COM column density ratios have often been used to gain insight into their formation mechanisms \citep[e.g.,][]{Bisschop07, Fontani07, Fayolle15, Law_G10p6, Nazari23_CC_destruction}. Recent surveys have identified nearly constant COM abundance ratios in sources across a wide range of bolometric luminosities, including in both high- and low-mass protostars \citep{Coletta20, vanGelder20, Nazari22_NCOMS, Chen23}, which suggests that COM abundances are likely set during the earlier, cold pre-stellar phases. For this to be the case, complex organics, or their precursors, must share a common formation environment, i.e., on pre-stellar ices. Detections of multiple COMs in pre-stellar cores \citep[e.g.,][]{Vastel14, Scibelli21} and on ices \citep{McClure23} lends further support to such an interpretation. Here, we compare column density ratios of several O-, N-, and S-bearing species in our sample to those derived in the literature to see if such a scenario is compatible with our observations.

Figure \ref{fig:Ncol_ratio_figure} shows column density ratios for a select set of molecules. Specifically, we chose a variety of O- and N-COMs, which are thought to be chemically related, e.g., CH$_3$OH/CH$_3$OCH$_3$/CH$_3$OCHO \citep[e.g.,][]{Garrod06, Balucani15} or HNCO/NH$_2$CHO \citep[e.g.,][]{LopezSelp15, Quenard18}, or have previous observational evidence of correlations, e.g., CH$_3$OH/CH$_3$CN \citep[][]{Gieser21, Law_G10p6}. We also include several S-bearing molecules in this comparison. The typical 1$\sigma$ range in abundance ratios reported in the literature, which are shown as shaded lines in Figure \ref{fig:Ncol_ratio_figure}, were computed using a sample of both high- and low-mass protostars \citep{Nazari22_NCOMS, Chen23}, with the exception of the CH$_3$OCHO/CH$_3$OCH$_3$ ratio and those of the S-bearing molecules, which were derived from only high-mass sources \citep{Li15_S, Coletta20}.

In general, we find remarkably consistent N$_{\rm{col}}$ ratios across our sample, which are largely in agreement with literature abundance ratios. Among our sample, the largest source-to-source differences (${\approx}$30$\times$) are those found for the N$_{\rm{col}}$ ratio of CH$_3$CN/CH$_3$OH, but this is perhaps not surprising. While these molecule have shown empirical correlations in prior surveys, there is no known chemical link between them. If we exclude this ratio, then both the O- and N-COM abundance ratios show source-to-source variations of no more than a factor of 5. As can be seen in Figure \ref{fig:Ncol_ratio_figure}, all of these variations are well within the uncertainties. The S-bearing molecules show modestly higher source-to-source differences, up to a factor of nearly 10, but are again consistent within errors. Moreover, all O- and N-bearing COMs shown in Figure \ref{fig:Ncol_ratio_figure} are in agreement with literature abundance ratios, as are the S-bearing species with the exception of the SO/OCS and CS/SO N$_{\rm{col}}$ ratios. The consistent discrepancies related to the SO abundance in all sources, namely, an elevated SO/OCS and lower-than-expected CS/SO ratio relative to the trends from \citet{Li15_S}, is likely an optical depth effect. The SO column densities reported in \citet{Li15_S} were derived using a single transition, but SO is typically optically thick in massive star-forming regions and these values are likely underestimated. Since we cover multiple lines of the $^{33}$SO and $^{34}$SO isotopologues, we are able to derive more robust column densities based on optically-thin lines. Thus, overall, the nearly constant N$_{\rm{col}}$ ratios measured in our sample support the notion that COM abundances must share similar formation environments.

\section{Conclusions} \label{sec:conlcusions}

We present a wideband (${\approx}$32~GHz) chemical survey of four massive star-forming regions with the SMA at sub-arcsecond angular resolution. We conclude the following: 

\begin{enumerate}
    \item We establish a comprehensive molecular inventory of all sources and identify over 60 molecules, including numerous COMs and isotopologues. We also detect several bright RRLs, indicating the presence of ionized gas.
    \item The sources in our sample demonstrate a wide range of chemical richness that is consistent with an evolutionary sequence from the line-rich hot cores and HC~\ion{H}{2} regions of G28 and G20 to the more chemically-modest UC~\ion{H}{2} regions in G35, followed by the molecule-poor \ion{H}{2} region W33.
    \item Our sample shows complex, often molecule-specific, emission morphologies on multiple spatial scales from several 1000~au up to ${\sim}$1~pc. In all sources, we identify diffuse, extended emission, indicating outflow-induced chemistry, as well as compact emission components that are driven by hot, gas-phase chemistry near the central protostars. Mutual spatial offsets among O-, N-, and S-bearing molecules and the millimeter continuum are observed in all four massive star-forming regions.
    \item We compute rotational temperatures and column densities for a large number of molecules at a few positions of interest in G28, G20, and G35. The inclusion of multiple isotopologues and lines with a wide range of excitation conditions in our observations enable robust estimates on the gas physical properties.
    \item We find potential evidence for carbon grain sublimation around the central protostar in G28. N-bearing molecules are co-located with the continuum and RRL peaks and have systemically warmer rotational temperatures (${\approx}$210~K), while O-bearing species are spatially offset and have lower rotational temperatures (${\approx}$140~K).     
    \item We also find evidence of an elemental oxygen gradient in G35, with an O-rich environment close to the two central UC~\ion{H}{2} regions and an O-poor chemistry at larger radii. One possible explanation for this gradient is vigorous thermal sublimation of nearby O-rich ices, i.e., water ice, due to high gas temperatures while at larger radii, freeze-out has already halted much of the gas-phase O-chemistry. G35 also hosts a relatively rich S-chemistry, which may be intrinsic or could point to an additional dynamical mechanism, e.g., shocks, outflows, outbursts.
    \item We observe remarkably constant COM abundance ratios among our sources that agree with literature values in other low- and high-mass protostars. Overall, our sample supports the idea that COM formation occurs in similar environments during the earlier pre-stellar phases, i.e., likely in ices.
\end{enumerate}

The authors thank the anonymous referee for valuable comments that improved both the content and presentation of this work. The authors wish to recognize and acknowledge the very significant cultural role and reverence that the summit of Maunakea has always had within the indigenous Hawaiian community. We are most fortunate to have had the opportunity to conduct observations from this mountain.

Support for C.J.L. was provided by NASA through the NASA Hubble Fellowship grant No. HST-HF2-51535.001-A awarded by the Space Telescope Science Institute, which is operated by the Association of Universities for Research in Astronomy, Inc., for NASA, under contract NAS5-26555. H.B.L. is supported by the National Science and Technology Council (NSTC) of Taiwan (Grant Nos. 111-2112-M-110-022-MY3). L.I.C. acknowledges support from NASA ATP 80NSSC20K0529. R.G.M. acknowledges support from UNAM-PAPIIT project IN108822 and from CONACyT Ciencia de Frontera project ID 86372. 

%

\facilities{SMA}


\software{Astropy \citep{astropy_2013,astropy_2018,Astropy22}, \texttt{bettermoments} \citep{Teague18}, CASA \citep{McMullin_etal_2007, CASATeam20}, \texttt{cmasher} \citep{vanderVelden20},  \texttt{emcee} \citep{Foreman13}, Matplotlib \citep{Hunter07}, MIR (\url{https://lweb.cfa.harvard.edu/~cqi/mircook.html}), NumPy \citep{vanderWalt_etal_2011}, SciPy \citep{Virtanen_etal_2020}}



\appendix

\section{Large-scale Emission in W33~Main} \label{sec:appendix:w33}

\begin{figure*}[b]
\centering
\includegraphics[width=0.965\linewidth]{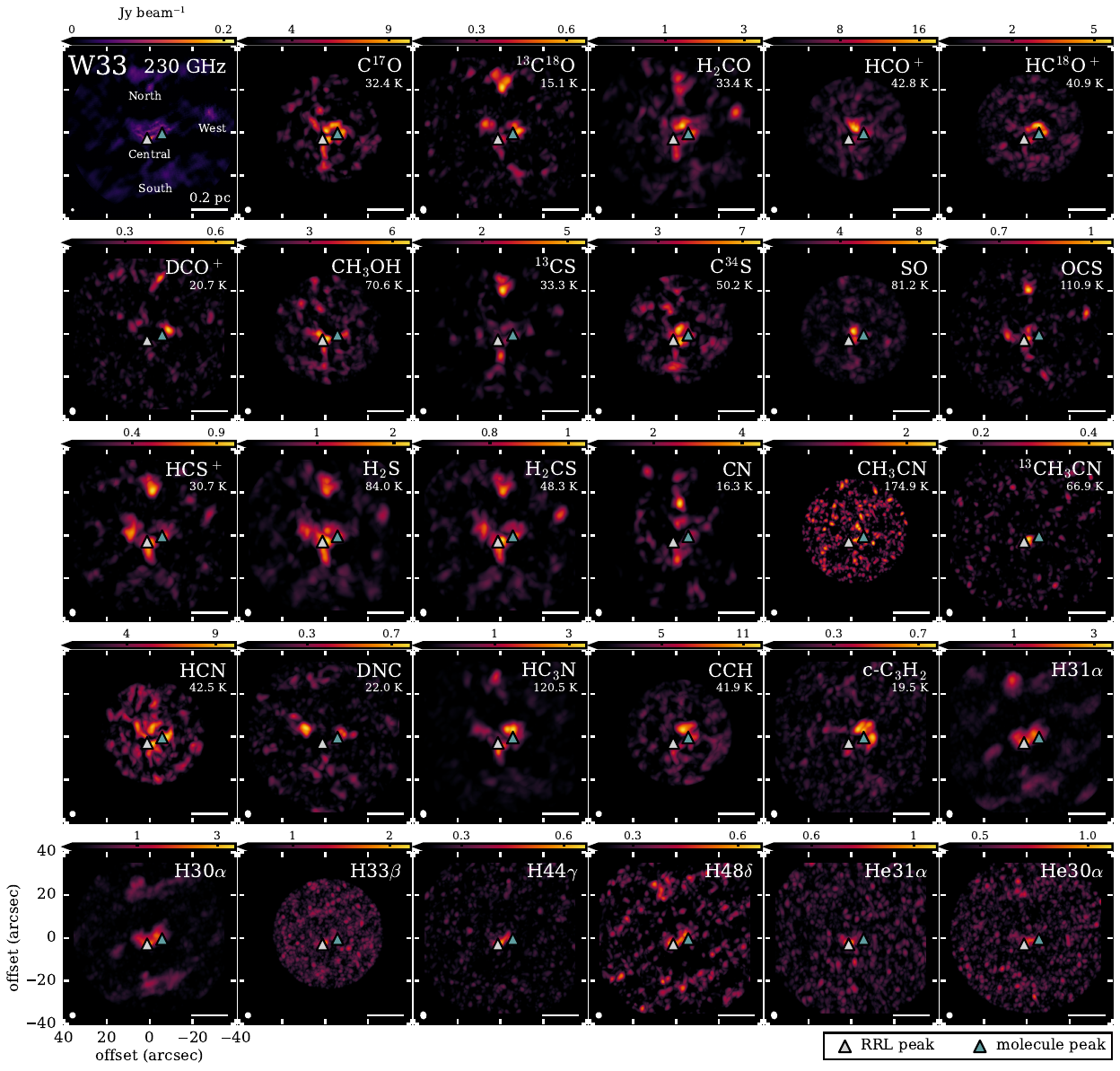}
\caption{230~GHz continuum image and gallery of peak line intensity maps of W33 in an extended field of view. The location of the cores (W33 Main-Central, W33 Main-North, W33 Main-West, W33 Main-South) initially identified in  \citet{Immer14} are labeled in the top left panel.}
\label{fig:W33_peak_intensity}
\end{figure*}

Figure \ref{fig:W33_peak_intensity} provides a gallery of line peak intensity maps for all detected molecules in W33. These maps demonstrate the complex morphology of the continuum and molecular line emission present in the central few 0.1~pc of W33. In addition to the main core (``W33 Main-Central"), we also label the location of the additional cores reported in \citet{Immer14}, i.e., W33 Main-North, W33 Main-West, and W33 Main-South. The spectral extraction positions in Figures \ref{fig:w33_rx240_ls}-\ref{fig:w33_rx345_us} are indicated by crosses.

\section{Full Spectra and Line Identification} \label{sec:full_spectra}

\begin{figure*}[h]
\centering
\includegraphics[width=.95\linewidth]{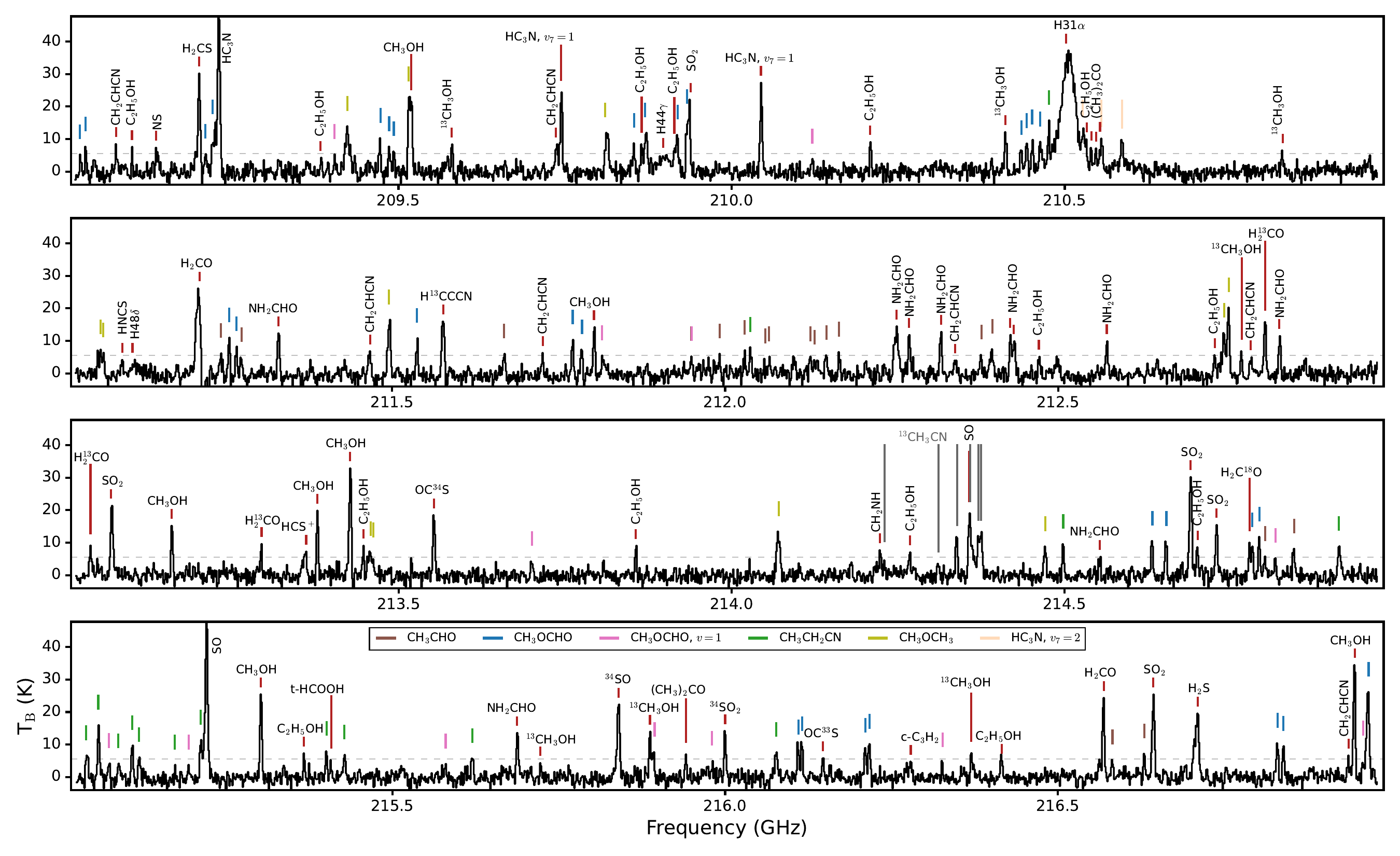}
\caption{Zoomed-in and labeled SMA spectra of G28 from ${\approx}$209-217~GHz. The gray dashed line shows 3$\times$RMS.}
\label{fig:g28_rx240_ls}
\end{figure*}

\begin{figure*}[h]
\centering
\includegraphics[width=.95\linewidth]{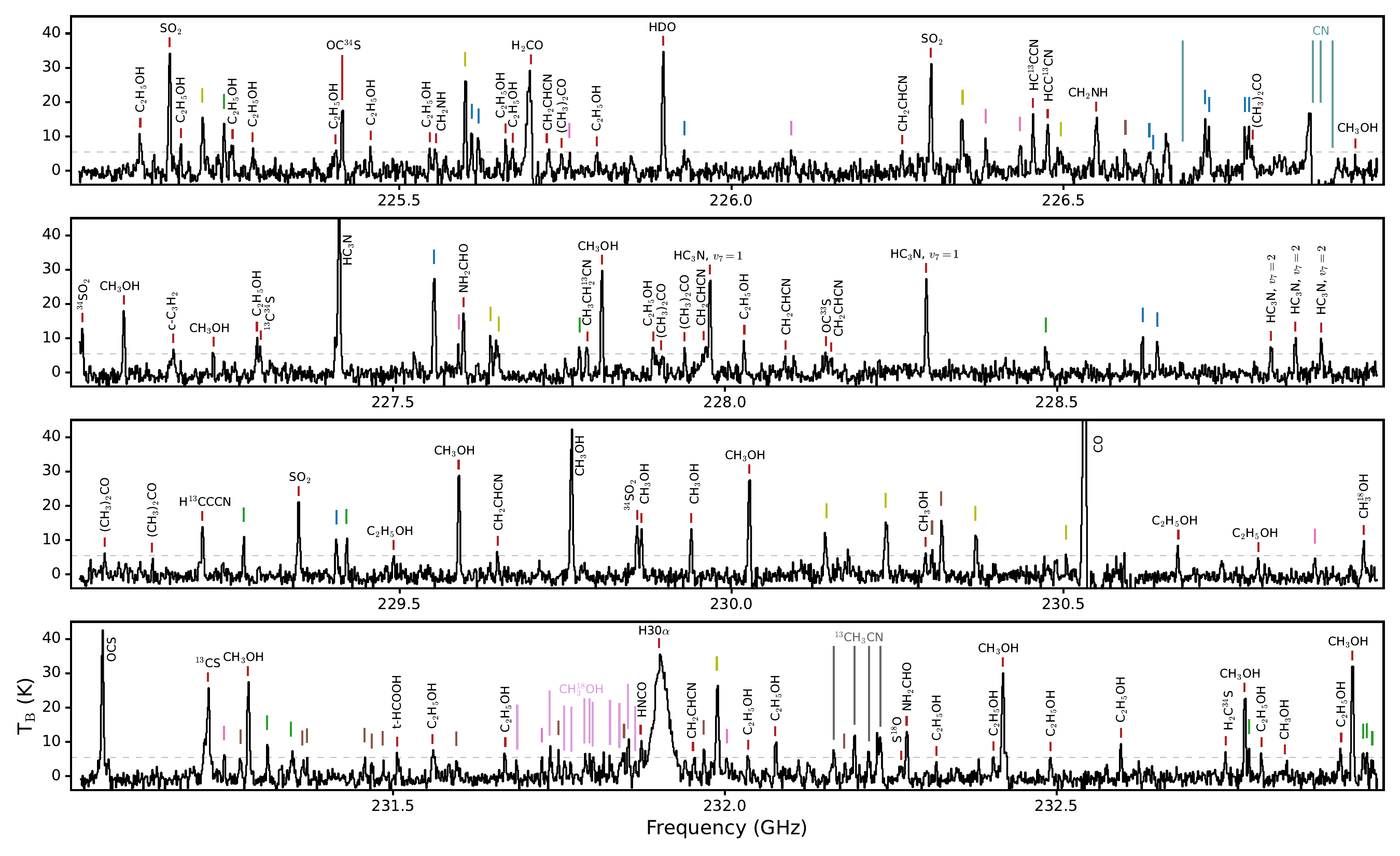}
\caption{Zoomed-in and and labeled SMA spectra of G28 from ${\approx}$225-233~GHz. The gray dashed line shows 3$\times$RMS.}
\label{fig:g28_rx240_us}
\end{figure*}

\begin{figure*}[p!]
\centering
\includegraphics[width=.95\linewidth]{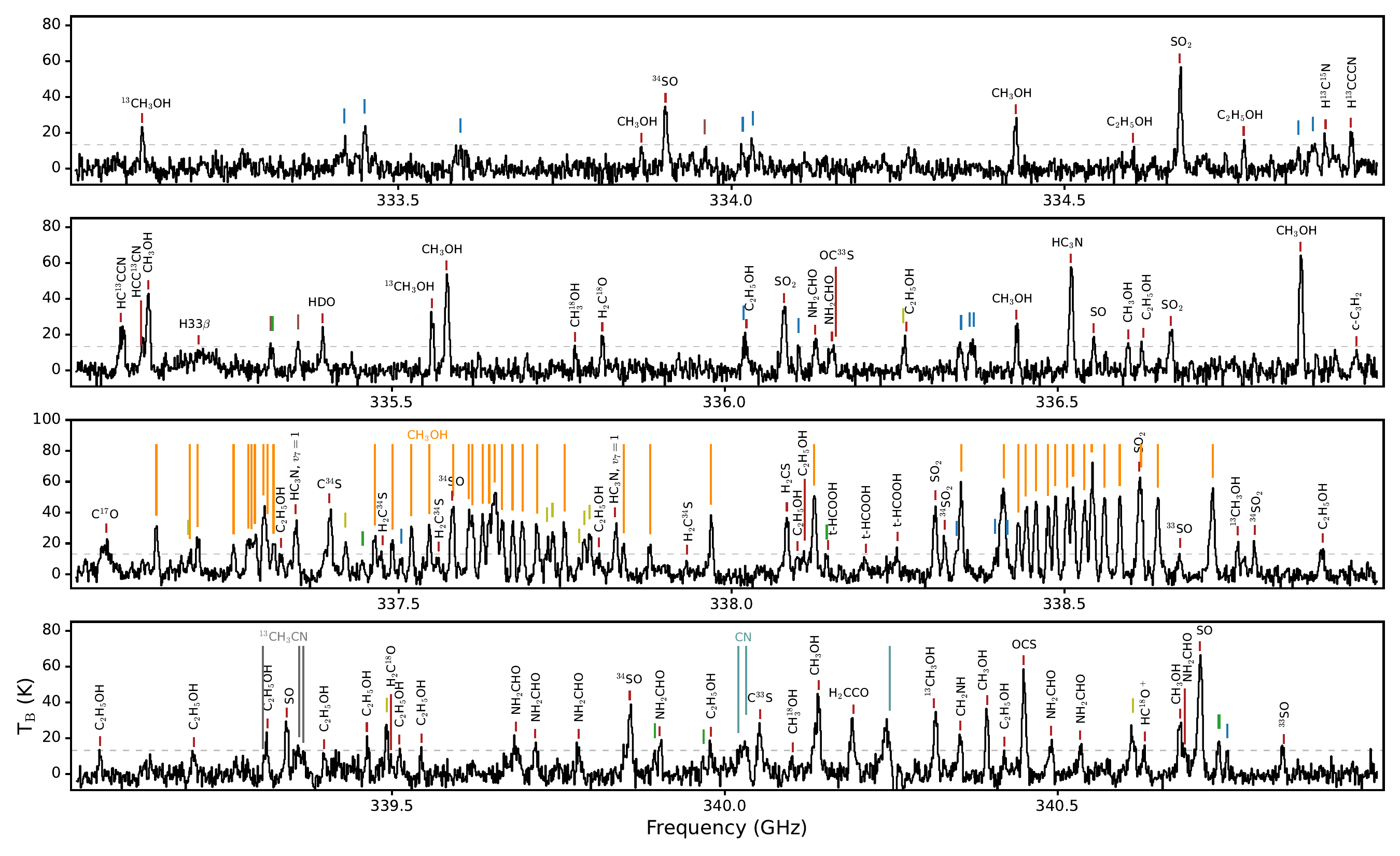} 
\caption{Zoomed-in and labeled SMA spectra of G28 from ${\approx}$333-341~GHz. The gray dashed line shows 3$\times$RMS.}
\label{fig:g28_rx345_ls}
\end{figure*}

\begin{figure*}[p!]
\centering
\includegraphics[width=.95\linewidth]{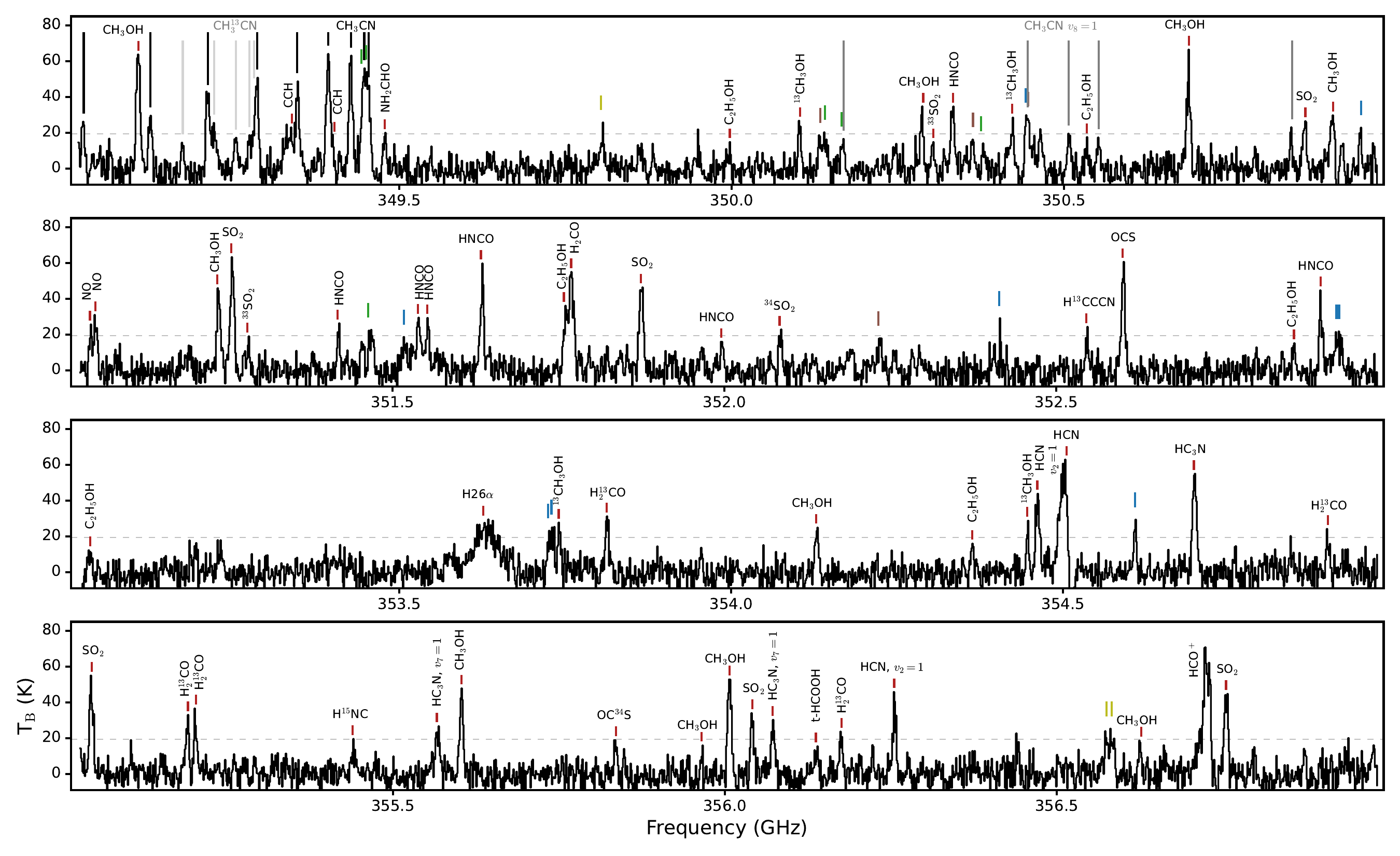} 
\caption{Zoomed-in and labeled SMA spectra of G28 from ${\approx}$349-357~GHz. The gray dashed line shows 3$\times$RMS.}
\label{fig:g28_rx345_us}
\end{figure*}

\begin{figure*}[p!]
\centering
\includegraphics[width=.95\linewidth]{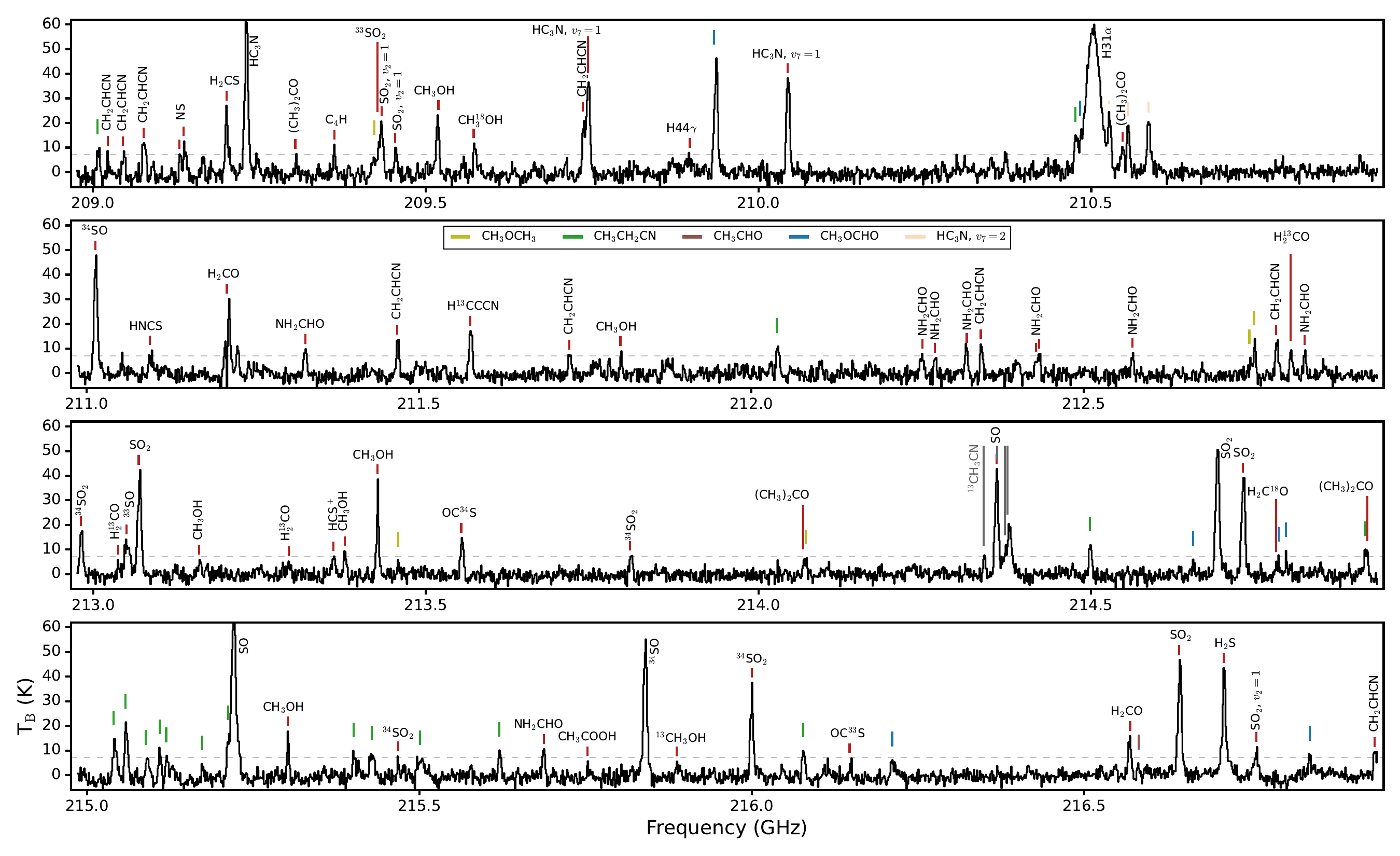} 
\caption{Zoomed-in and labeled SMA spectra of G20 from ${\approx}$209-217~GHz. The gray dashed line shows 3$\times$RMS.}
\label{fig:g20_rx240_ls}
\end{figure*}

\begin{figure*}[p!]
\centering
\includegraphics[width=.95\linewidth]{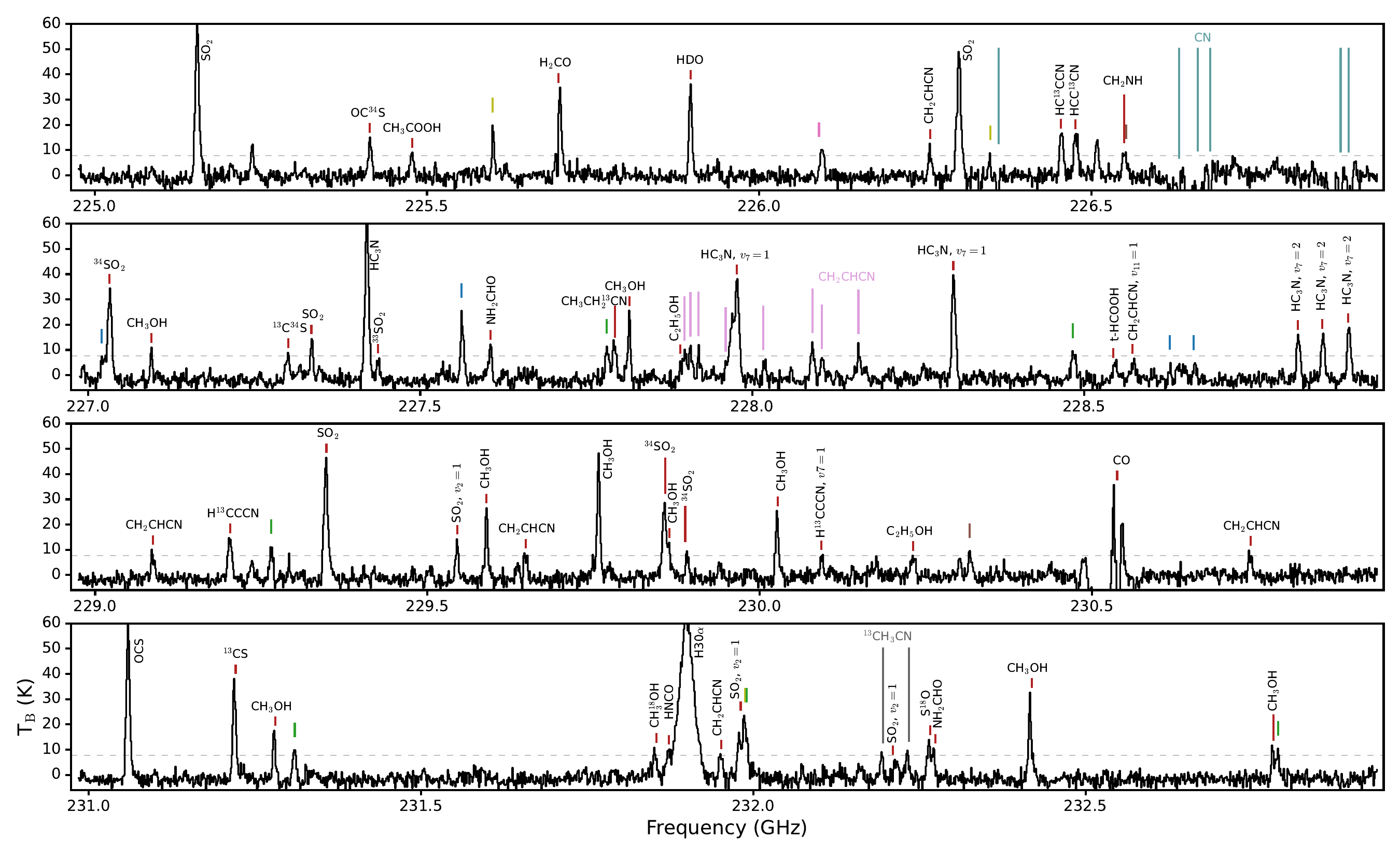} 
\caption{Zoomed-in and labeled SMA spectra of G20 from ${\approx}$225-233~GHz. The gray dashed line shows 3$\times$RMS.}
\label{fig:g20_rx240_us}
\end{figure*}

\begin{figure*}[p!]
\centering
\includegraphics[width=.95\linewidth]{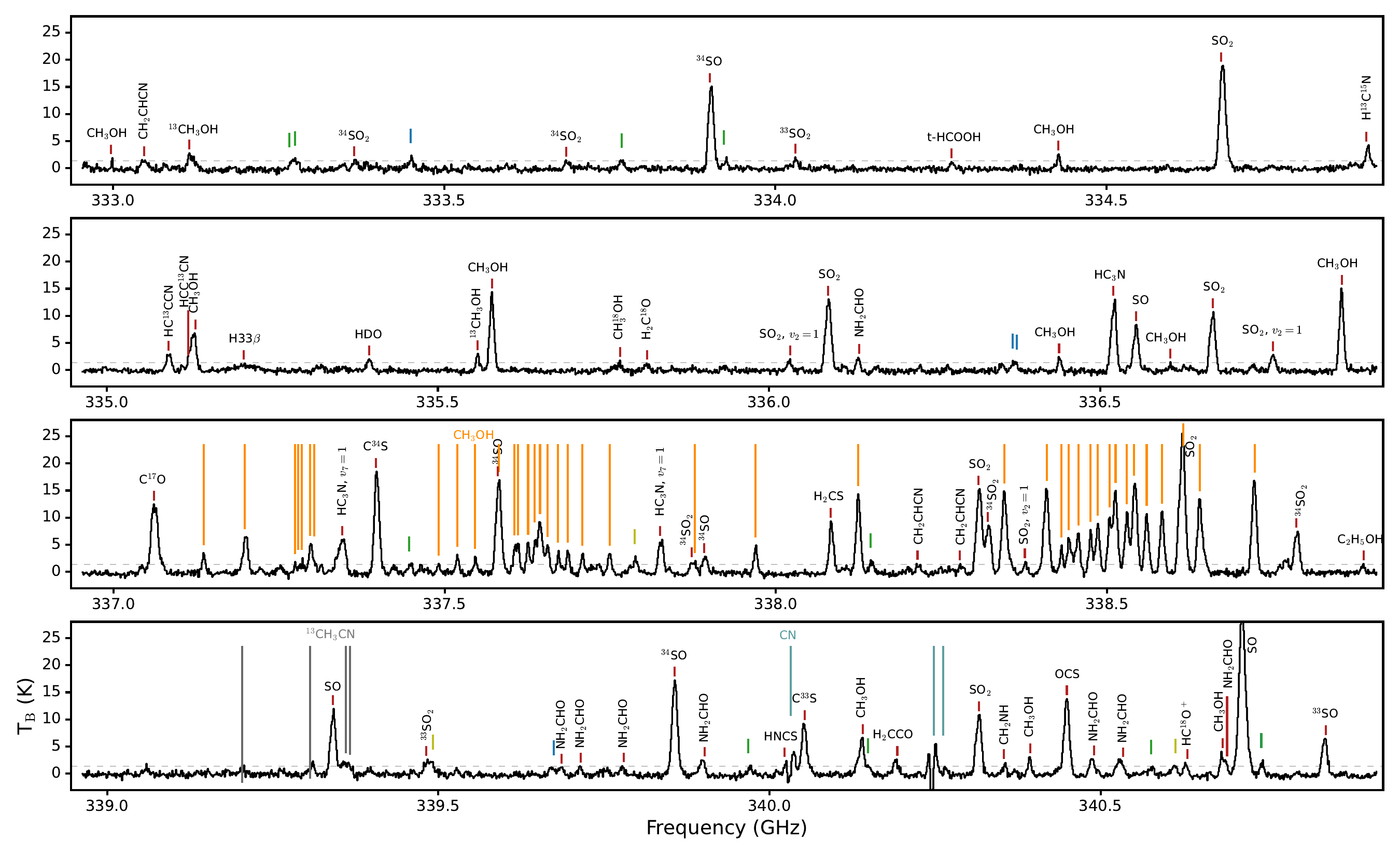} 
\caption{Zoomed-in and labeled SMA spectra of G20 from ${\approx}$333-341~GHz. The gray dashed line shows 3$\times$RMS.}
\label{fig:g20_rx345_ls}
\end{figure*}

\begin{figure*}[p!]
\centering
\includegraphics[width=.95\linewidth]{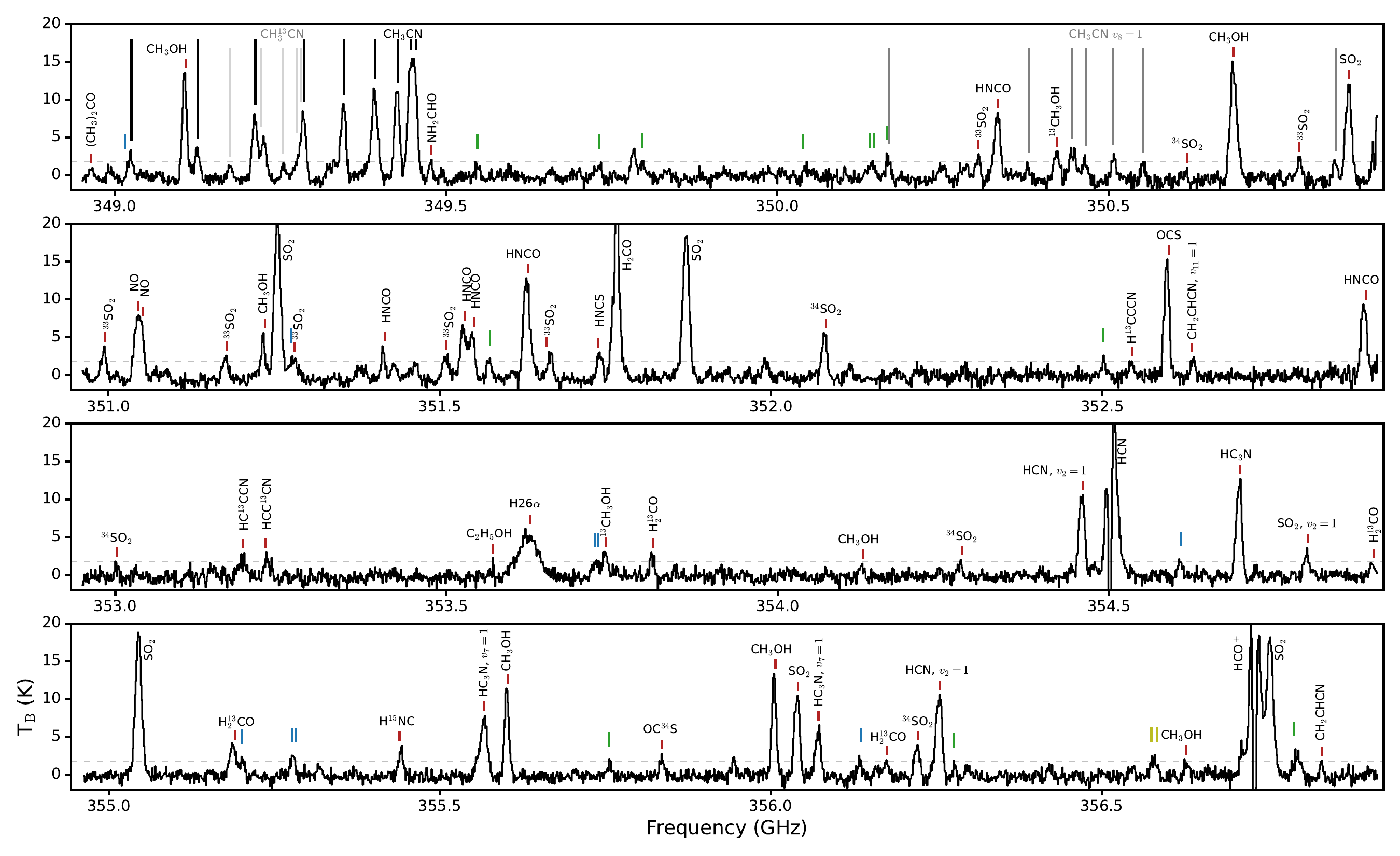} 
\caption{Zoomed-in SMA and labeled spectra of G20 from ${\approx}$349-357~GHz. The gray dashed line shows 3$\times$RMS.}
\label{fig:g20_rx345_us}
\end{figure*}

\begin{figure*}[p!]
\centering
\includegraphics[width=.95\linewidth]{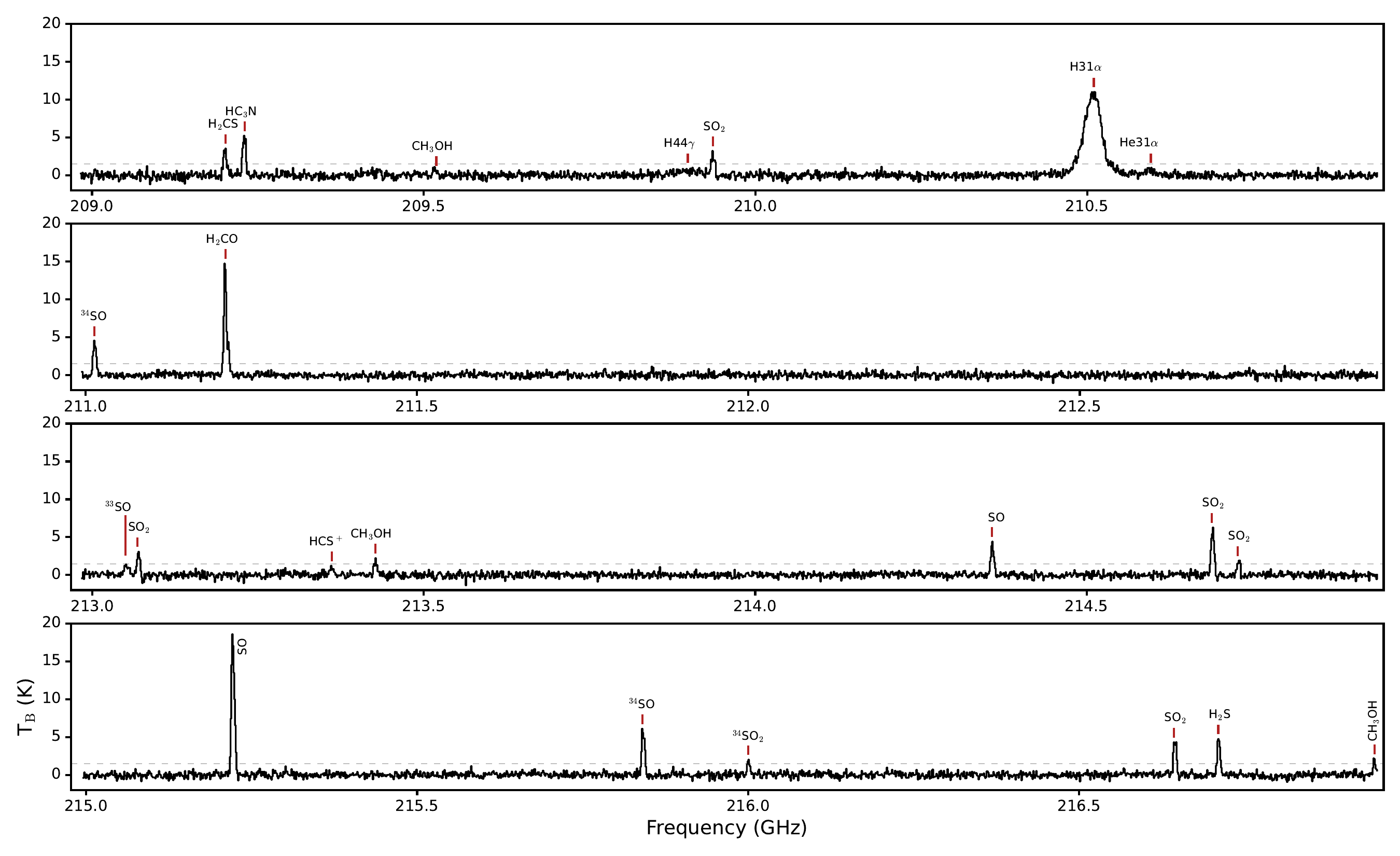} 
\caption{Zoomed-in and labeled SMA spectra of G35 from ${\approx}$209-217~GHz. The gray dashed line shows 3$\times$RMS.}
\label{fig:g35_rx240_ls}
\end{figure*}

\begin{figure*}[p!]
\centering
\includegraphics[width=.95\linewidth]{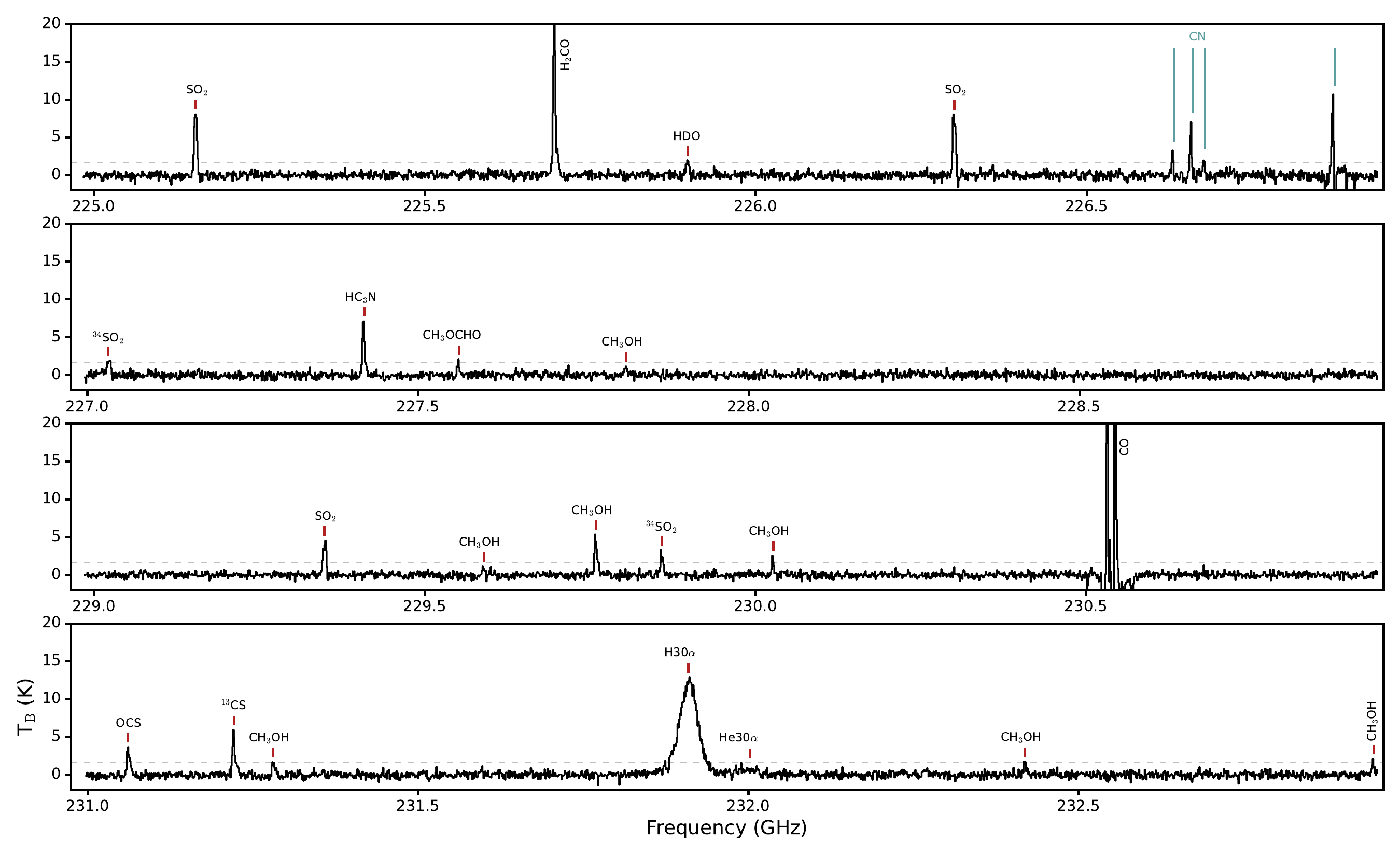} 
\caption{Zoomed-in and labeled SMA spectra of G35 from ${\approx}$225-233~GHz. The gray dashed line shows 3$\times$RMS.}
\label{fig:g35_rx240_us}
\end{figure*}

\begin{figure*}[p!]
\centering
\includegraphics[width=.95\linewidth]{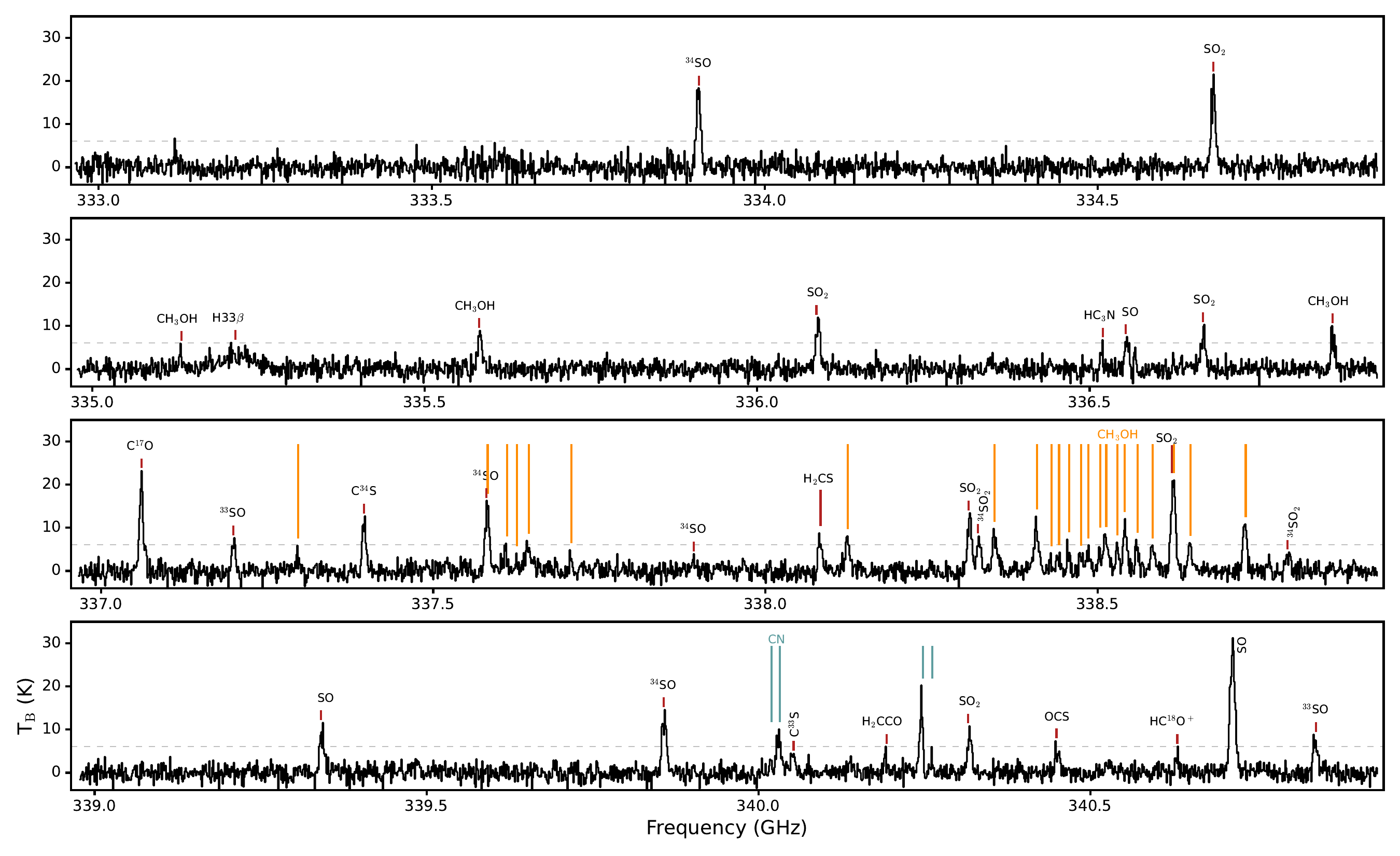}
\caption{Zoomed-in SMA and labeled spectra of G35 from ${\approx}$333-341~GHz. The gray dashed line shows 3$\times$RMS.}
\label{fig:g35_rx345_ls}
\end{figure*}

\begin{figure*}[p!]
\centering
\includegraphics[width=.95\linewidth]{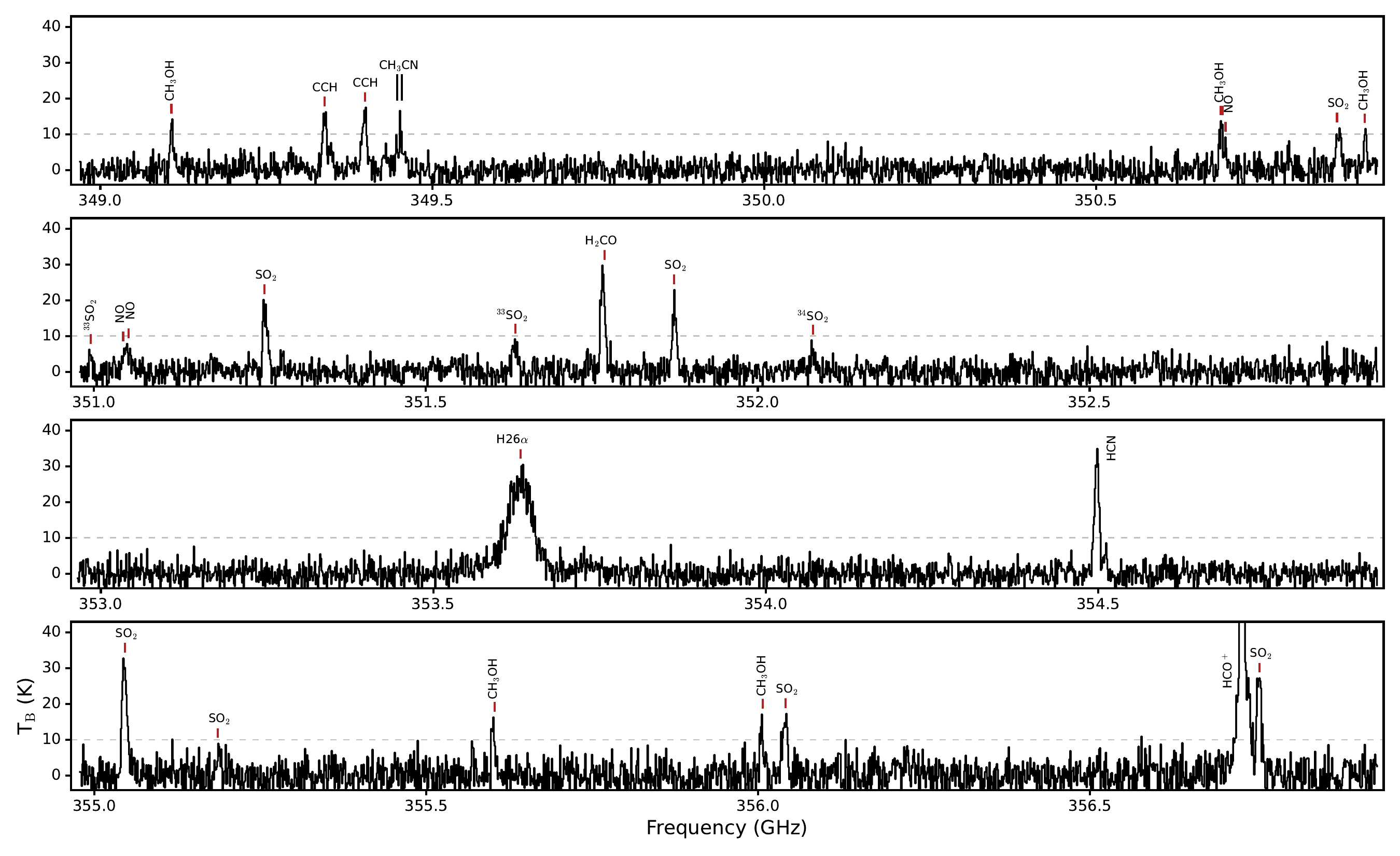}
\caption{Zoomed-in and labeled SMA spectra of G35 from ${\approx}$349-357~GHz. The gray dashed line shows 3$\times$RMS.}
\label{fig:g35_rx345_us}
\end{figure*}

\begin{figure*}[p!]
\centering
\includegraphics[width=.95\linewidth]{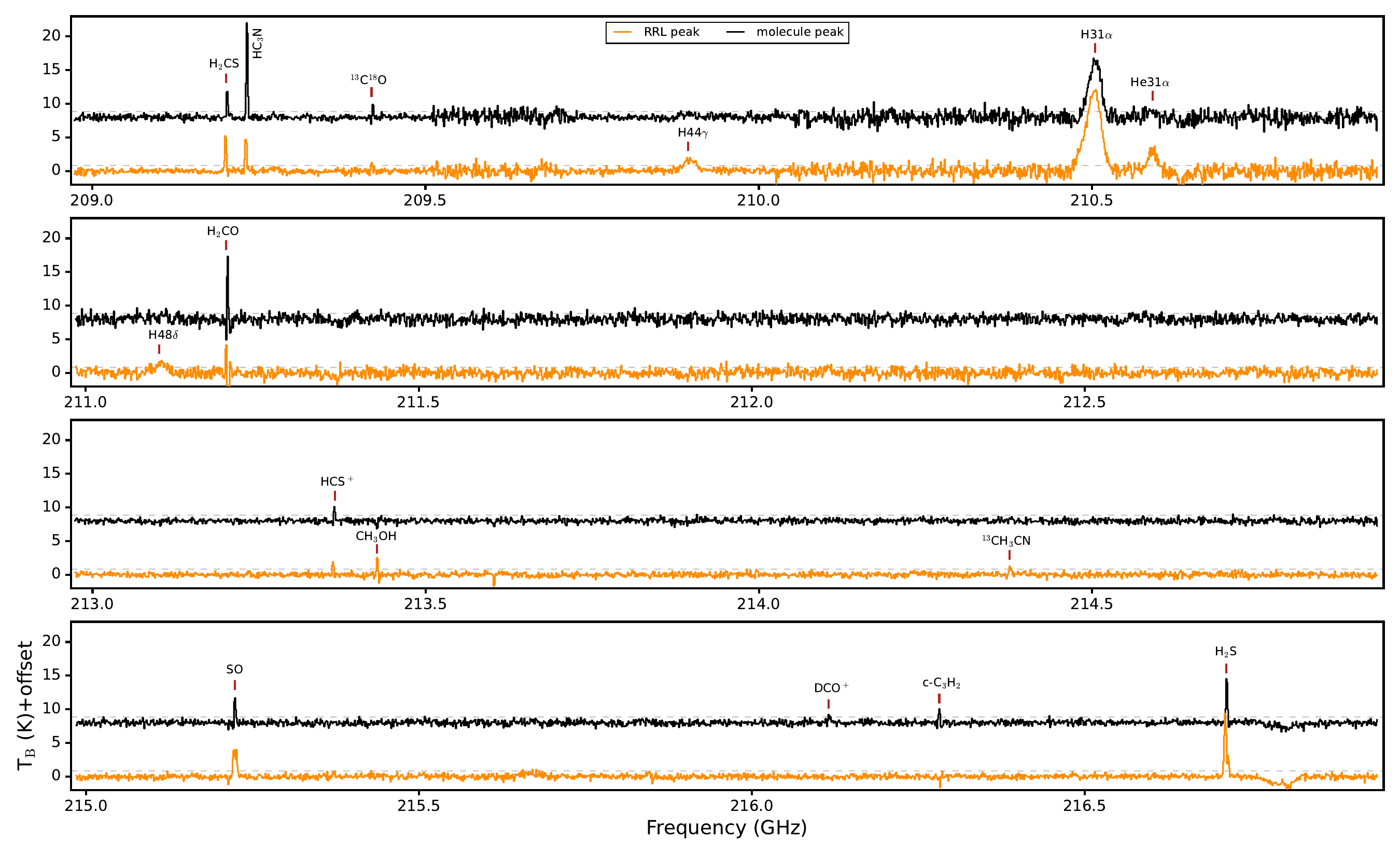} \vspace{-3mm}
\caption{Zoomed-in and labeled SMA spectra of W33 from ${\approx}$209-217~GHz. Spectra are vertically offset for visual clarity and the gray dashed lines shows 3$\times$RMS.}
\label{fig:w33_rx240_ls}
\end{figure*}

\begin{figure*}[p!]
\centering
\includegraphics[width=.95\linewidth]{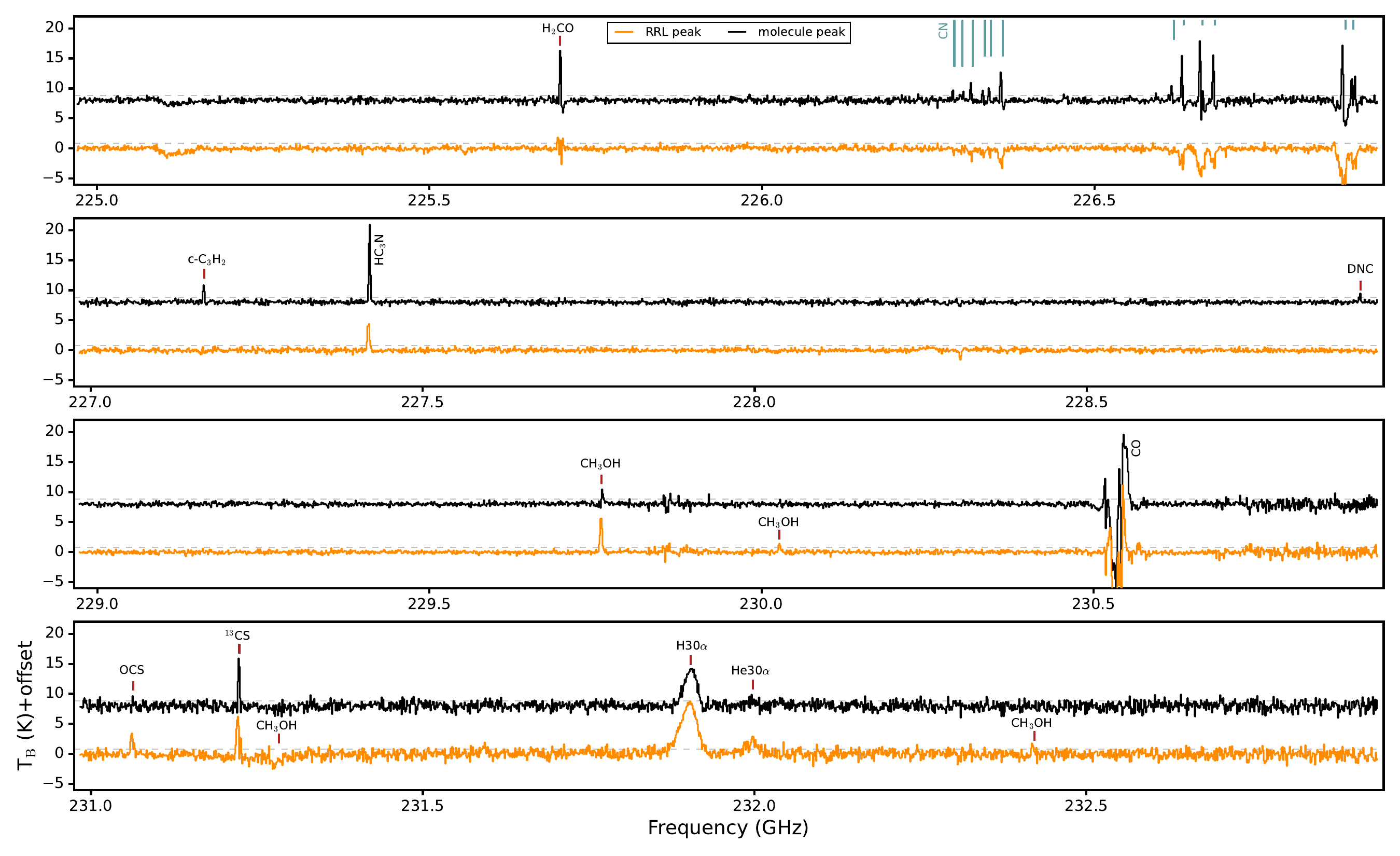} \vspace{-3mm}
\caption{Zoomed-in and labeled SMA spectra of W33 from ${\approx}$225-233~GHz. Spectra are vertically offset for visual clarity and the gray dashed lines shows 3$\times$RMS.}
\label{fig:w33_rx240_us}
\end{figure*}

\begin{figure*}[p!]
\centering
\includegraphics[width=.95\linewidth]{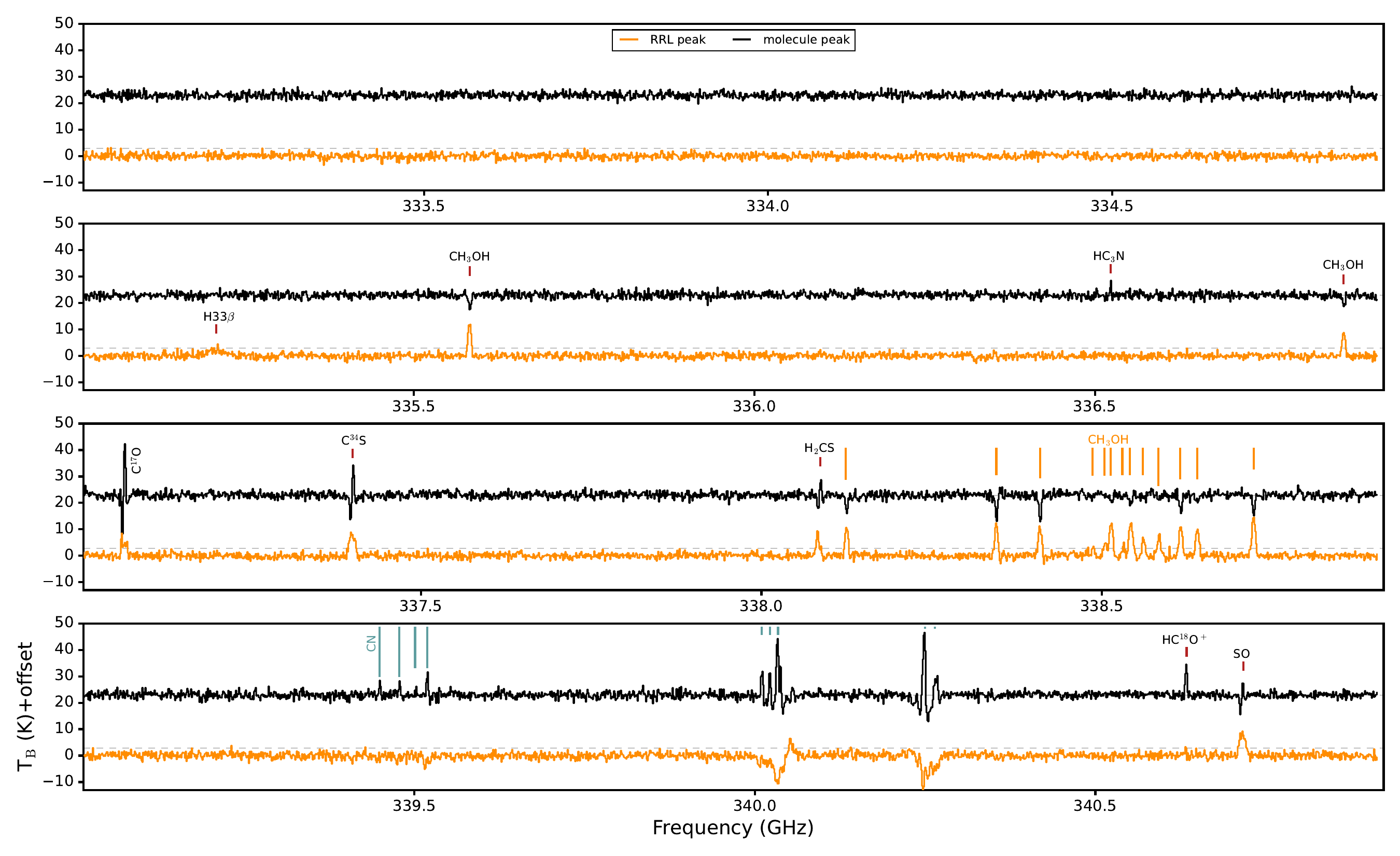} \vspace{-3mm}
\caption{Zoomed-in and labeled SMA spectra of W33 from ${\approx}$333-341~GHz. Spectra are vertically offset for visual clarity and the gray dashed lines shows 3$\times$RMS.}
\label{fig:w33_rx345_ls}
\end{figure*}

\begin{figure*}[p!]
\centering
\includegraphics[width=.95\linewidth]{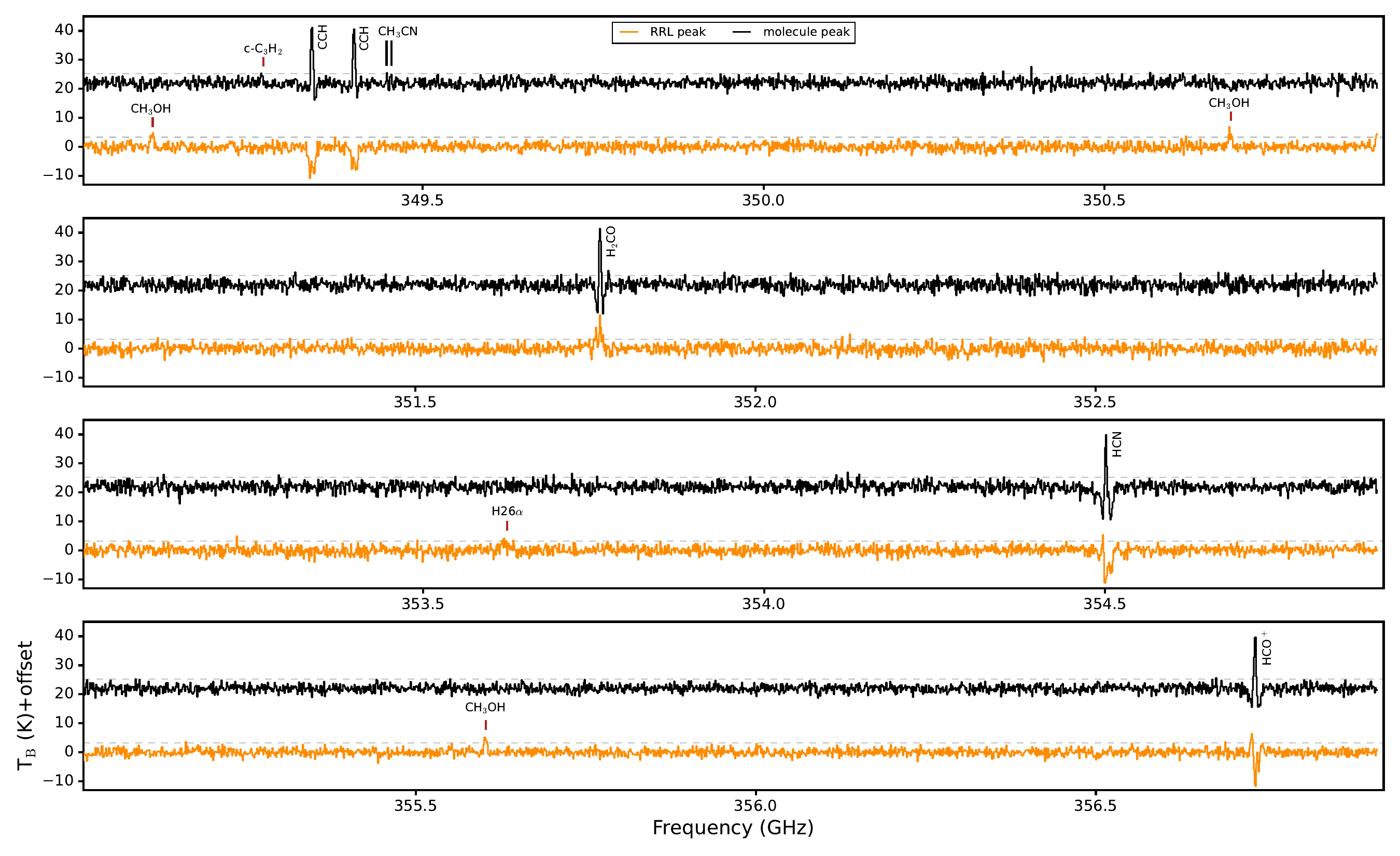}\vspace{-3mm}
\caption{Zoomed-in and labeled SMA spectra of W33 from ${\approx}$349-357~GHz. Spectra are vertically offset for visual clarity and the gray dashed lines shows 3$\times$RMS.}
\label{fig:w33_rx345_us}
\end{figure*}

The full spectra for each source labeled with all detected lines used in this work is shown in Figure Set \ref{fig:g28_rx240_ls} (Figures \ref{fig:g28_rx240_ls}-\ref{fig:w33_rx345_us}) The spectra for G28 and G20 are extracted at the continuum peak, while G35 shows the spectra of the East UC~\ion{H}{2} region. In W33, we show spectra extracted from two different positions corresponding to a molecule-rich region and the peak of the RRL emission, which is comparatively molecule-poor (see Figure \ref{fig:W33_peak_intensity}).

\section{Gallery of Zeroth Moment Maps} \label{sec:appendix_mom0_maps}

\begin{figure*}[h!]
\centering
\includegraphics[width=\linewidth]{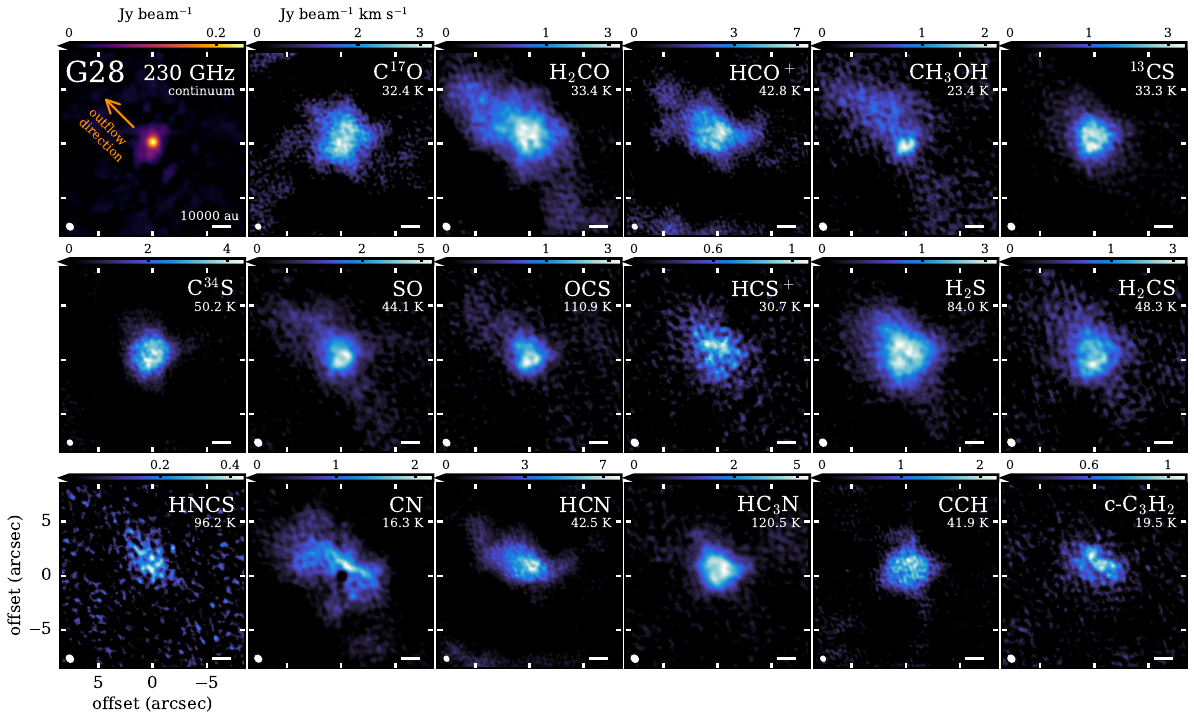}
\caption{Zeroth moment maps of representative lines of the spatially-extended molecules in G28. The upper state energy of each transition is labeled in the upper right. The synthesized beam and a scale bar indicating 10000~au is shown in the lower left and right corner, respectively, of each panel. The outflow direction, as traced by SiO emission \citep{Gorai23}, is marked in the continuum image.}
\label{fig:g28_mom0_gallery_extended}
\end{figure*}

\begin{figure*}[p!]
\centering
\includegraphics[width=\linewidth]{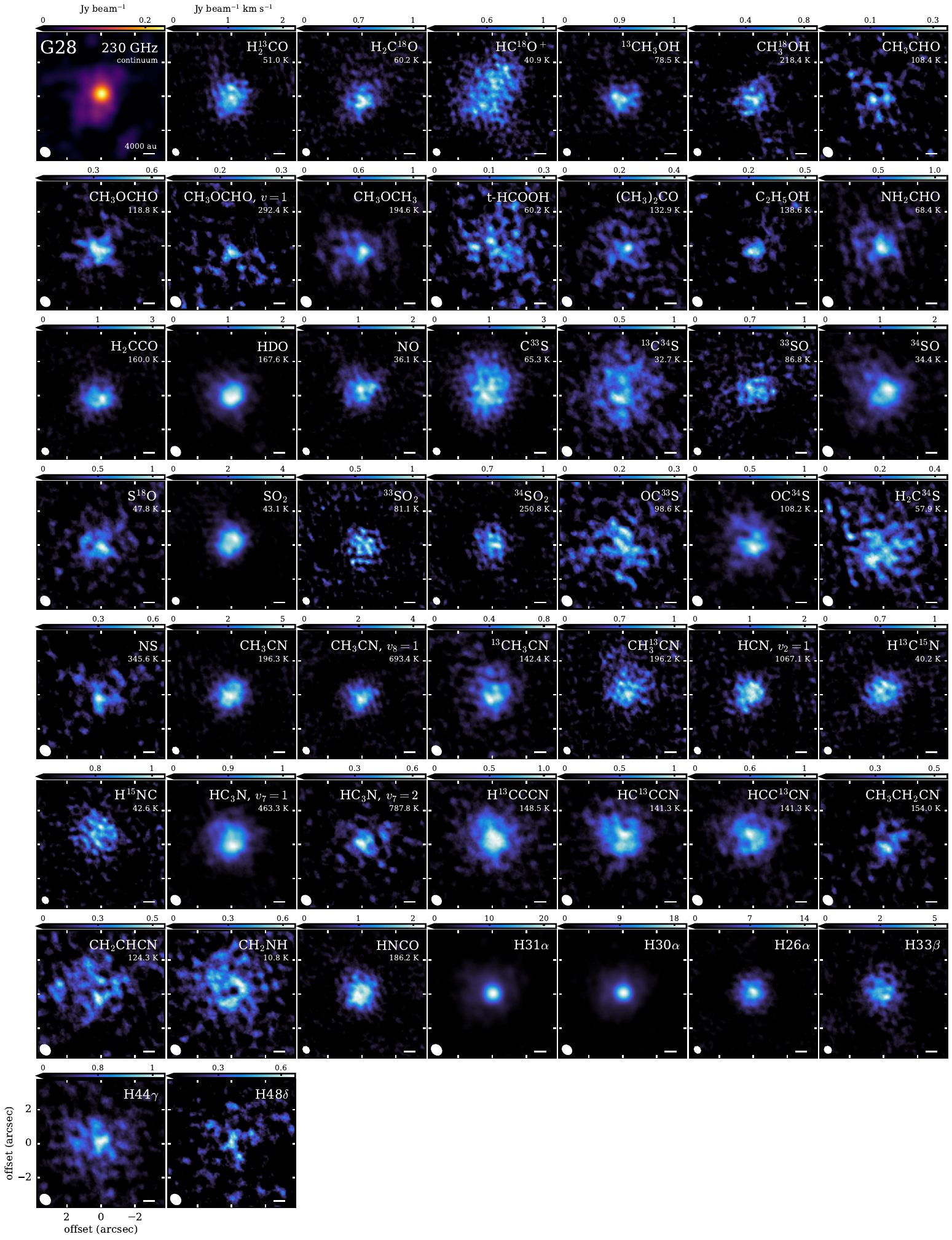}
\caption{Zeroth moment maps of representative lines of the spatially-compact molecules in G28. The synthesized beam and a scale bar indicating 4000~au is shown in the lower left and right corner, respectively, of each panel.}
\label{fig:g28_mom0_gallery_compact}
\end{figure*}

\begin{figure*}[p!]
\centering
\includegraphics[width=\linewidth]{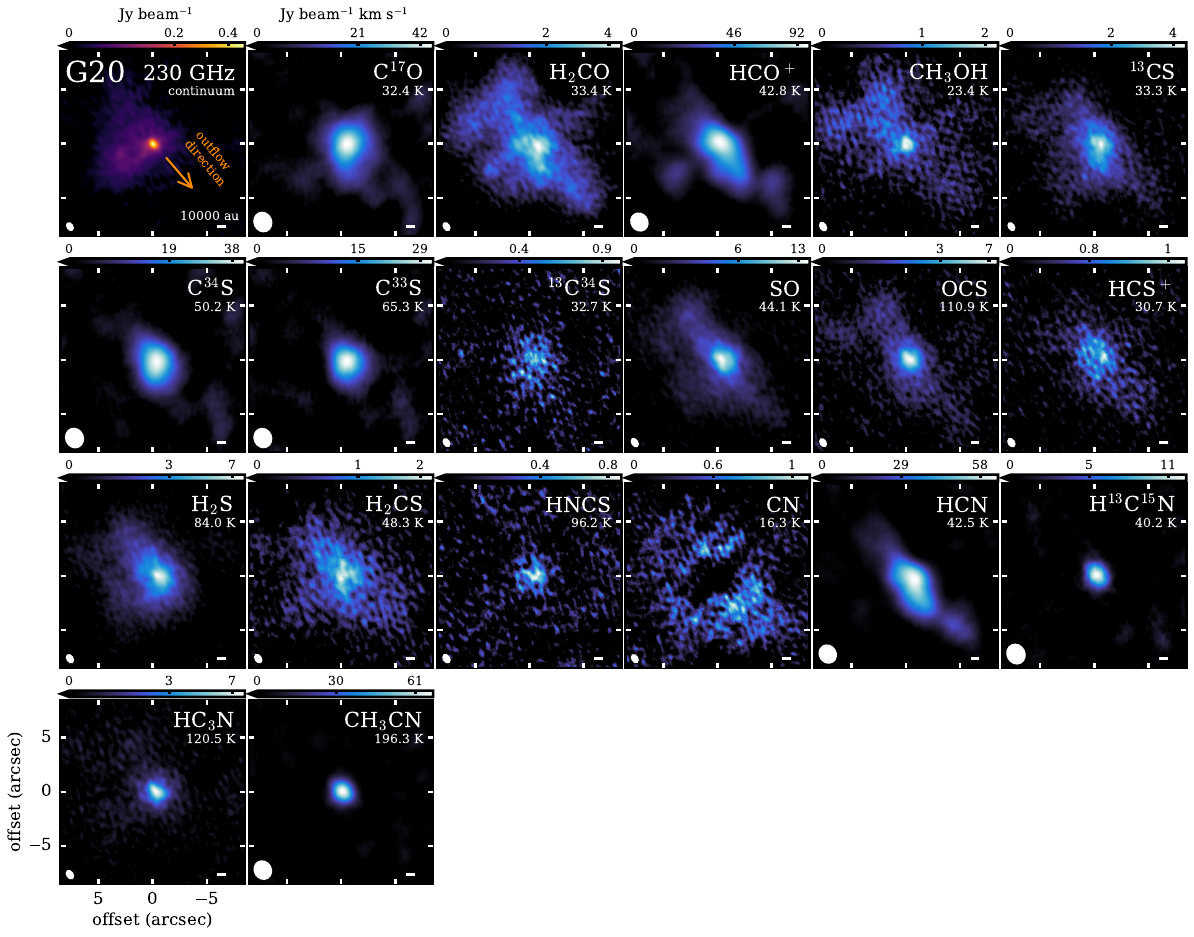}
\caption{Gallery of spatially-extended lines in G20. The outflow direction, as traced by SiO emission \citep{Xu13_G20}, is marked in the continuum image. Otherwise, as in Figure \ref{fig:g28_mom0_gallery_extended}.}
\label{fig:g20_mom0_gallery_extended}
\end{figure*}

\begin{figure*}[p!]
\centering
\includegraphics[width=\linewidth]{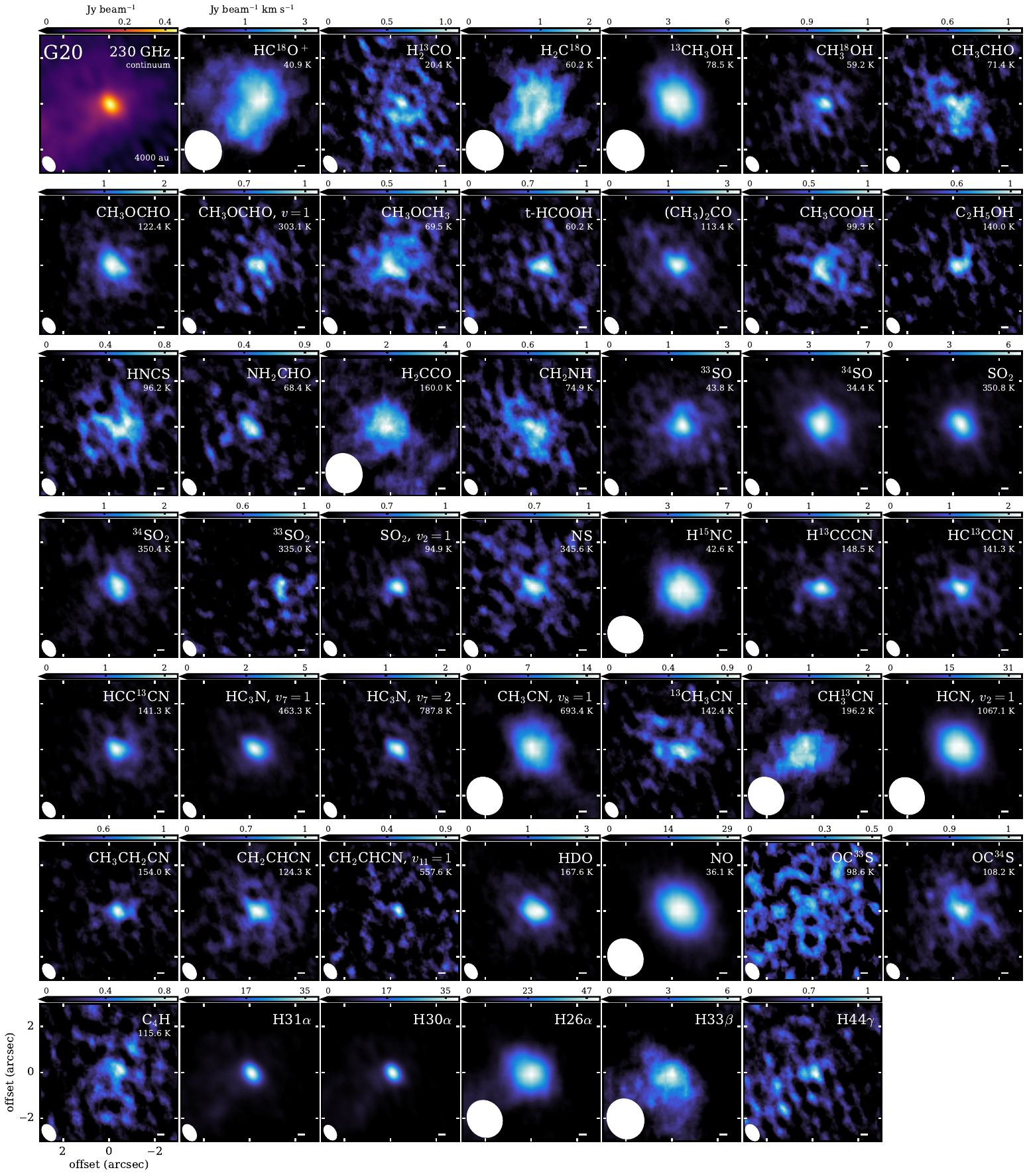}
\caption{Gallery of spatially-compact lines in G20. Otherwise, as in Figure \ref{fig:g28_mom0_gallery_compact}.}
\label{fig:g20_mom0_gallery_compact}
\end{figure*}

\begin{figure*}[h!]
\centering
\includegraphics[width=\linewidth]{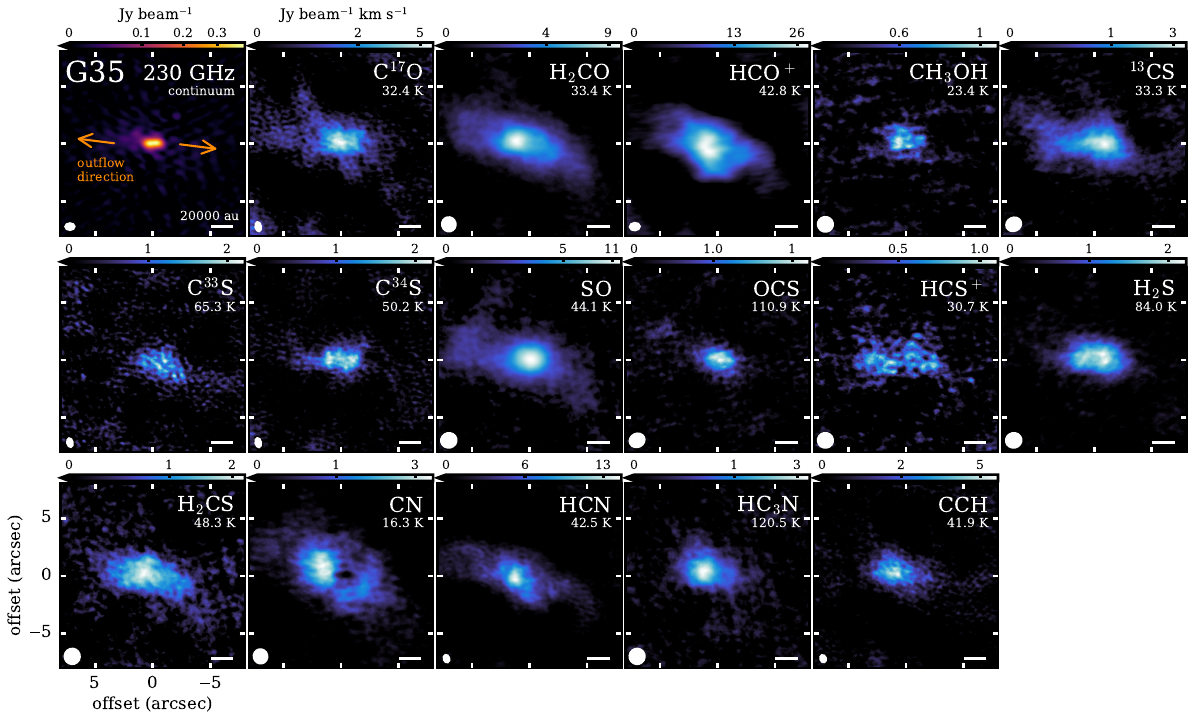}
\caption{Gallery of spatially-extended lines in G35. The outflow direction, as traced by high-velocity $^{12}$CO emission \citep{Zhang14}, is marked in the continuum image. Otherwise, as in Figure \ref{fig:g28_mom0_gallery_extended}.}
\label{fig:g35_mom0_gallery_extended}
\end{figure*}

\begin{figure*}[b!]
\centering
\includegraphics[width=\linewidth]{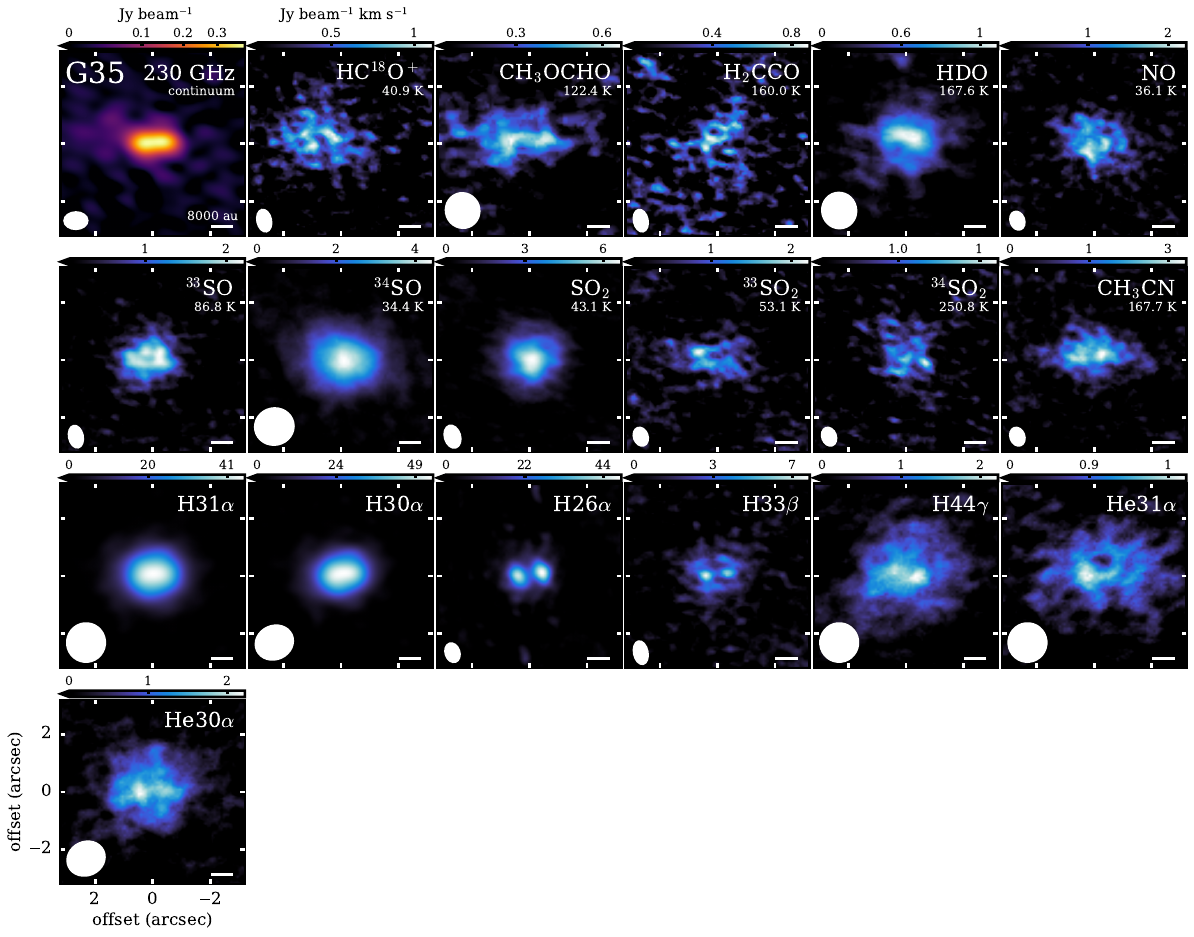}
\caption{Gallery of spatially-compact lines in G35. Otherwise, as in Figure \ref{fig:g28_mom0_gallery_compact}.}
\label{fig:g35_mom0_gallery_compact}
\end{figure*}

We provide zeroth moment map galleries of all detected molecules toward G28 (Figures \ref{fig:g28_mom0_gallery_extended}-\ref{fig:g28_mom0_gallery_compact}), G20 (Figures \ref{fig:g20_mom0_gallery_extended}-\ref{fig:g20_mom0_gallery_compact}), and G35 (Figures \ref{fig:g35_mom0_gallery_extended}-\ref{fig:g35_mom0_gallery_compact}). For ease of visualization, we categorize each molecule as either extended or spatially-compact. In the case of G28, the ring-like structure, which manifests on intermediate scales, is apparent in molecules in both categories.

\section{Spectroscopic Data} \label{sec:spec_data}

Table \ref{tab:spec_data_table} shows a complete set of spectroscopic details for the molecules included in this study. Care was taken to select, when possible, spectroscopic data for partition functions that included any non-negligible contributions from vibrational and torsional states relevant at the high gas temperatures of massive star-forming regions. However, this was not possible for all species and thus, we applied a range of temperature-dependent corrections (C$_{\rm{vib}}$) to the derived column densities of the following molecules:

The spectroscopic data for C$_2$H$_5$OH is from CDMS and the vibrational correction factor is 2.8 at 300~K \citep{Durig75}.

The spectroscopic data for NH$_2$CHO is taken from CDMS and the vibrational correction factor is 1.5 for 300~K \citep{Kryvda09, Motiyenko12}.

The spectroscopic data for OCS, OC$^{34}$S, and OC$^{33}$S were obtained from CDMS. For OCS, the vibrational correction factor is 1.2 at 300~K \citep{Saupe96, Murtz00}. We adopt the same C$_{\rm{vib}}$ for OC$^{34}$S, and OC$^{33}$S.

The spectroscopic data for CH$_3$CN, $^{13}$CH$_3$CN, and CH$_3^{13}$CN are from CDMS. While the partition function of CH$_3$CN includes vibrational contributions, those of the isotopologues do not. Thus, we adopt the vibrational correction factor of CH$_3$CN for $^{13}$CH$_3$CN, and CH$_3^{13}$CN, which is 1.5 at 300~K \citep{Bocquet88, Tolonen93}.

The spectroscopic data for H$^{15}$NC is taken from JPL and we adopt the same vibrational correction factor of HNC, which is 1.5 at 300~K \citep{Burkholder87, Maki01}.

The spectroscopic data for HC$_3$N, H$^{13}$CCCN, HC$^{13}$CCN, and HCC$^{13}$CN are taken from CDMS. The vibrational correction factor for HC$_3$N is 2.6 at 300~K \citep{Mallinson76}. We adopt the same C$_{\rm{vib}}$ for H$^{13}$CCCN, HC$^{13}$CCN, and HCC$^{13}$CN.

The spectroscopic data for CH$_3$CH$_2$CN was  obtained from CDMS. For CH$_3$CH$_2$CN, the vibrational correction factor ranges 1.1 at 100~K to 3.5 at 300~K \citet{Heise81}.

\begin{deluxetable*}{llllc}[ht!]
\tabletypesize{\scriptsize}
\tablecaption{References for Spectroscopic Information Used in this Work\label{tab:spec_data_table}}
\tablehead{
\colhead{Name} & \colhead{Molecule} & \colhead{Catalog}  & \colhead{References} }
\startdata
Carbon monoxide & C$^{17}$O & CDMS & \citealt{Klapper03} \\
Formaldehyde & H$_2$CO & CDMS & \citealt{Bocquet96, Mendoza23}  \\
 & H$_2^{13}$CO & CDMS & \citealt{Muller00_H213CO, Mendoza23}   \\
 & H$_2$C$^{18}$O & CDMS & \citealt{Muller00_H2C18O, Mendoza23}   \\
Formylium & HCO$^+$ & CDMS & \citealt{Tinti07}  \\
 & HC$^{18}$O$^+$ & CDMS &  \citealt{Schmid04}  \\
Methanol & CH$_3$OH & CDMS & \citealt{Xu08_CH3OH}  \\
 & $^{13}$CH$_3$OH & CDMS & \citealt{Xu97_CH3OH}  \\
 & CH$_3^{18}$OH & CDMS & \citealt{Fisher07_CH3OH}  \\
Acetaldehyde & CH$_3$CHO & JPL & \citealt{Kleiner96}  \\
Methyl formate & CH$_3$OCHO & JPL & \citealt{Ilyushin09, Carvajal19} \\
 & CH$_3$OCHO, $v=1$ & JPL & \citealt{Ilyushin09}  \\
Dimethyl ether & CH$_3$OCH$_3$ & CDMS & \citealt{Endres09}  \\
Formic acid & t-HCOOH & CDMS & \citealt{Winnewisser02}   \\
Acetone & (CH$_3$)$_2$CO & JPL & \citealt{Groner02}  \\
Acetic acid & CH$_3$COOH & CDMS & \citealt{Ilyushin13_acetic_acid, Xue19}   \\
Ethanol & C$_2$H$_5$OH & CDMS & \citealt{Pearson08, Muller16}  \\
Formamide & NH$_2$CHO & CDMS & \citealt{Hirota74, Motiyenko12}  \\
Ketene & H$_2$CCO & CDMS & \citealt{Brown90, Johns92}  \\
Water & HDO & JPL & \citealt{Messer84}  \\
Nitric oxide & NO & CDMS & \citealt{Muller15_NO}   \\
Carbon monosulfide & $^{13}$CS & CDMS & \citealt{Bogey82_CS}   \\
 & C$^{34}$S & CDMS & \citealt{Gottlieb03}   \\
 & C$^{33}$S & CDMS & \citealt{Bogey81_CS_iso}   \\
 & $^{13}$C$^{34}$S & CDMS & \citealt{Bogey81_CS_iso}   \\
Sulfur monoxide & SO & CDMS & \citealt{Tiemann82, Bogey97_SO}  \\
 & S$^{18}$O & CDMS & \citealt{Tiemann74}  \\
 & $^{34}$SO & CDMS & \citealt{Tiemann74, Klaus96}  \\
 & $^{33}$SO & CDMS & \citealt{Klaus96}  \\
Sulfur dioxide & SO$_2$ & CDMS & \citealt{Alekseev96, Muller05_SO2}  \\
 & SO$_2$, $v_2 = 1$ & CDMS & \citealt{Alekseev96, Muller05_SO2}  \\
 & $^{34}$SO$_2$ & CDMS  & \citealt{Alekseev96, Belov98}  \\
 & $^{33}$SO$_2$ & CDMS & \citealt{Muller00_SO2}  \\
Carbonyl sulfide & OCS & CDMS & \citealt{Dubrulle80,Golubiatnikov05} \\
 & OC$^{34}$S & CDMS & \citealt{Dubrulle80,Vanek89}  \\
 & OC$^{33}$S & CDMS & \citealt{Dubrulle80,Burenin81}  \\
Thioformylium & HCS$^+$ & CDMS & \citealt{Margules03}  \\
Hydrogen sulfide & H$_2$S & CDMS & \citealt{Belov95}  \\
Thioformaldehyde & H$_2$CS & CDMS & \citealt{Beers72, Maeda08}  \\
 & H$_2$C$^{34}$S & CDMS & \citealt{Mcnaughton93}  \\
Nitric sulfide & NS & CDMS & \citealt{Lee95}  \\
Isothiocyanic acid & HNCS & CDMS & \citealt{Niedenhoff95_HNCS}  \\
Methyl cyanide & CH$_3$CN & CDMS & \citealt{Muller09_CH3CN}   \\
 & CH$_3$CN, $v_8=1$ & JPL & \citealt{Bocquet88}  \\
 & $^{13}$CH$_3$CN & CDMS & \citealt{Muller09_CH3CN} \\
 & CH$_3^{13}$CN & CDMS & \citealt{Muller09_CH3CN}  \\
Hydrogen cyanide & HCN, $v_2=1$ & CDMS & \citealt{Thorwirth03_HCN_vib}  \\ 
 & H$^{13}$C$^{15}$N & CDMS & \citealt{Fuchs04}  \\ 
Hydrogen isocyanide & H$^{15}$NC & JPL & \citealt{Creswell76, Pearson76}  \\
Cyanoacetylene & HC$_3$N & CDMS & \citealt{Thorwirth00}  \\
 & HC$_3$N, $v_7=1$ & CDMS & \citealt{Yamada86,Thorwirth00}  \\
 & HC$_3$N, $v_7=2$ & CDMS & \citealt{Yamada86,Thorwirth00}  \\
 & H$^{13}$CCCN & CDMS & \citealt{Thorwirth01}  \\
 & HC$^{13}$CCN & CDMS & \citealt{Thorwirth01}  \\
 & HCC$^{13}$CN & CDMS & \citealt{Thorwirth01}  \\
Ethyl cyanide & CH$_3$CH$_2$CN & CDMS & \citealt{Brauer09}  \\
Vinyl cyanide & CH$_2$CHCN & CDMS & \citealt{Muller08_CH2CHCN}  \\
 & CH$_2$CHCN, $v_{11}$=1 & CDMS & \citealt{Muller08_CH2CHCN}  \\
Methanimine & CH$_2$NH & CDMS & \citealt{Dore12}  \\
Isocyanic acid & HNCO & CDMS & \citealt{Lapinov07, Carvajal19}  \\
Ethynyl radical & CCH & CDMS & \citealt{Muller00_CCH,Padovani09}  \\
Cyclopropenylidene & c-C$_3$H$_2$ & CDMS & \citealt{Bogey86_C3H2, Bogey87}  \\
Butadiynyl & C$_4$H & CDMS & \citealt{Guelin82,Gottlieb83_C4H} \\
\enddata
\tablecomments{All spectroscopic data were obtained from the JPL \citep{Pickett98} and CDMS \citep{Muller01, Muller05, Endres16} catalogs.}
\end{deluxetable*}

\section{Isotopic Ratios Used in This Work} \label{sec:isotopic_ratio_used_appendix}

When available, we infer column densities from minor isotopologues to mitigate potential high line optical depths of the main isotopologues. To do so, we use the following isotope ratios for $^{12}$C/$^{13}$C \citep{Milam05}, $^{16}$O/$^{18}$O \citep{Wilson94}, $^{32}$S/$^{33}$S \citep{Yan23}, and $^{32}$S/$^{34}$S \citep{Yu20}:

\begin{equation} \label{eqn:eqn1}
    (^{12}\rm{C} / ^{13}\rm{C}) = (6.21 \pm 1.00) \rm{D}_{\rm{GC}} + (18.71 \pm 7.37)
\end{equation}

\begin{equation}
    (^{16}\rm{O} / ^{18}\rm{O}) = (58.8 \pm 11.8) \rm{D}_{\rm{GC}} + (37.1 \pm 82.6)
\end{equation}

\begin{equation}
    (^{32}\rm{S}/^{33}\rm{S}) = (2.64 \pm 0.77)\rm{D}_{\rm{GC}} + (70.80 \pm 5.57)
\end{equation}

\begin{equation} \label{eqn:eqn4}
    (^{32}\rm{S}/^{34}\rm{S}) = (1.56 \pm 0.17)\rm{D}_{\rm{GC}} + (6.75 \pm 1.22).
\end{equation}

\noindent Table \ref{tab:isotopic_prop} lists the adopted isotopic ratios for each of our sources.

\clearpage

\begin{deluxetable*}{cccccc}[h]
\tablecaption{Assumed Isotopic Ratios\label{tab:isotopic_prop}}
\tablewidth{0pt}
\tablehead{
\colhead{Source} &  \colhead{D$_{\rm{GC}}$} & \colhead{$^{12}$C/$^{13}$C} & \colhead{$^{16}$O/$^{18}$O} & \colhead{$^{32}$S/$^{33}$S} & \colhead{$^{32}$S/$^{34}$S}   \\ 
\colhead{} & \colhead{(kpc)}  &  & 
}
\startdata
G28.20-0.05    & 4.6 &  47~$\pm$~5 & 308~$\pm$~54 & 83~$\pm$~4 & 14~$\pm$~1 \\
G20.08-0.14~N  & 5.3 &  52~$\pm$~5 & 349~$\pm$~63 & 85~$\pm$~4 & 15~$\pm$~1 \\
G35.58-0.03    & 6.0 &  56~$\pm$~6 & 390~$\pm$~71 & 87~$\pm$~5 & 16~$\pm$~1\\
\enddata
\tablecomments{Following the approach of \citet{Nazari22_NCOMS, Chen23}, we do not include the intercept error in the error propagation, as only the slope in Equations \ref{eqn:eqn1}-\ref{eqn:eqn4} contains the information of the trend between the isotopic ratio and D$_{\rm{GC}}$.}
\end{deluxetable*}


\bibliography{SMA_MYSO}{}
\bibliographystyle{aasjournal}



\end{document}